\documentclass[11pt]{article}
\usepackage{latexsym,amsmath,amsfonts,amssymb}
\usepackage[american]{babel}
\usepackage{color}
\usepackage[unicode]{hyperref}
\usepackage{enumerate}
\usepackage{subcaption}
\DeclareMathOperator{\arccot}{arccot}

\usepackage{graphicx}
\usepackage{amsmath,euscript,array,mathrsfs,amsfonts}
\usepackage{hyperref}
\hypersetup{
colorlinks = true,
linkcolor = red,
linktocpage = true,
citecolor = blue
}

\usepackage{verbatim}

\makeatletter\def\l@subsubsection#1#2{}%
\makeatother

\renewcommand\theequation{\arabic{section}.\arabic{equation}} 

\newcommand{\beqs}{\begin{eqnarray}}
\newcommand{\eeqs}{\end{eqnarray}}
\newcommand{\bal}{\begin{aligned}}
\newcommand{\eal}{\end{aligned}}
\makeatletter
\newcommand\setItemnumber[1]{\setcounter{enum\romannumeral\@enumdepth}{\numexpr#1-1\relax}}
\makeatother
\begin{document}
\baselineskip=15.5pt
\pagestyle{plain}
\setcounter{page}{1}



\def\andrea#1{\textcolor{green}{#1}}
\def\aldo#1{\textcolor{blue}{#1}}
\def\francesco#1{\textcolor{red}{#1}}

\newcommand{\ber}{\begin{eqnarray}}
\newcommand{\eer}{\end{eqnarray}}



\newfont{\namefont}{cmr10}
\newfont{\addfont}{cmti7 scaled 1440}
\newfont{\boldmathfont}{cmbx10}
\newfont{\headfontb}{cmbx10 scaled 1728}

\numberwithin{equation}{section}



%

%
\setcounter{footnote}{0}
\renewcommand{\theequation}{{\rm\thesection.\arabic{equation}}}

\begin{titlepage}

\begin{center}

\vskip .5in 
\noindent

{\Large \bf{Cosmic Topological Defects from Holography} }
\bigskip\medskip

Francesco Bigazzi$^a$\footnote{bigazzi@fi.infn.it}, Aldo L. Cotrone$^{a,b}$\footnote{cotrone@fi.infn.it} and Andrea Olzi$^{a,b}$\footnote{andrea.olzi@unifi.it}  \\

\bigskip\medskip
{\small 
$^a$INFN, Sezione di Firenze, Via G. Sansone 1, I-50019 Sesto Fiorentino (Firenze), Italy.
\\
$^b$ Dipartimento di Fisica e Astronomia, Universit\'a di Firenze, Via G. Sansone 1, I-50019 Sesto Fiorentino (Firenze), Italy.
}

\vskip .5cm 
\vskip .9cm 
     	{\bf Abstract }\vskip .1in
\end{center}

\noindent
This work investigates cosmic topological defects in gauge theories, focusing on models with an $SU(N)$ gauge group coupled with a single flavor, explored through a holographic framework. At low energies, the effective theory is described by an axion-like particle resulting from the spontaneous breaking of the axial $U(1)_A$ flavor symmetry. As the Universe cools below a critical temperature, the chiral symmetry is broken, and non-trivial vacuum configurations form, resulting in the creation of cosmic strings and domain walls. We provide a UV description of these defects in a particular holographic theory, the Witten-Sakai-Sugimoto model, as probe D6-branes. We show the presence of a first-order phase transition separating string loop from domain wall solutions. String loops charged under the baryon symmetry and with angular momentum - vortons - can be understood as excitations of a topological phase of matter given by a Chern-Simons theory living on the D6-brane world volume. Finally, we provide an effective description of string loops and vortons in terms of degrees of freedom living on the flavor brane, \textit{i.e.}~mesonic modes. 

 \noindent
\vskip .5cm
\vskip .5cm
\vfill
\eject

\end{titlepage}

\setcounter{footnote}{0}

\small{
\tableofcontents}

\normalsize

\newpage
\renewcommand{\theequation}{{\rm\thesection.\arabic{equation}}}
%
\section{Introduction}
Topological defects in gauge theories often arise as remnants of phase transitions in the early Universe \cite{Kibble:1976sj,Vilenkin:1984ib,Vachaspati:1984dz,Vachaspati:1986cc,Carter:1993wu,Vilenkin:2000jqa}. As a playground theory, consider a hidden sector composed by an
$SU(N)$ gauge theory coupled with a single massless flavor. At large $N$ and low energies, this theory can be effectively described by an axion,\footnote{This particle could also serve as an axion or an axion-like particle (ALP) for the Standard Model. Nevertheless, in this paper we do not take into account interactions with the visible sector.} which is the pseudo-Goldstone boson of the spontaneously broken axial $U(1)_A$ flavor symmetry,  as it happens in Kim-Shifman-Vainshtein-Zakharov-like axionic models \cite{Kim:1979if,Shifman:1979if}. 
This scenario and its variants have been widely studied in cosmology due to their rich phenomenological implications. In particular, the cosmological evolution of these theories often leads to the formation of various topological defects, such as cosmic strings and domain walls (DWs). 
A key feature of the model we consider is the presence of a residual vector $U(1)_V$ symmetry, associated with the baryonic charge of the single massless flavor. 

In holographic QCD models or their dark sector analogs, these defects typically emerge from the spontaneous breaking of symmetries such as chiral or axial symmetries. In our context, the phase transition responsible for generating topological defects is closely tied to the breaking of chiral symmetry. As the Universe cools below a critical temperature, non-trivial vacuum configurations form, resulting in the creation of the above-mentioned cosmic strings and domain walls.

The Witten-Sakai-Sugimoto (WSS) model \cite{Witten:1998zw,Sakai:2004cn,Sakai:2005yt} offers a concrete holographic framework to study such topological defects. At low energies, the WSS model describes a non-supersymmetric $SU(N)$ gauge theory in four dimensions coupled with $N_f$ fundamental flavors, and a tower of massive Kaluza-Klein adjoint fields. The Yang-Mills plus adjoint field sector is realized as the low energy theory of a stack of $N$ D4-branes wrapped on a circle. At strong coupling and in the planar limit, it is holographically described by a dual Type IIA supergravity theory on a specific background, which accounts for confinement and mass gap formation in the dual field theory. Fields in the fundamental representation, \textit{i.e.}~quarks, are obtained by introducing $N_f$ pairs of D8/$\overline{\text{D}8}$-branes, which, in the $N_f\ll N$ limit, can be treated as probes on the above-mentioned background. To minimize their energy, these branes form a unique U-shaped configuration, geometrically realizing the spontaneous breaking of chiral symmetry. 

To mimic a single flavor $SU(N)$ theory in a cosmological scenario, we study the $N_f=1$ WSS model at finite temperature, in its deconfined phase, which is characterized by significant changes in the background geometry, notably the appearance of a horizon. This phase occurs at high temperatures, analogously to the deconfinement transitions observed in Yang-Mills theories. In this phase, the WSS model can be seen as effectively describing a non-supersymmetric $SU(N)$ dark gauge theory coupled with a dark flavor sector, where the gauge fields and flavor degrees of freedom interact in a high-temperature regime. This dark flavor sector plays a crucial role in the study of topological defects, which form when the axial symmetry is spontaneously broken. 
In this model, the scale of condensation of the extra flavor is set by the strongly coupled version of a non-local quartic Nambu-Jona-Lasinio interaction \cite{Antonyan:2006vw}.
Thus, it can have a chiral symmetry-breaking phase in the deconfined regime.

In \cite{Bigazzi:2022ylj} the straight axionic string of the WSS model in the deconfined phase was described as a probe D6-brane wrapped on a four-cycle $S^4$ of the background geometry. 
The effective four-dimensional action for this string can be derived from the Dirac-Born-Infeld (DBI) action of the D8-branes, whose gauge fields are sourced by the D6-brane. This construction provides a physical interpretation for the otherwise problematic short-distance divergence in the effective action of global cosmic strings: the D6-brane, acting as a hard core, regularizes the small-distance behavior of the string.

In a realistic cosmological scenario, axionic strings can break and form string loops. Moreover, when DWs are formed, they can end on cosmic strings. In this work, we study these objects in detail as described as probe D6-branes, wrapped on $S^4$, with a circular boundary on the flavor branes. In particular, we find that at fixed temperature the string loop and the DW solutions are separated by a first-order transition as their size is varied.\footnote{As we will see, these are not equilibrium configurations, so the transition is not the usual first-order phase transition between two equilibrium phases.} 

Charged string loops with an angular momentum are commonly referred to as \textit{vortons} in the literature. In the scenario we are considering,  the conserved charge is associated with the baryon symmetry $U(1)_V$, and a crucial role is played by the Chern-Simons (CS) theory that lives on the world-volume of the wrapped D6-brane. This is effectively a (2+1)-dimensional $U(1)_N$ CS theory, which is famously the theory that describes the bulk physics of the fractional quantum Hall effect (FQHE). As a general review, see for instance \cite{Tong:2016kpv}. 
In this paper we show that vortons in the WSS model are described by spinning D6-branes and can be thought of as excitations of the CS theory, satisfying the famous relation ``angular momentum $\sim$ charge$^2\, $", which is the hallmark of Hall-like anyonic systems. 
Our findings strengthen the interpretation of such topological defects in $SU(N)$ gauge theories with flavors as excitations of a topological phase described by a CS theory, as pointed out in \cite{Gaiotto:2017tne,Komargodski:2018odf,Ma:2019xtx,Karasik:2020pwu,Ma:2020nih,Karasik:2020zyo,Kitano:2020evx,Nastase:2022pts,Lin:2023qya,Rho:2024ihu}.

We also provide the effective, low-energy description of string loops and vortons, in terms of the degrees of freedom living on the flavor branes, \textit{i.e.} mesonic modes. We provide the four-dimensional effective action for the uncharged string loop and, by turning on the electric potential on the flavor branes, we study the stability of the defect when charged under the baryon symmetry. 

The role of the vectorial $U(1)_V$ symmetry, associated with baryonic charge, is essential for the stability properties of these defects. In particular, the baryonic charge could prevent some decay mechanisms that would typically collapse the defects. 
However, we find that vortons with small charge cannot be stable, but could be (meta)stable if the charge is huge. Charged DWs could be (meta)stable with a very suppressed decay channel, leading to new candidates of dark matter, recently predicted in \cite{Bigazzi:2022ylj} as \textit{axionic-baryons} (a-baryons). 

The paper is organized as follows. In Section \ref{sec:model} we briefly introduce the WSS model in the deconfined phase.

In section \ref{sec:D6} we study the embeddings of the probe D6-brane representing axionic topological defects. First, we review the straight axionic string. Then, we study the string loop and DW embeddings considering a D6-brane with a circular boundary on the flavor branes. We show that these two embeddings, which are both unstable, cannot be realized for every value of the  D6-brane boundary radius but are separated by a first-order phase transition. The critical radius for the transition depends on the temperature. We conclude this section by studying the string loop solution charged under the baryon symmetry and with angular momentum. We obtain the scaling behavior of the stability radius, which turns out to be independent on $N$, and we find that there are no vortons with a small baryon charge.

Section \ref{sec:D8} is devoted to the effective description of the axionic string loop, and its charged version, the vorton, in terms of the flavor degrees of freedom living on the D8-brane. 
We provide the mesonic profile and the four-dimensional effective action of the uncharged string loop. 
Then, we study the charged string loop near the D8/$\overline{\text{D}8}$-branes tip where the space is approximately flat, finding an estimate for the stability radius which is consistent with the local D6-brane analysis. Finally, we extend the solution to the spacetime boundary where exact solutions to the linearized equations of motion can be found, providing the vector meson profiles describing the string charge.

We will conclude with a summary and some observations in section \ref{sec:conclusions}.


\section{The holographic model}\label{sec:model}
The WSS model \cite{Witten:1998zw,Sakai:2004cn,Sakai:2005yt} is the top-down holographic theory closest to large $N$ QCD and it has been very successful in
modeling aspects of its strong coupling dynamics. 
It is a non-supersymmetric (3+1)-dimensional Yang-Mills theory with gauge group $SU(N)$, coupled to $N_f$ quarks and a tower of Kaluza-Klein (KK) matter fields. This theory arises at low energy from a setup involving $N$ D4-branes wrapping a circle $S^1_{x_4}$ - with $x_4\sim x_4 + 2\pi/M_{KK}$ the compact coordinate - and $N_f$ D8/$\overline{\text{D}8}$ ``flavor branes" orthogonal to it. In the following, we will focus on the $N_f=1$ case with zero quark mass. 

The large $N$ strong coupling regime $\lambda\gg1$ of the model, where $\lambda$ is the 't Hooft coupling at the KK mass scale $M_{KK}$, is holographically described by a type IIA supergravity background sourced by the D4-branes and probed by the flavor branes. The holographic model captures key non-perturbative features of the dual field theory, from confinement and mass gap to spontaneous chiral symmetry breaking. The mass scale of the glueballs is set by the dimensionful parameter $M_{KK}$. Spontaneous chiral symmetry breaking is geometrically realized by the joining, in a unique U-shaped configuration, of the two stacks of  D8 and $\overline{\text{D}8}$ branes, whose asymptotic distance along $x_4$ will be denoted by $L\leq \pi/M_{KK}$. 
Mesons and baryons arise as flavor brane gauge field small excitations and instantons, respectively.

The WSS model exhibits a rich phase structure, featuring both a first-order confinement/{decon-finement} transition and a first-order transition for the restoration of chiral symmetry. These two transitions can take place at distinct critical temperatures, depending on the specific parameters of the model. Notably, within the regime outlined earlier, these phenomena can be captured and analyzed in a precise manner using the dual holographic description. The confinement/deconfinement phase transition occurs at the critical temperature
$T_c = M_{KK}/(2\pi)$. In the dual gravity picture, it corresponds to a Hawking-Page transition
between a solitonic background, dual to the confining phase, and a black brane solution dual to the deconfined one. If the flavor branes are placed at antipodal points on the compactification circle, \textit{i.e.}~when
$L M_{KK} = \pi$, chiral symmetry breaking and confinement occur at the same energy scale.
In other words, when $T<T_c$ chiral symmetry is broken and the theory confines, while at $T>T_c$
the theory enters a deconfined phase with chiral symmetry restoration being realized by two disconnected stacks of  D8/$\overline{\text{D}8}$ branes crossing the black hole horizon. However, for non-antipodal configurations with $L M_{KK}\lesssim 0.97$, an intermediate phase with deconfinement but
broken chiral symmetry arises \cite{Aharony:2006da}. This is the scenario on which we want to focus in the present work.

The black brane background corresponding to the deconfined phase of the WSS model is given, in the string frame, as
\begin{align}\label{eq:bkgmetric}
     &ds^{2} =\left(\frac{u}{R}\right)^{3/2} \left[-f_{T}(u)dt^{2} + dx^i dx^i + d x_4^2\right] +\left(\frac{R}{u}\right)^{3/2}\left[\dfrac{du^{2}}{f_{T}(u)} + u^{2}d\Omega_4^2\right]\,,\\ \nonumber
     &f_{T}(u) = 1 - \dfrac{u_{T}^{3}}{u^{3}}\,,\qquad e^{\phi} = g_{s}\left(\frac{u}{R}\right)^{3/4}\,, \qquad F_{4} = \dfrac{2\pi N}{V_4}\omega_4\,, \qquad R^{3} = \pi g_{s}Nl_{s}^{3}\,,
\end{align}
where $t$, $x^{i}\, (i=1,2,3)$ and $x^4$ span the D4-branes worldvolume, $u\in [u_T, \infty)$ is the radial coordinate holographically related to the renormalization group energy scale of the dual field theory, $\omega_4$ is the volume form of the transverse four sphere $S^4$, with volume $V_4=8\pi^2/3$, $\phi$ is the dilaton and $F_4$ is a Ramond-Ramond form.\footnote{Here and in the following we will adopt the conventions of \cite{Sakai:2004cn} for what concerns Ramond-Ramond forms and Chern-Simons terms in the D-brane actions. In particular, the Ramond-Ramond forms are scaled with
respect to the standard notation as
\begin{equation}
C_{p+1} \rightarrow \frac{k_0^2 \tau_{6-p}}{\pi}C_{p+1}\,,
\end{equation}
where $k_0^2=2^6\pi^7l_s^8$ and $\tau_p=(2\pi)^{-p}l_s^{-(p+1)}$.} 
The parameter $u_T$ is related to the black brane temperature $T$ (which corresponds to the field theory temperature) through
\begin{equation}\label{eq:uT}
9u_T = 16\pi^2 R^3 T^2\,.
\end{equation}
The 't Hooft coupling $\lambda$ is related to the string theory parameters through\footnote{Here, again, we stick to the conventions in \cite{Sakai:2004cn} according to which $\lambda = N g_{YM}^2 = 2\pi g_s N l_s M_{KK}$.}
\begin{equation}\label{eq:lambda}
    \dfrac{R^3}{l_s^2} = \dfrac{1}{2}\dfrac{\lambda}{M_{KK}}\,.
\end{equation}
D-brane configurations probing the above background and wrapping the transverse $S^4$ play a crucial role in the dual field theory. Apart from the already mentioned D8-branes providing the flavor degrees of freedom, let us mention the identification of baryon vertices with wrapped D4-branes and that of domain walls with wrapped D6. We will come back to these issues in the following.

\subsection{Non-antipodally embedded D8-branes}
In this section, we present the axionic (flavor) D8-brane embedding at finite temperature and we review the holographic first-order phase transition, occurring in the deconfined phase, between the chiral symmetry broken phase (corresponding to the connected configuration) and the chirally symmetric one (corresponding to the disconnected configuration). 

We start by writing the induced metric on the D8-brane's world volume, in terms of the inverse embedding $u = u(x_{4})$
  \begin{equation*}
  ds^{2}_\text{D8} =\left(\frac{u}{R}\right)^{3/2} \left[-f_{T}(u)dt^{2} + dx^i dx^i + \left(1 + \left(\frac{R}{u}\right)^{3}\frac{u'(x_4)^{2}}{f_{T}(u)}\right)d x_4^2\right] +\left(\frac{R}{u}\right)^{3/2}u^{2}d\Omega_4^2\,,
  \end{equation*}
where $u'(x_4) = du/dx_4$. 
The DBI action (with no gauge fields) for the D8-branes reads
\begin{equation*}
S_\text{D8}^\text{DBI} = - T_{8}V_{3+1}V_{4}\int dx_{4}\,u^{4}\sqrt{f_{T}(u) + \left(\frac{R}{u}\right)^{3}u'(x_4)^{2}}\,,
\end{equation*}
with $T_8 = (2\pi)^{-8}l_s^{-9}$ the D8-brane tension and $V_{3+1}$ the volume of Minkowski space. The action has a first integral
\begin{equation*}
\dfrac{\partial\mathcal{L}}{\partial u'}u' - \mathcal{L} = -\dfrac{u^{4}f_{T}(u)}{\sqrt{f_{T}(u) + \left(\frac{R}{u}\right)^{3}u'(x_4)^{2}}} = \text{constant}\,.
\end{equation*}
The U-shaped connected configuration has a tip at $u=u_J$ where $u'=0$. Thus,
\begin{equation*}
u^{4}_{J}\sqrt{f_{T}(u_{J})} = \dfrac{u^{4}f_{T}(u)}{\sqrt{f_{T}(u) + \left(\frac{R}{u}\right)^{3}u'(x_4)^{2}}}\,.
\end{equation*}
We can extract $u'$ to get
\begin{equation}\label{eq:u'}
u'(x_4)^{2} = \left(\frac{u}{R}\right)^{3}\dfrac{f_{T}(u)}{u^{8}_{J}f_{T}(u_{J})}\left(u^{8}f_{T}(u) - u^{8}_{J}f_{T}(u_{J})\right)\,.
\end{equation}
Then, the profile of the D8-brane reads
\begin{align}\label{x4}
x_{4}^{\pm} &= \pm\int du \left(\frac{R}{u}\right)^{3/2}\dfrac{u^{4}_{J}\sqrt{f_{T}(u_{J})}}{\sqrt{f_{T}(u)}\sqrt{u^{8}f_{T}(u) - u^{8}_{J}f_{T}(u_{J})}}\,.
\end{align}
We can now integrate the equation \eqref{x4} to get the asymptotic brane separation $L$, which reads
\begin{align}\label{eq:LRuJ}
L = \int dx_{4} = 2\int\dfrac{du}{u'} = J_{T}(\tilde{b})\dfrac{R^{3/2}}{u_J^{1/2}}\,,
\end{align}
where 
\begin{equation}
    \tilde{b} \equiv \dfrac{u_T}{u_J}\,,
\end{equation}
and
\begin{equation}
    J_T(\tilde{b})= \dfrac{2}{3}\sqrt{1-\tilde{b}^3}\int^1_0dy\dfrac{\sqrt{y}}{\sqrt{\left(1-\tilde{b}^3y\right)\left(1-\tilde{b}^3y-\left(1-\tilde{b}^3\right)y^{8/3}\right)}}\,.
\end{equation}
Using (\ref{eq:uT}) we see that $u_{J}$ becomes a non trivial function of $T$:
\begin{equation}
    LT = \dfrac{3}{4\pi}J_T(\tilde{b})\sqrt{\tilde{b}}\,.
\end{equation}

\begin{figure}
\center
\includegraphics[height = 5 cm]{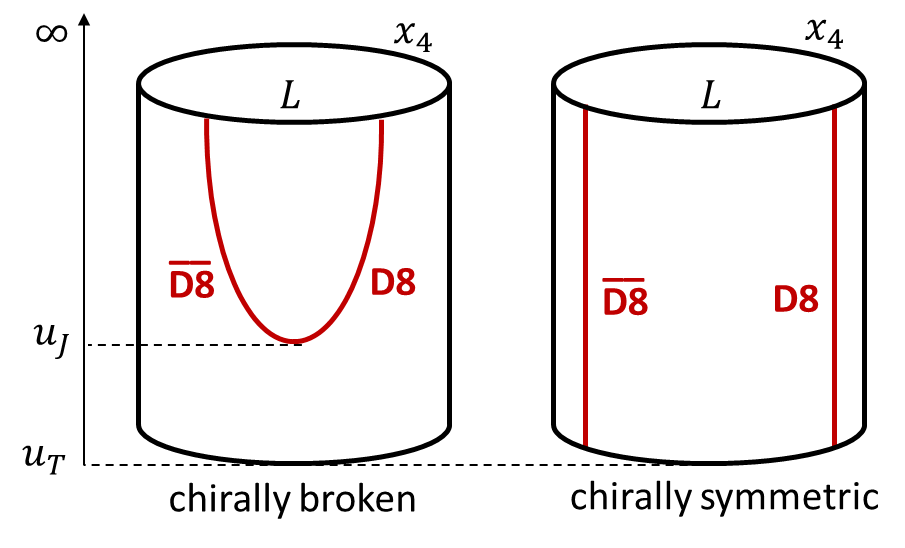}\caption{Schematic picture of the chirally broken (left) and chirally symmetric (right) phases in the deconfined phase of the Witten-Sakai-Sugimoto model. 
The $x_4$-direction is compactified on a circle with radius $1/M_{KK}$,
and $u$ is the holographic coordinate with $u =\infty$ being the boundary where the dual field theory lives. If the asymptotic distance $L$ between the flavor branes is sufficiently small, a deconfined, chirally broken phase (left figure) becomes possible. In this phase, the connected flavor branes are embedded non-trivially in the background according to a definite function $x_4(u)$.}
\label{fig:wssphases}
\end{figure}

The connected configuration is not the only possible solution satisfying the requirement that the asymptotic separation of the flavor brane branches is $L$. Another solution exists with $u'(x_4)=\pm\infty$, which describes two disconnected branches extended along the radial direction $u$ till the horizon. At 
\begin{equation}
T=T_a \simeq \frac{0.154}{L}\,,
\end{equation}
there is a first-order phase transition between the connected solution, holding at $T<T_a$, and the disconnected one. 
Hence for $T<T_a$ chiral symmetry is broken, while for $T>T_a$ chiral symmetry is restored. See figure \ref{fig:wssphases} for a pictorial representation of these two phases in the $(x_4,u)$ subspace. 

Non-antipodally embedded D8-branes in the chirally-broken, deconfined phase of WSS describe a dual theory that shares key properties with the theory presented in the introduction: an $SU(N)$ gauge theory coupled to a massless flavor at finite temperature. At low energy, the physics is captured by D8-brane excitations whose zero mode is the axion associated with the pseudo-Goldstone boson of the chiral symmetry breaking \cite{Bigazzi:2019eks}.

We conclude this section with some useful formulas. Using \eqref{eq:u'} we can rewrite the D8-brane induced metric as
 \begin{align}
  &ds^{2}_\text{D8} =\left(\frac{u}{R}\right)^{3/2} \left[-f_{T}(u)dt^{2} + dx^i dx^i + \left(\frac{R}{u}\right)^{3}\dfrac{du^{2}}{\gamma_{T}(u)}\right] +\left(\frac{R}{u}\right)^{3/2}u^{2}d\Omega_4^2\,,\label{ushapmet}\\
  &\sqrt{-\det g_\text{D8}} = \left(\frac{R}{u}\right)^{3/4}u^{4}\sqrt{\dfrac{f_{T}(u)}{\gamma_{T}(u)}}\,,
  \end{align} 
where we have introduced the function 
\begin{equation}
    \gamma_{T}(u) = \dfrac{u^{8}f_{T}(u) - u^{8}_{J}f_{T}(u_{J})}{u^{8}}\,.
\end{equation}

Turning on the gauge field on the world volume of the D8-brane we get a DBI action which, up to quadratic order in $\alpha'$, can be expanded as
\begin{align}\label{sd8}
S_\text{D8}^\text{DBI} &= -T_{8}\int d^{9}x\,e^{-\phi}\sqrt{-\det(g_\text{D8} + (2\pi\alpha')\mathcal{F})} \approx \nonumber\\
&\approx -\dfrac{T_{8}}{4}\dfrac{V_{4}}{g_{s}}(2\pi\alpha')^{2}R^{3/2}\int d^{4}x\,du\,u^{5/2}\sqrt{\dfrac{f_{T}(u)}{\gamma_{T}(u)}}g^{MN}g^{PQ}\mathcal{F}_{MP}\mathcal{F}_{NQ} = \nonumber\\ 
&= -\dfrac{T_{8}}{4}\dfrac{V_{4}}{g_{s}}(2\pi\alpha')^{2}R^{3/2}\int d^{4}x\,du\,u^{5/2}\sqrt{\dfrac{f_{T}(u)}{\gamma_{T}(u)}}\Biggl[\dfrac{1}{2}\left(\frac{R}{u}\right)^{3}F_{ij}^2 - \left(\frac{R}{u}\right)^{3}\dfrac{F_{ti}^2}{f_{T}(u)} + \nonumber \\
& \hspace{9cm} - \dfrac{\gamma_{T}(u)}{f_{T}(u)}F_{tu}^2 + \gamma_{T}(u)F_{iu}^2 \Biggr]\,,
\end{align}
where we have used that $\mathcal{F} = F/\sqrt{2}$ and the indices $M,N,P,Q$ run over $(t,x^i,u)$.
Upon reduction on the four-sphere, the Chern-Simons part of the action reads 
\begin{align}\nonumber
    S_{CS} &= \dfrac{N}{24\pi^2}\int \mathcal{A}\wedge\mathcal{F}\wedge\mathcal{F}= \dfrac{N}{48\pi^2\sqrt{2}}\int A\wedge F\wedge F =\\ \label{cscompleto}
    &= \dfrac{N}{24\pi^2\sqrt{2}}\epsilon^{tMNPQ}\int d^5x\left(\dfrac{1}{4}A_tF_{MN}F_{PQ} - A_M F_{tN}F_{PQ}\right),
\end{align}
where the convention for the anti-symmetric tensor is that $\epsilon^{t1234} = 1$.

It is often convenient to work with a dimensionless holographic coordinate $z$, defined as
\begin{equation}
    u(z) = u_J(1+z^2)^{1/3} = u_Jk(z)^{1/3}\,,
\label{uzeta}
\end{equation}
where we have defined
\begin{equation}
    k(z) = 1+z^2.
\end{equation}
Notice that $z\in(-\infty,+\infty)$ where the two extremal values correspond to the two D8/$\overline{\text{D8}}$-branches asymptotically sitting at different points on the $x_4$ circle. Using this coordinate, the action \eqref{sd8} becomes\footnote{Even if not specified, every $u$ has to be intended as a function of $z$.}
\begin{align}\nonumber
    S_\text{D8}^\text{DBI} = &-\dfrac{T_{8}}{4}\dfrac{V_{4}}{g_{s}}(2\pi\alpha')^{2}R^{3/2} 
\int d^4x\,dz \,u^{5/2}\sqrt{\dfrac{f_T(u)}{\gamma_T(u)}}\bigg|\dfrac{\partial u}{\partial z}\bigg|\Bigg[\dfrac{1}{2}\left(\dfrac{R}{u}\right)^{3}F_{ij}^2 - \left(\dfrac{R}{u}\right)^{3}\dfrac{F_{ti}^2}{f_T(u)} +\\ \label{eq:d8action}
    & \hspace{7cm} - \bigg|\dfrac{\partial u}{\partial z}\bigg|^{-2}\dfrac{\gamma_T(u)}{f_T(u)}F_{tz}^2 + \bigg|\dfrac{\partial u}{\partial z}\bigg|^{-2}\gamma_T(u)F_{iz}^2\Bigg]\,.
\end{align}

\section{D6-brane description of the defects}\label{sec:D6}
In this section, we study the various D6-brane embeddings that could holographically describe axionic topological defects, such as the straight axionic string, the axionic string loop, and the axionic DW. In the usual cosmological scenarios, the axionic DWs arise in the confined phase of a QCD-like theory, around the confinement/deconfinement transition, and they are not present at higher temperatures. In our model, DWs can be formed earlier, leading to a novel cosmological feature that can have interesting phenomenological consequences. In particular, DWs charged under the baryon symmetry can form in the early Universe and be (meta)stable leading to new candidates of dark matter, recently predicted in \cite{Bigazzi:2022ylj} as \textit{axionic-baryons} (a-baryons). 

We will start by presenting the D6-brane embedding associated with the straight axionic string (already described in \cite{Bigazzi:2022ylj}) as a warmup; then we will study the D6-brane embeddings which describe the axionic string loop and the axionic DW. Surprisingly these two embeddings cannot always be realized, but there is a first-order phase transition between the two depending on the radius of the string loop/DW at the tip of the U-shaped D8-branes. As far as we know, this phase transition has not been investigated in the literature.

Within the WSS model, the role played by a D6-brane wrapping the $S^4$ cycle of the background has been investigated recently in \cite{Argurio:2018uup,Bigazzi:2022luo,Bigazzi:2022ylj}. 
A key feature of this object is the presence of a $U(1)_N$ Chern-Simons theory on its world-volume. This comes from the CS term of the D6-brane
\begin{equation}
    \dfrac{1}{8\pi^2}\int C_3\wedge da\wedge da = \dfrac{1}{8\pi^2}\int F_4\wedge a\wedge da = \dfrac{N}{4\pi}\int a \wedge da\,,
\end{equation}
where $a$ is the D6-brane gauge field and we have used $\int_{S^4} F_4 = 2\pi N$. 

We will begin by turning off the D6-brane gauge field. This will correspond to describing uncharged topological defects. Then we will consider the charged case, mainly focusing on the charged axionic string loop, also known as vorton, where the above CS term will play a crucial role.

\subsection{Straight axionic string embedding}
\label{subsec:straight}
\begin{figure}
\center
\includegraphics[height = 5 cm]{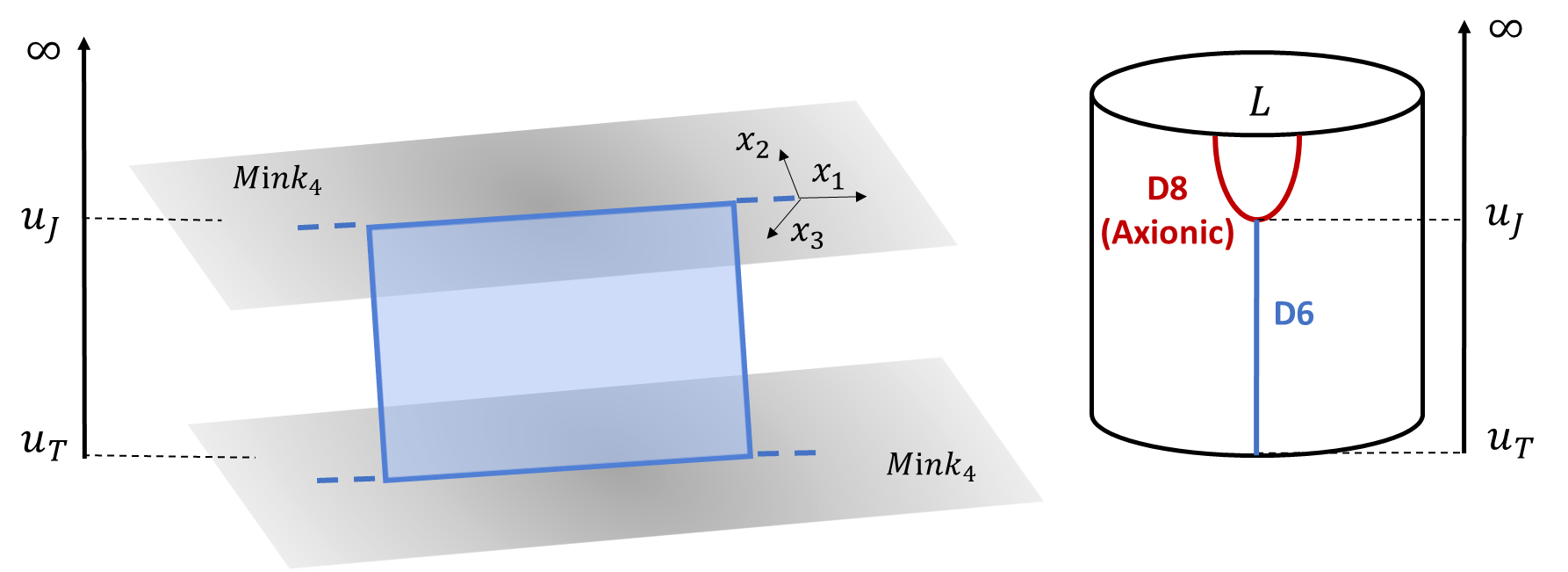}\caption{The D6-brane embedding for the straight axionic string is presented from the Minkowski space point of view (left) and in the $(x_4,u)$ cylinder sub-space (right). The D6-brane is attached to the 
D8-brane tip and it ends on the horizon.}
\label{holostring}
\end{figure}

The straight axionic string can be described holographically by a D6-brane wrapping $S^4$, attached to the tip of the U-shaped D8-brane, ending on the black-hole horizon. The D6-brane is also extended along, say, the $x_1$-coordinate of Minkowski space. See figure \ref{holostring} for a pictorial representation.
The D6-brane, in general, could have a non-trivial profile say in the $(x_2,u)$-plane\footnote{Of course, one could equivalently consider the $x_3$ direction or both.} and it is described by the embedding function $x_2 = x = x(u)$. 
The induced metric on the D6-brane reads
\begin{equation}
    ds^2_\text{D6} = \left(\dfrac{u}{R}\right)^{3/2}\left[-f_T(u)dt^2 + dx_1^2 +\left(x'(u)^2 + \left(\dfrac{R}{u}\right)^{3}\dfrac{1}{f_T(u)}\right)du^2\right] + R^{3/2}u^{1/2}d\Omega_4^2\,,
\end{equation}
where $x'(u) = dx/du$. The DBI action for the D6-brane with the world-volume gauge field switched off is
\begin{align}\nonumber
    &S_\text{D6}^\text{DBI} = -\dfrac{T_6V_4R^3V_{x_1}}{g_s}\int dt du\,u\,\sqrt{1 + \left(\dfrac{u}{R}\right)^3f_T(u)x'(u)^2} = V_4 V_{x_1} \int dtdu\,\mathcal{L}\,,\\
    &\text{where}\,\,\,\,\,\,\mathcal{L} = -\dfrac{T_6 R^3}{g_s}u\,\sqrt{1 + \left(\dfrac{u}{R}\right)^3f_T(u)x'(u)^2}\,.
\end{align}
The Euler-Lagrange equation for the profile $x(u)$ reads
\begin{equation}
  x''(u) + \dfrac{5 u^3 - 2u_T^3}{2R^3 u}x'(u)^3 + \dfrac{4u^3-u_T^3}{u(u^3-u_T^3)}x'(u)  = 0\,,
\end{equation}
and admits $x'=0$ as a solution. This is referred to as the straight axionic string configuration (saxs).

The tension of the straight axionic string can be easily computed as
\begin{align}\nonumber
&S_\text{D6}^\text{DBI} = T_\text{saxs}\int dt\,dx_1,\,\,\,\,\,\,\,\,\,\,\,\,\text{where}\,\,\,\,\,\,\,\,\,\,\,\,T_\text{saxs} = \dfrac{T_6V_4R^3}{g_s}\dfrac{u_J^2-u_T^2}{2}\,.
\end{align}
In \cite{Bigazzi:2022ylj} the interested reader can find many more details on this defect. 
 
\subsection{Axionic string loop $\&$ axionic domain wall embeddings}
\label{subsec:axldw}
\begin{figure}
\center
\includegraphics[height = 5 cm]{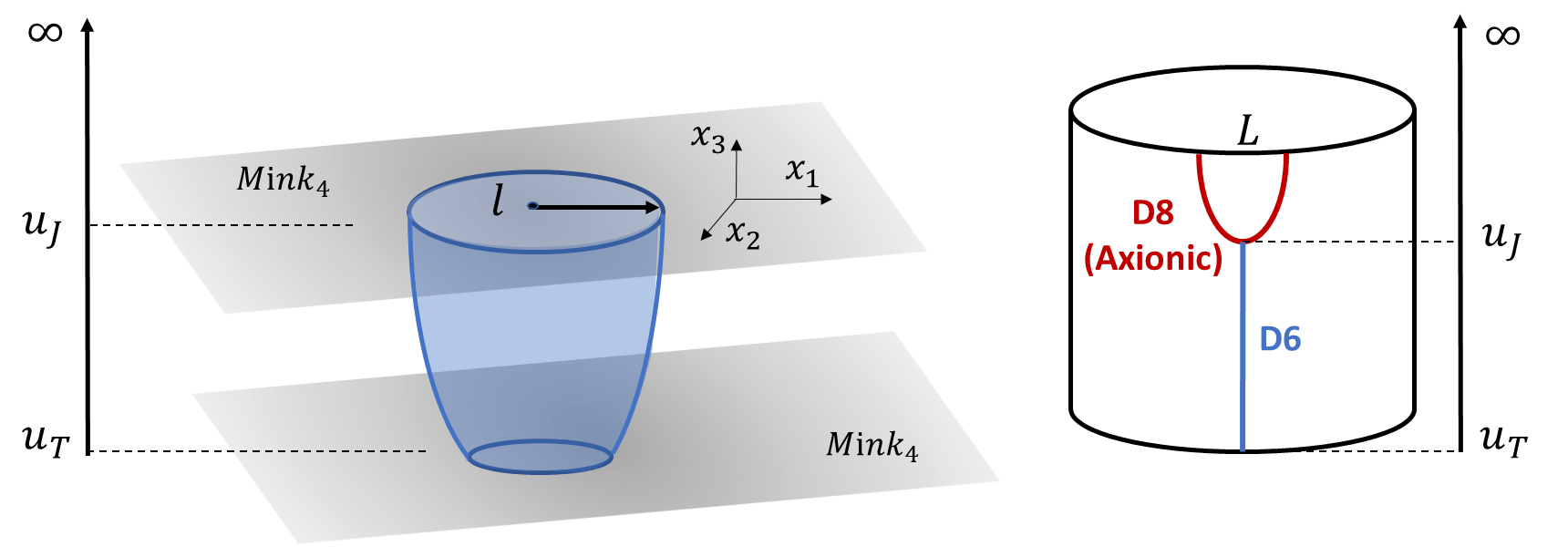}\caption{Pictorial representation of the D6-brane embedding describing holographically the axionic string loop. The D6-brane has a radius $l$ at $u=u_J$ where it is attached to the D8-brane, while it ends on the horizon with a non-trivial profile.}
\label{holovorton}
\end{figure}

\begin{figure}
\center
\includegraphics[height = 5 cm]{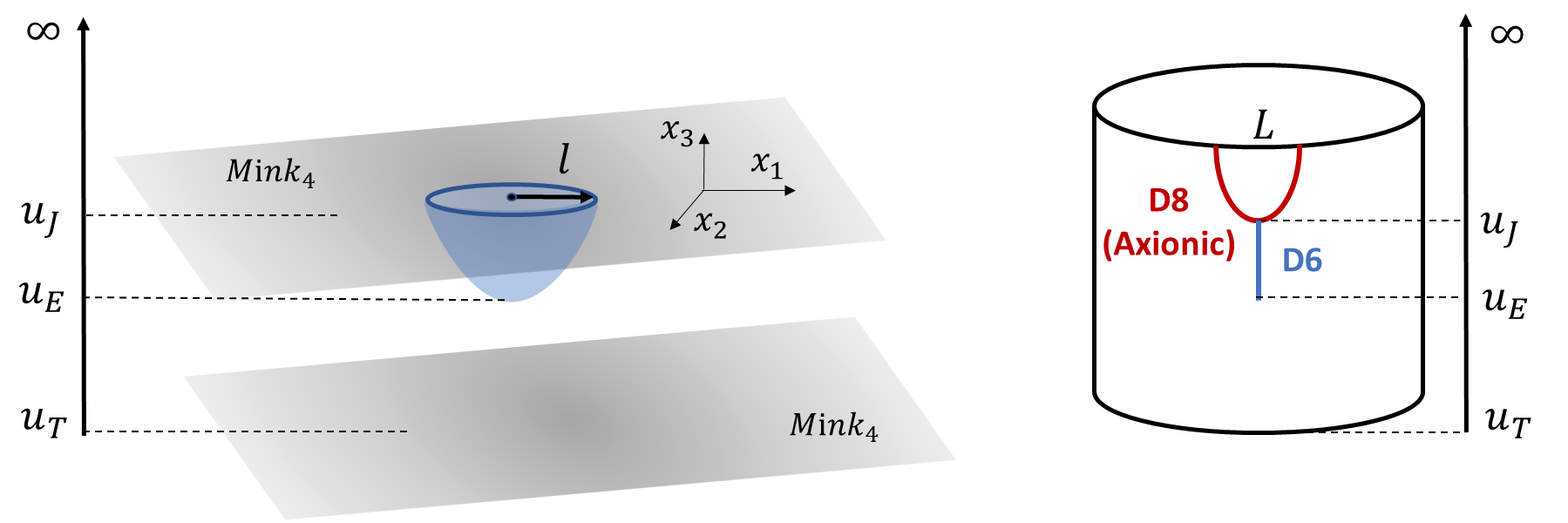}\caption{Pictorial representation of the D6-brane embedding describing the axionic domain wall ending on a string loop. The domain wall has a radius $l$ at $u=u_J$ where it is attached to the D8-brane, and it shrinks to zero size at $u=u_E$, with $u_T<u_E<u_J$.}
\label{holopot}
\end{figure}

The axionic string loop and the axionic domain wall configurations correspond to two different profiles of a D6-brane, wrapped on $S^4$, ending with a circular boundary of radius $l$ (say in the $(x_1,x_2)$ plane) at the tip of the U-shaped D8-brane. The two profiles correspond to two different solutions for the embedding function $\rho= \sqrt{x_1^2+x_2^2}= \rho(u)$ satisfying the boundary condition $l=\rho(u_J)$. The induced metric on the D6-brane now reads
\begin{equation}
    ds^2_\text{D6} = \left(\dfrac{u}{R}\right)^{3/2}\left[-f_T(u)dt^2+\rho^2(u)d\psi^2 + \left(\rho'(u)^2+ \left(\dfrac{R}{u}\right)^3\dfrac{1}{f_T(u)}\right)du^2\right] + R^{3/2}u^{1/2}d\Omega_4^2\,,
\end{equation}
where $\rho'(u) = d\rho/du$ and the $(x_1, x_2)$ plane has been expressed in terms of polar coordinates $(\rho, \psi)$. 
The DBI action then reads
\begin{equation}
S_\text{D6}^\text{DBI} = -C\int dt du\,\rho(u)\,u\sqrt{1+\dfrac{u^3}{R^3}f_T(u)\rho'(u)^2} = -C\int dt du\,\rho(u)\,u\, D_0(u) \,, 
\label{eqactd6}
\end{equation}
where
\begin{equation}
    D_0(u) = \sqrt{1+\dfrac{u^3}{R^3}f_T(u)\rho'(u)^2}\,,
\end{equation}
and
\begin{equation}
C\equiv\dfrac{T_6}{g_s}R^{3}2\pi V_4 = \frac{N}{3(2\pi\alpha')^2}=
 \dfrac{1}{3\cdot 2^4\pi^2}\dfrac{\lambda^2 N}{M_{KK}^2R^6}\,.
\label{defC}
\end{equation}
The Euler-Lagrange equation for $\rho(u)$ derived from the action above
\begin{equation} 
\label{equncr}
 \partial_u\left(\dfrac{u\,\rho(u)}{D_0(u)}\left(\dfrac{u}{R}\right)^3f_T(u)\rho'(u)\right)=u\,D_0(u)\,, 
 \end{equation}
has to be solved numerically and turns out to admit (at least) two solutions: one which describes a D6-brane that extends from $u=u_J$ to the horizon $u=u_T$ (see figure \ref{holovorton}); another which gives a D6-brane profile going from $u_J$ to some intermediate coordinate $u_E(>u_T)$ at which the configuration shrinks to zero size (see figure \ref{holopot}). The first possibility corresponds to an uncharged \textit{axionic string loop} while the second one describes an uncharged \textit{axionic domain wall} ending on a string loop. It is important to notice that, due to the absence of a charge which might counterbalance the tension of these objects, both configurations are, not unexpectedly, \textit{unstable}. 
In fact, for both of them, it turns out that
\begin{equation}
\rho'(u_J) \neq 0\,,
\end{equation}
which means that the D6-brane does not end orthogonally to the D8, as it would be required by the zero-force condition.\footnote{From equation \eqref{equncr} it follows that if we impose $\rho'(u_J)=0$, we get $\rho''(u_J)>0$ which implies that $u_J$ is a local minimum for the radius $\rho(u)$. This is opposite to the behavior of the unstable solutions shown in figures \ref{holovorton} and \ref{holopot}.}

As we will see, at any fixed value of $\tilde{b}= u_T/u_J$ in the chiral symmetry broken phase, the string loop configuration exists only for some $l\geq l_{\text{MIN}}$, while the domain wall exists only for $l\leq l_{\text{MAX}}$, where $l_{\text{MIN}}, l_{\text{MAX}}$ are critical values for the D6-brane radii at the tip of the U-shaped D8-brane. This suggests the possible occurrence of a ``phase transition" between the two configurations.\footnote{Here and in the following we will loosely use the term ``phase transitions" although, properly speaking, this should be referred to transitions between phases that are stable on certain domains at equilibrium, while our uncharged configurations are unstable.} As we will see, this is the case.

In order to find a numerical solution of \eqref{equncr} we adopt dimensionless coordinates, rescaling 
\begin{equation}
    \rho = L\hat{\rho}\,,\,\,\,\,\,\,\,\,\,\,\,\,\,u = u_J\,\hat{u}\,.
\end{equation}
Dropping the hats for simplicity and using \eqref{eq:LRuJ} the action can be rewritten as
\begin{eqnarray}
    &S_\text{D6}^\text{DBI} = -C u_J^2 L\int dtdu\,\rho(u)\,u\,{\hat D}_0(u)\,,\nonumber \\
     &{\hat D}_0(u)\equiv\sqrt{1+J_T(\tilde{b})^2f_T(u)u^3\rho'(u)^2}\,,\,\,\,\,\,\,\,\,\,\,\,f_T(u)=1-\dfrac{\tilde{b}^3}{u^3}\,,
\label{tildeDunch}
\end{eqnarray}
and the equation for $\rho(u)$ reads
\begin{align}\label{eomrho}
J_T(\tilde{b})^2\partial_u\left(\dfrac{u^4\,\rho(u)}{\hat{D}_0(u)}f_T(u)\rho'(u)\right)- u\,\hat{D}_0(u)=0\,.
\end{align}

The free energy of a D6-brane configuration solving the above equation reads
\begin{equation}
\label{d6energy}
E_\text{D6}=C u_J^2 L \int^1_{u^*}du\,\rho(u)\,u\,\hat{D}_0(u)=\dfrac{\lambda^2 N}{3\cdot2^6 \pi^2}\dfrac{J_T(\tilde{b})^4T_a^3}{(0.154)^3M_{KK}^2}\int^1_{u^*}du\,\rho(u)\,u\,\tilde{D}_0(u)\,.
\end{equation}
The integral over $u$ runs from $u^*=\tilde{b}$ for the string loop solution and from $u^*=u_E/u_J$ for the domain wall one.

\subsubsection*{String loop solution}
Examining equation \eqref{eomrho} we can immediately notice that a cylindrical-shaped string loop solution, satisfying $\rho'(u) = 0$ for every $u$, does not exist. This does not mean that the D6-brane cannot end at the horizon. It does, but with a non-trivial shape that we are going to determine. Since the D6-brane ends at the horizon, we must carefully choose the boundary conditions there. We use a series expansion to give initial values for the solution and its first derivative at the horizon.\footnote{In our numerics we employ a cutoff at the horizon of $\epsilon=10^{-4}$.} We insert in equation \eqref{eomrho} the expansion
\begin{equation}\label{eq:seriesexpvorton}
    \rho(u) = \sum\limits_{n=0}^{N_b} c_n(u-\Tilde{b})^n\,,
\end{equation}
with $N_b = 6$ and we expand it around $u = \Tilde{b}$. All the coefficients $c_n$ are determined given $c_0$ which is the non-zero radius at the horizon. For example, the first coefficients of the expansion are
\begin{equation*}
    c_1 = \dfrac{1}{3\tilde{b}J_T(\tilde{b})^2c_0}\,,\quad c_2 = -\dfrac{1+18\tilde{b}J_T(\tilde{b})^2c_0^2}{72\tilde{b}^4J_T(\tilde{b})^4c_0^3}\,,\quad...
\end{equation*}
This series expansion is consistent with the D6-brane entering orthogonally at the horizon. This can be better seen switching to Euclidean space: regularity of the background metric \eqref{eq:bkgmetric} around $u=u_T$ is immediately realized, after the condition \eqref{eq:uT} is imposed, by introducing the coordinate $v\sim\sqrt{u-u_T}$, which renders the metric in the sub-space $(t_E,v)$ explicitly flat. In terms of this coordinate, the near-horizon equation for the embedding reads
\begin{equation}
    (v-\rho(v)\rho'(v))(1+\rho'(v)^2)=v\rho(v)\rho''(v)\,.
\end{equation}
The orthogonality condition
\begin{equation}
    \rho(v=0) \neq 0\,,\hspace{1cm}\rho'(v=0)=0\,,
\end{equation}
is consistently realized by the series expansion \eqref{eq:seriesexpvorton}.

\begin{figure}
\center
\includegraphics[height = 4 cm]{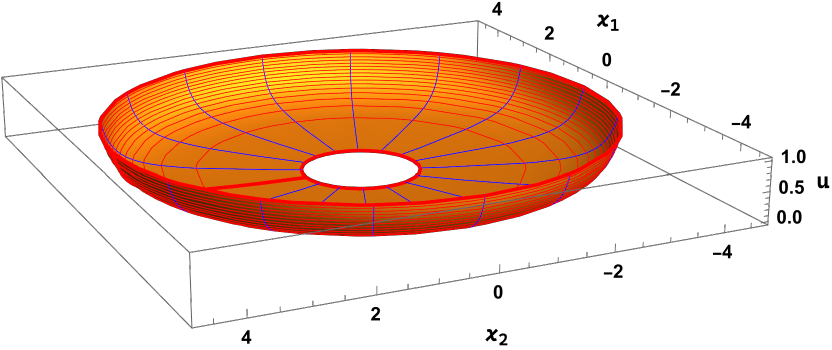}\caption{Revolution plot around the $u$ axis of the non-trivial embedding $\rho(u)$ describing the string loop configuration, for $\tilde{b} = 0.1$ and $c_0=1$. The $u$ variable runs from the horizon $\tilde{b}=0.1$ to the D8-brane tip at $u=1$.}
\label{mathevorton}
\end{figure}

The corresponding solution describes the string loop configuration that ends at the horizon, realizing the axionic string loop in the (3+1)-dimensional Minkowski space. A particular case with $c_0=1$ and $\tilde{b}=0.1$ is presented in figure \ref{mathevorton}. A numerical study of the solutions leads us to observe that for each value of $\Tilde{b}$ we cannot reach an arbitrarily small radius $l$ at $u_J$. Therefore, the string loop configuration, at fixed $\Tilde{b}$, does not exist for arbitrarily small radii $l$. Another observation is that for large values of the radius at the horizon ($c_0$), the radius $l$ is almost equal to $c_0$, so we have a quasi-cylinder configuration.

\begin{figure}
    \centering
    \begin{subfigure}{0.45\textwidth}
        \centering
        \includegraphics[width=\textwidth]{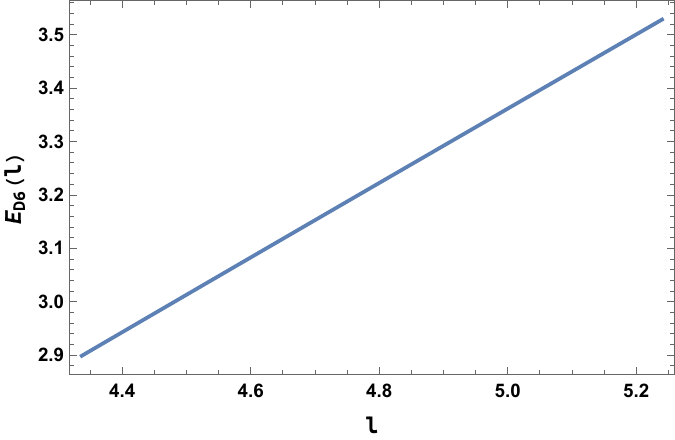}
        \caption{Plot of the D6-brane free energy in the string loop phase for $\tilde{b} = 0.1$.}
        \label{fig:Evorton}
    \end{subfigure}
    \hfill
    \begin{subfigure}{0.45\textwidth}
        \centering
        \includegraphics[width=\textwidth]{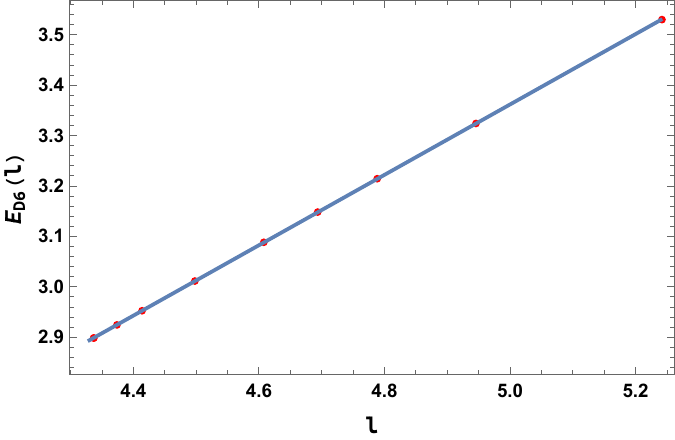}
        \caption{The red points represent values of the D6-brane free energy for $\tilde{b}=0.1$, while the blue line is a fit with a linear function.}
        \label{fig:Evortonfit}
    \end{subfigure}
    \caption{Plots of the D6-brane free energy in the string loop phase as a function of $l$ and the fit with a linear function for  $\tilde{b} = 0.1$.}
\end{figure}

The free energy \eqref{d6energy} of the string loop configurations at fixed $\Tilde b$ turns out to be linear in $l$, as expected by the scaling of their tension with the length. This can be seen, for instance, from the plots in figures \ref{fig:Evorton} and \ref{fig:Evortonfit}, where the free energy at $\Tilde b =0.1$ is given in units of $C u_J^2 L$. 
Figure \ref{fig:Evortonfit} shows the fit of the energy with a linear function of $l$. Several values of the energy as a function of $l$ (the red dots) are well approximated by the linear function $y(l) = c\,l$ with $c\approx 0.7$ for $\tilde{b} = 0.1$ (the blue line). We have checked that this linear behavior persists also for other values of $\tilde{b}$.

\begin{figure}
\center
\includegraphics[height = 8 cm]{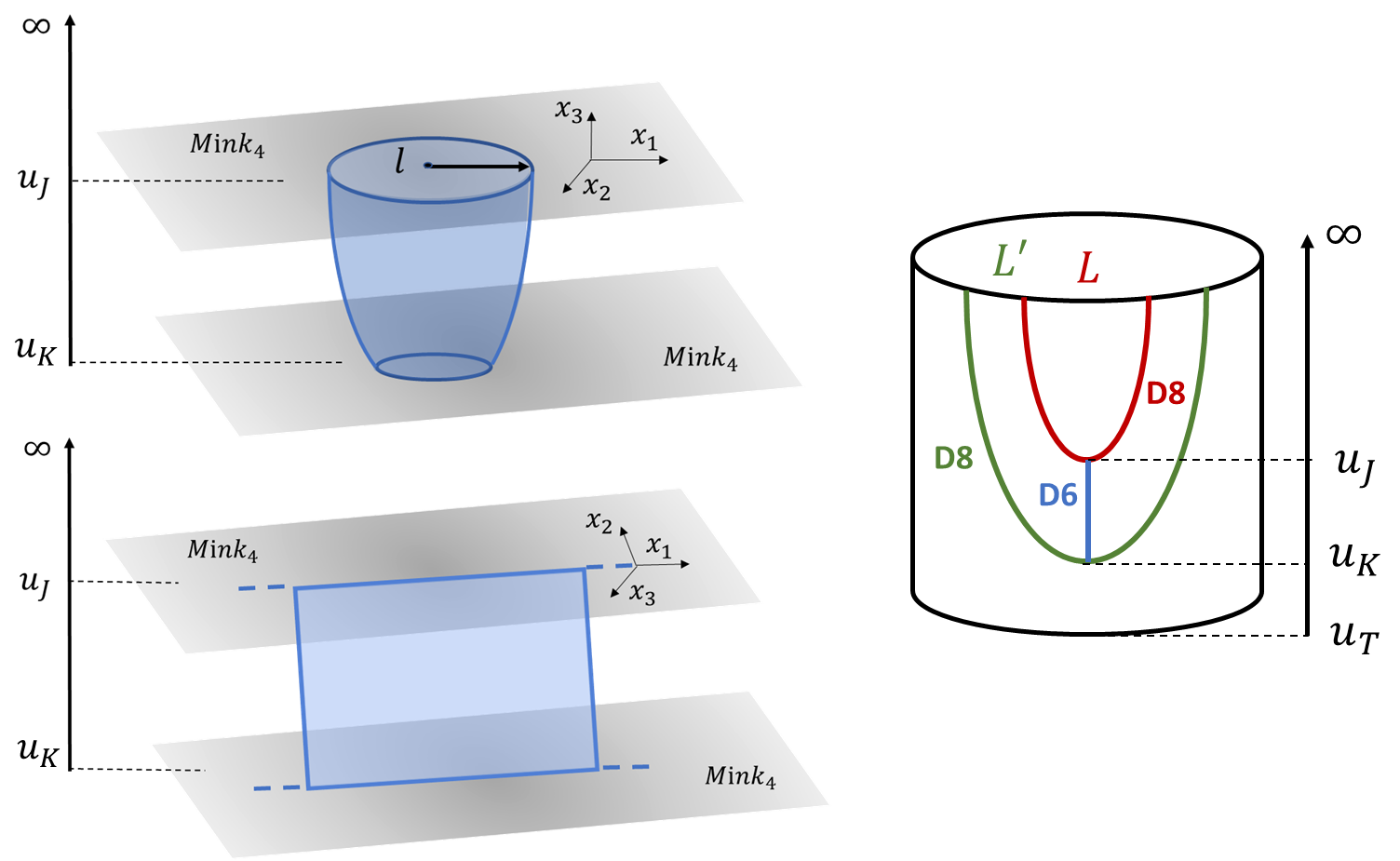}\caption{Straight axionic string and string loop configurations between two types of D8-branes. The D6-brane does not terminate at the horizon but has two boundaries attached to the two flavor branes.}
\label{doubled8}
\end{figure}

Before discussing the DW solution, we mention another possibility for the axionic string and string loop solutions. We have seen that the D6-brane in figures \ref{holostring} and \ref{holovorton} has a boundary on the D8-brane and terminates in the horizon. However, the D6-brane can extend from two different flavor branes as in figure \ref{doubled8}. 
The second flavor brane can be thought of as another dark flavor that condenses at a temperature $T_a'$ set by $u_K$ (the radial variable of the tip of the corresponding D8-brane), distinct from $T_a$. 
Analogously, if we consider the confined phase of the theory, the second flavor brane can represent a QCD flavor. 
We have such a possibility because the asymptotic separation between a D8-brane and a $\overline{\text{D8}}$-brane $L$, of a given flavor, is a free parameter in the model and hence we can tune $L$ to get the desired $T_a$ at which the chiral symmetry is broken. This ``sandwich" structure has been studied in the literature \cite{Gabadadze:2000vw,Forbes:2000et,Zhitnitsky:2002qa,Liang:2016tqc}, in particular between topological defects composed by the axion and the $\eta'$ field, realizing composite objects of dark excitations and Standard Model excitations, leading to interactions between the two sectors. 

\subsubsection*{Domain wall solution}

\begin{figure}
\center
\includegraphics[height = 4 cm]{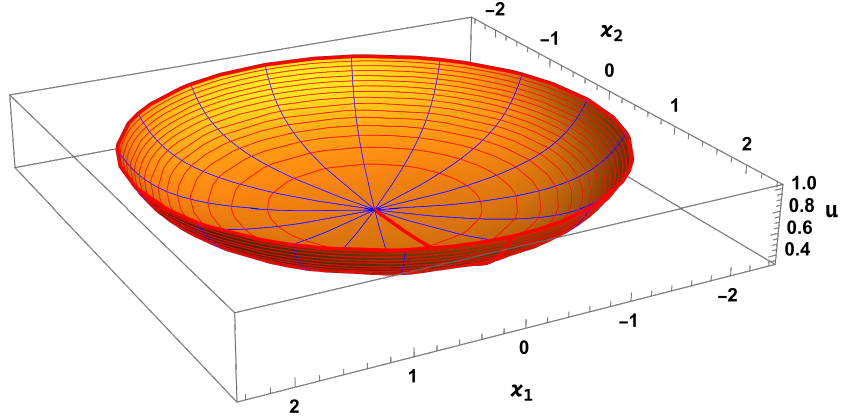}\caption{Revolution plot around the $u$ axis of the non-trivial embedding $\rho(u)$ describing the domain wall configuration for $\tilde{b} = 0.1$. The $u$ coordinate runs from $u_E/u_J = 0.35$, where the D6-brane has its tip, to the D8-brane tip at $u_J=1$.}
\label{mathedw}
\end{figure}

The domain wall configuration is a solution of the equation of motion \eqref{eomrho} with boundary conditions: $\rho(u_E/u_J) = 0$ and $\rho'(u_E/u_J) = \infty$, \textit{i.e.}~the wall closes smoothly at $u_E$. Notice that the equation depends only on $\tilde{b}$ so, once we have fixed it, by choosing the value of the holographic coordinate $u_E$, the value of the radius $l = \rho(u_J)$ is determined.  A particular solution for $\tilde{b}=0.1$ is presented in figure \ref{mathedw}. For a given value of $\tilde{b}$, no matter which value of $u_E$ we pick, the radius $l$ will be limited from above. We cannot get arbitrarily big radii $l$ if we want the configuration not to reach the horizon.

\begin{figure}
\center
\includegraphics[height = 5cm]{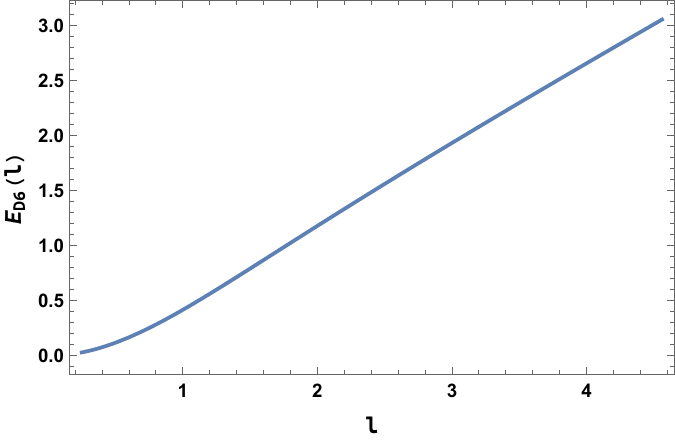}\caption{The D6-brane free energy of the domain wall configuration as a function of $l$ is plotted for $\tilde{b} = 0.1$.}
\label{Ed6conn}
\end{figure}

\begin{figure}
    \centering
    \begin{subfigure}{0.45\textwidth}
        \centering
        \includegraphics[width=\textwidth]{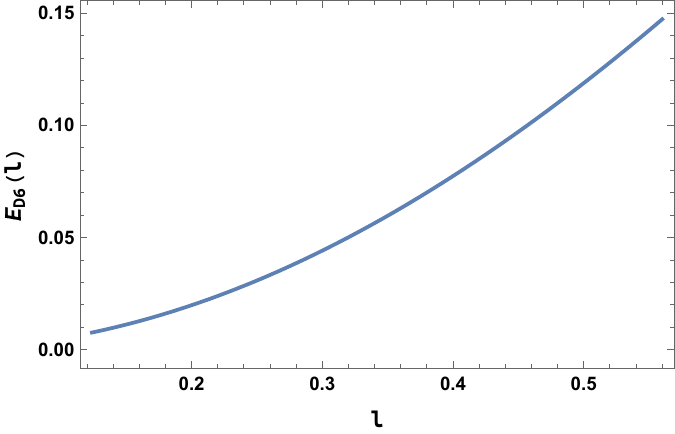}
        \caption{Plot of the D6-brane free energy in the domain wall phase for small $l$ and $\tilde{b} = 0.1$.}
        \label{fig:Edw}
    \end{subfigure}
    \hfill
    \begin{subfigure}{0.45\textwidth}
        \centering
        \includegraphics[width=\textwidth]{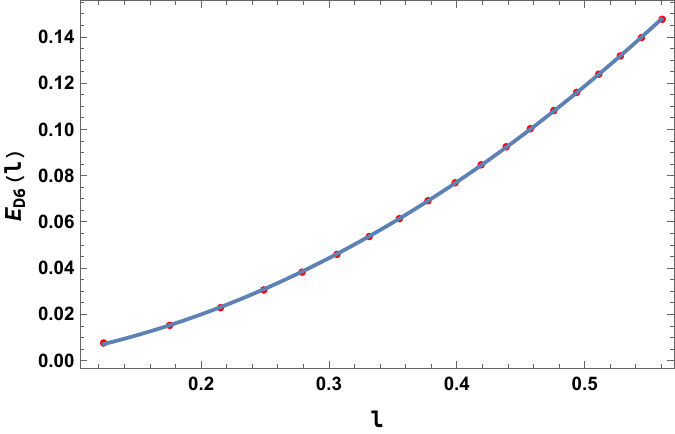}
        \caption{The red points represent values of the D6-brane free energy for $\tilde{b}=0.1$, while the blue line is a fit with a parabolic function.}
        \label{fig:Edwfit}
    \end{subfigure}
    \caption{Plots of the D6-brane free energy in the domain wall phase  as a function of $l$ and the fit with a parabolic function for small $l$ and $\tilde{b} = 0.1$.}
\end{figure}

The free energy \eqref{d6energy} of the domain wall configuration shows the following features. In figure \ref{Ed6conn} we examine it (in units of $C u_J^2 L$) for $\tilde{b} = 0.1$. For large values of $l$, the energy scales linearly with $l$ (as the energy of the string loop configuration), while for small $l$, the energy scales quadratically with $l$. Figure \ref{fig:Edw}  shows the analysis of the D6-brane energy for small radii $l$. A fit of the energy with a second-degree polynomial is shown in figure \ref{fig:Edwfit}. We see that a set of values of the energy at a given $l$ (the red dots) are well-approximated by the parabolic curve $p(l) = a\,l^2$ with $a \approx 0.5$ for $\tilde{b} = 0.1$ (blue line).

The reason for this behavior is that at small radii the D6-brane is close to the D8-brane and very flattened, so the energy is coming from a disk shape of almost constant tension.
Instead, for large radii the D6-brane has a region where the embedding is almost vertical in the $u$ direction (resembling a string configuration) before turning inside to close near to the horizon, where its tension is small; so its energy is dominated by its string-like portion.
In other words, from the holographic QFT dual point of view, the tension of the domain wall is reduced in the region far from the string at its boundary. So at large radii, the string dominates the energy, while at small radii the DW world volume is never far from the string, the tension is never small and the whole world volume contributes to the energy.

\subsubsection{Analysis of the first-order phase transition}
\label{subsubphase}
\begin{figure}
\center
\includegraphics[height = 6 cm]{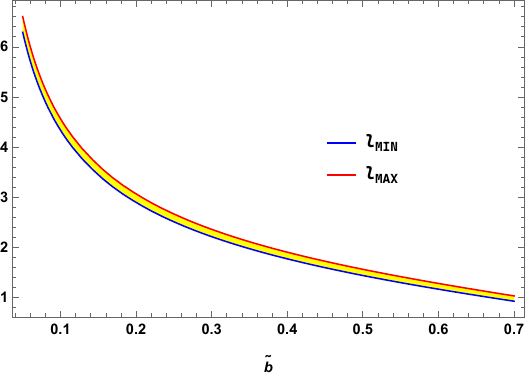}\caption{The blue line is the minimal allowed radius  $l_\text{MIN}$ for string loop configuration, while the red line is $l_\text{MAX}$, the maximal radius allowed for the domain wall configuration. Both are functions of $\tilde{b}$. At a given value of $\tilde{b}$, for $l <  l_\text{MIN}$ only the DW phase exists, while for  $l >  l_\text{MAX}$ only the string loop one is present. Inside the \textit{yellow region},  $l_\text{MIN}< l <  l_\text{MAX}$, both phases coexist.}
\label{comparison}
\end{figure}

In this section, we analyze the competition between the two configurations studied above. As we said, for the string loop solution the asymptotic radius $l$ is limited from below, while for the domain wall, $l$ is limited from above. We can therefore compute the minimal allowed radius for the string loop phase and the maximal allowed radius for the domain wall phase as functions of $\Tilde{b}$ and compare them. The result is shown in figure \ref{comparison} where the blue line is the minimal allowed radius for the string loop configuration $l_\text{MIN}$, while the red line is $l_\text{MAX}$, the maximal radius allowed for the domain wall configuration. At a given value of $\tilde{b}$, for $l <  l_\text{MIN}$ only the DW phase exists, while for  $l >  l_\text{MAX}$ only the string loop one is present. Inside the yellow region,  $l_\text{MIN}< l <  l_\text{MAX}$, both phases coexist.

\begin{figure}
    \centering
    \begin{subfigure}{0.45\textwidth}
        \centering
        \includegraphics[width=\textwidth]{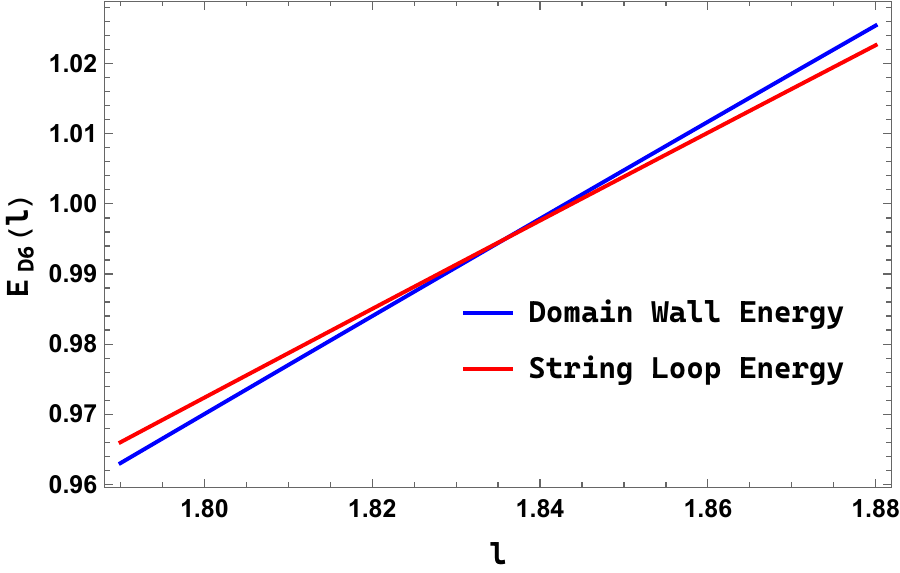}
        \caption{Plot of the free energy of the DW (blue) and string loop (red) configurations in the range of $l$ where both phases coexist. We analyze the model at $\Tilde{b} = 0.4$.}
        \label{fig:energycomparison}
    \end{subfigure}
    \hfill
    \begin{subfigure}{0.45\textwidth}
        \centering
        \includegraphics[width=\textwidth]{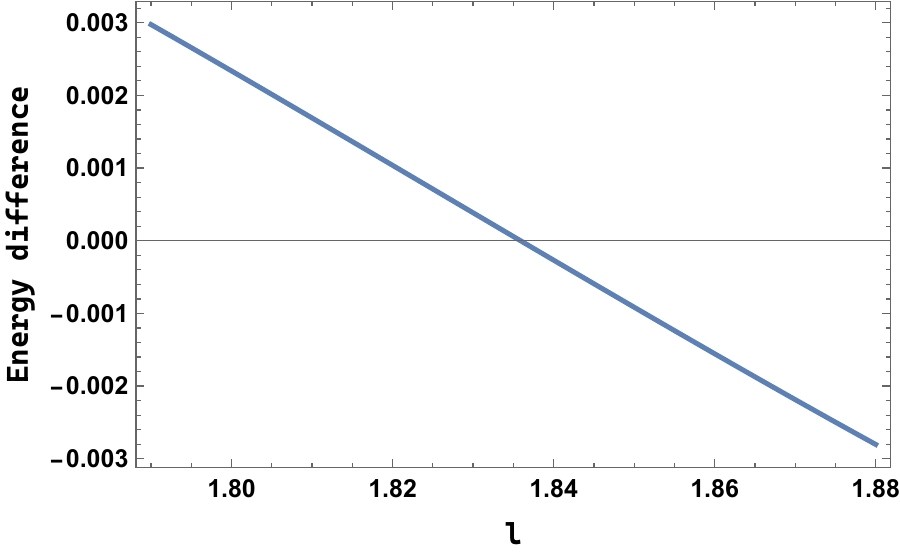}
        \caption{Plot of the difference of free energies, $(E_\text{D6}^{\text{string loop}}-E_\text{D6}^{\text{DW}})(l)$ as a function of $l$, for $\Tilde{b} = 0.4$.}
        \label{fig:energydifference}
    \end{subfigure}
    \caption{Analysis and comparison of the free energy balance in the string loop and DW phases for $\tilde{b} =0.4$.}\label{fig:PTenergy}
\end{figure}

In the small region where the two phases compete, we can compare the free energies of the associated configurations using \eqref{d6energy}. The results are shown in figure \ref{fig:PTenergy}. 
The plot in figure \ref{fig:energycomparison} shows the energy of the DW solution (in blue) and the string loop (in red) configurations. For small $l$ the DW phase is less energetic, while for bigger values of $l$ the string loop phase is less energetic. Despite the difference in energy (see figure \ref{fig:energydifference}) being very small, it shows the characteristic behavior of a first-order phase transition, whose occurrence is ubiquitous in holographic scenarios like the one presented in this work. Even if this analysis refers to the model at temperature $\tilde{b} = 0.4$, the results apply to every value of $\tilde{b}$.\footnote{The behavior of this energy difference with the temperature is not trivial but instead grows if $\tilde{b}$ grows.}

\begin{figure}
    \centering
    \begin{subfigure}{0.45\textwidth}
        \centering
        \includegraphics[width=\textwidth]{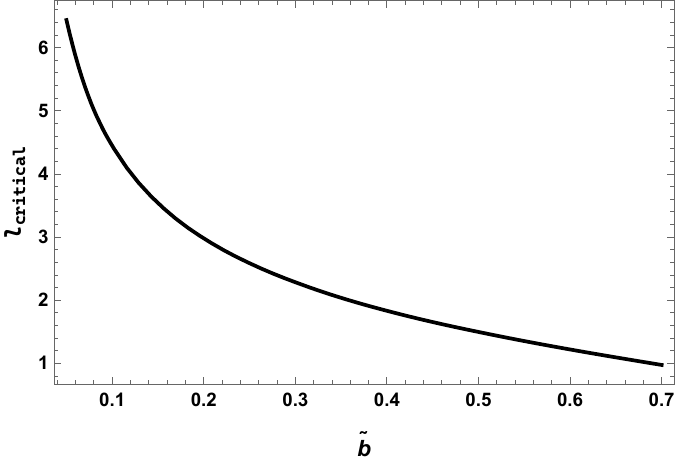}
        \caption{Plot of $l_{\text{critical}}$ as a function of $\tilde{b}$ with $\tilde{b}\in(0.05,0.7)$.}
        \label{fig:lcritico}
    \end{subfigure}
    \hfill
    \begin{subfigure}{0.45\textwidth}
        \centering
        \includegraphics[width=\textwidth]{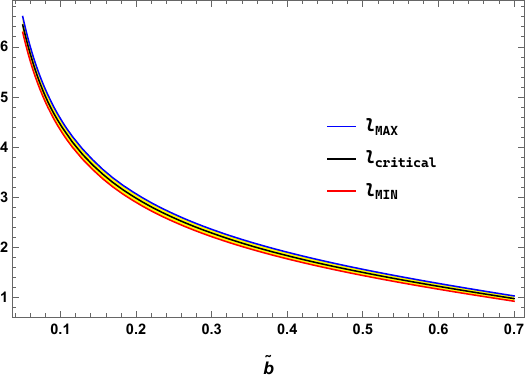}
        \caption{Plot of $l_{\text{critical}}$,  $l_\text{MAX}$ and $l_\text{MIN}$ as a function of $\tilde{b}$ with $\tilde{b}\in(0.05,0.7)$.}
        \label{fig:lminmaxcritico}
    \end{subfigure}
    \caption{Plot of the critical radius as a function of $\tilde{b}$ with the comparison with $l_\text{MAX}$ and $l_\text{MIN}$.}\label{fig:lcr}
\end{figure}

All in all, in the \textit{yellow region} of figure \ref{comparison} where both phases are allowed, we find that there is a critical radius $l_\text{critical}$ such that for $l< l_\text{critical}$ the DW phase dominates while for $l> l_\text{critical}$ the string loop phase dominates over the DW one. The value of $l_\text{critical}$ depends on $\tilde{b}$ and it is the critical value at which the system exhibits a first-order phase transition. We can compute numerically  $l_\text{critical}$ as a function of $\tilde{b}$ and the result is shown in figure \ref{fig:lcritico} and in figure \ref{fig:lminmaxcritico} where the critical radius is plotted together with $l_\text{MIN}$ and $l_\text{MAX}$. 

We can also describe the evolution of the system by varying $\tilde{b}$ or in other words, the temperature of the background. Given an intermediate value of $l$, looking at figure \ref{fig:lminmaxcritico}, we can start for example at a very high temperature (we are thinking about the thermal evolution of an expanding Universe), that is very high $\tilde{b}$, where only the string loop phase exists. 
The D6-brane cannot close off, realizing the DW configuration, because the horizon is very close to the D8-brane tip. 
Then, as the temperature is reduced we enter the \textit{yellow region}, where the DW configuration exists but the string loop phase still prevails until we reach $l_{\text{critical}}$. A first-order phase transition occurs and the DW phase starts to dominate because it is energetically favored. Eventually, by lowering further the temperature, we get out of the competing region and only the DW configuration exists from now on.
That is, in the thermal history of this theory, string loops produced during the chiral symmetry-breaking phase transition can survive up to the confinement phase transition only if they are very large and do not have the time to decay significantly (a phenomenon which reduces their size); in the other cases, reached a certain temperature they undergo a transition where a DW forms inside the loop.

\subsection{Vortons}\label{subsec:spinning}
The uncharged configurations studied in the previous section are unstable. 
The fact that the embeddings do not end orthogonally at the D8-brane is a signal that the D6-branes are going to shrink in size due to their tension.
In fact, both the string loops and the DWs can decay at least by emitting mesons (mostly axions) on the D8-brane. 
It is thus relevant to ask whether it is possible to get (meta)stable solutions turning on the electric and magnetic fields on the D6-brane, \textit{i.e.}~considering charged solutions. 
The charge could counter-balance the tension, and charge conservation would prevent the defects to decay by simple direct axion emission.
In this section we will try to answer this question focusing on the string loop case, leaving the interesting study of the charged DW one for the future \cite{JL}. 

Our D6-brane wrapped on $S^4$ behaves effectively as a D2-brane. Now, it is very well known, in the context of supersymmetric D2-brane supertube solutions in flat spacetime \cite{Mateos:2001qs,Emparan:2001ux,Mateos:2001pi,Kruczenski:2002mn,Bena:2004wt,Bena:2008dw} that the angular momentum induced by the electromagnetic field on the D6-brane can compete with its tension and stabilize it. The question we might ask is thus whether an analogous mechanism occurs in our context. As it will be clear in the following, the fact that our would-be cylindrical D6-brane is probing a curved black brane background \eqref{eq:bkgmetric} with non-zero $F_4$ flux along $S^4$ (which induces a Chern-Simons term in the effective action) crucially leads to several differences w.r.t.~the supersymmetric flat case. 

In fact, the action for our wrapped D6-brane with a non-zero world-volume gauge field $a$ is
\begin{equation}\label{eq:actionspinning}
    S_\text{D6} = -T_{6}V_4R^3\int d^3x\,u\,e^{-\phi}\sqrt{-\det(g_{\mu\nu}+2\pi\alpha' f_{\mu\nu})} + \dfrac{N}{4\pi}\epsilon^{\mu\nu\rho}\int d^3x\,a_\mu\partial_\nu a_\rho\,,
\end{equation}
where the determinant is taken over the indices $\mu,\nu = t,\psi,u$ of the effective, dimensionally reduced spacetime and $f_{\mu\nu} = \partial_\mu a_\nu - \partial_\nu a_\mu$. We use the convention $\epsilon^{t\psi u} = +1$. The D6-brane has two boundaries, one at the tip of the D8-brane and one at the horizon and we need to take care of them. First, the presence of a boundary breaks the gauge invariance of the CS term; therefore, we must pay attention to the boundary conditions for the gauge field. If we consider a variation of the gauge field $\delta a = d\chi$, the CS action variation reads
\begin{align*}
    \delta_\chi S_{CS} = \dfrac{N}{4\pi}\int d\chi\wedge da = \dfrac{N}{4\pi}\epsilon^{\mu\nu\rho}\int dt\,d\psi du\,\partial_\mu\chi\partial_\nu a_\rho = -\dfrac{N}{4\pi}\int_{\Sigma(u=u_J)} dt\,d\psi\,\chi\,f_{t\psi}\,.
\end{align*}
The gauge symmetry is preserved only if we restrict to the flat connection $f_{t\psi}=0$ at $u=u_J$. We will work with an ansatz consistent with this requirement. We have used the standard assumption that $\delta a = 0$ at the horizon \cite{Kobayashi:2006sb,Karch:2007pd,Fujita:2009kw}. Secondly, the  on-shell variation of the CS action gives
\begin{align}\label{eq:CSvariation}
    \delta S_{CS} = \dfrac{N}{4\pi}\int_{\Sigma(u=u_J)} dt\,d\psi\,(a_t\delta a_\psi - a_\psi\delta a_t)\,.
\end{align}
As we will see, this variation will be zero for our solutions.\footnote{In some cases the problem of the non-gauge invariance of the action can be resolved, following \cite{Elitzur:1989nr}, by adding a boundary term 
\begin{equation}
    S_{\text{bdry}} = \dfrac{N}{4\pi}\int_{\Sigma(u=u_J)}dt\,d\psi\,a_t\,a_\psi\,.\nonumber
\end{equation}
However, we will see that it is not necessary to introduce this term for our system. In more general contexts, even if this term does not affect the equations of motion, it is relevant if we want to compute the conductivity of the Hall state described by the D6-brane. Let us just remind that quantum Hall physics has been widely discussed in holographic contexts, see for example \cite{Fujita:2009kw,Hikida:2009tp,Jokela:2011eb}.}

We make the following ansatz for the gauge field components:\footnote{This ansatz is consistent with having the flat connection $f_{t\psi}=0$ at $u=u_J$.}
\begin{equation*}
    a_t(u)\neq 0\,,\,\,\,\,\,\,\,\,\,\,\,\,\,\,\,a_\psi(u)\neq 0\,,\,\,\,\,\,\,\,\,\,\,\,\,\,\,\,a_u = 0\,.
\end{equation*}
After some algebra, the action can be written as
\begin{align}
    S_\text{D6} = &-C\int dt\,du\,\rho(u)\,u\,\sqrt{1+f_T(u)\left(\dfrac{u}{R}\right)^3\rho'(u)^2 + \dfrac{f_T(u)}{\rho^2(u)}\,(\partial_u\bar{a}_\psi)^2-(\partial_u \bar{a}_t)^2} \,\,+\\ \nonumber
    &+ \dfrac{3C}{2}\int dt\,du\,(\bar{a}_\psi\partial_u \bar{a}_t -\bar{a}_t\partial_u \bar{a}_\psi)\,,\,\,\,\,\,\,\,\,\,\,\,
\end{align}
where $C$ has been defined in \eqref{defC} and the barred quantities are rescaled with $2\pi\alpha'$ as 
\begin{equation}
\bar{a} = 2\pi\alpha'\,a \,.
\end{equation}

Defining
\begin{equation}
    D(u) \equiv \sqrt{1+f_T(u)\left(\dfrac{u}{R}\right)^3\rho'(u)^2 + \dfrac{f_T(u)}{\rho^2(u)}\,(\partial_u\bar{a}_\psi)^2-(\partial_u \bar{a}_t)^2}\,,
\end{equation}
we can write the equations of motion derived from the action above as follows.
\begin{itemize}
    \item Equation of motion for $\rho(u)$
    \begin{align}\label{eq:eqrho}
  \partial_u\left(\dfrac{u\,\rho(u)}{D(u)}\left(\dfrac{u}{R}\right)^3f_T(u)\rho'(u)\right)- u\,D(u) + \dfrac{u\,f_T(u)}{D(u)}\dfrac{(\partial_u\bar{a}_{\psi})^2}{\rho^2(u)}=0\,.
    \end{align}
    \item Equation of motion for $\bar{a}_t(u)$
    \begin{align}\label{eq:eqat}
     \partial_u\left(\dfrac{u\,\rho(u)}{D(u)}\partial_u\bar{a}_{t}\right) +3\partial_u\bar{a}_\psi=0\,.
    \end{align}
     \item Equation of motion for $\bar{a}_\psi(u)$
    \begin{align}\label{eq:eqapsi}
     \partial_u\left(\dfrac{u}{D(u)}\dfrac{f_T(u)}{\rho(u)}\partial_u\bar{a}_{\psi}\right)+3\partial_u\bar{a}_t=0\,.
    \end{align}
\end{itemize}
The equations of motion for the gauge field components are total derivatives, so from the equations \eqref{eq:eqat} and \eqref{eq:eqapsi} we get
\begin{align}\label{eq:defktkpsi}
    &\dfrac{u\,\rho(u)}{D(u)}\partial_u\bar{a}_{t}+3\bar{a}_\psi = \bar{\mathbf{k}}_t\,,\\\label{eq:defG}
    &\dfrac{u}{D(u)}\dfrac{f_T(u)}{\rho(u)}\partial_u\bar{a}_{\psi}+3\bar{a}_t = \bar{\mathbf{k}}_\psi\,,  
\end{align}
where $\bar{\mathbf{k}}_t$ and $\bar{\mathbf{k}}_\psi$ are two constants of motion to be fixed. As for the gauge field, the barred constants are related to un-barred ones through $\bar{\mathbf{k}}_{t,\psi} \equiv 2\pi\alpha'\,\mathbf{k}_{t,\psi}$.

In the supertube case, a charged cylindrical D2-brane in flat space supports both a D0-brane charge and a fundamental string charge. Analogously, our wrapped charged D6-brane can support a charge for a D4-brane wrapped on $S^4$, hence a baryon charge $n_B$, and a fundamental string charge equal to the number of quarks building up the bound state described by the D6-brane. 

The induced baryon charge can be read from the D6-brane CS coupling
\begin{equation}
\frac{1}{2\pi}\int C_5\wedge f = -\int du\, \partial_ua_\psi\int C_5\,,
\end{equation}
so that
\begin{equation}
\label{eq:nbD6}
n_B = -\int du\, \partial_ua_\psi\,.
\end{equation}

The induced fundamental string charge $q_s$ is given by
\begin{equation}
    q_s = \dfrac{\partial\mathcal{L}_\text{D6}}{\partial(\partial_u a_t)} = (2\pi\alpha')^2C\left(\dfrac{u\,\rho(u)}{D(u)}\,\partial_u a_t+3a_\psi\right) = N\,\dfrac{\mathbf{k}_t}{3}\,,
\label{qscharge}    
\end{equation}
so that, if we are interested in an object with string charge $q_s = m N$ the integration constant must be set as $\mathbf{k}_t = 3m$.

The angular momentum of our configuration along the $u$ axis is given by
\begin{align}
    J = \int d^6x\,\sqrt{-g}\,g_{\psi\psi}T^{t\psi},
\end{align}
where 
the energy-momentum tensor density associated with the action \eqref{eq:actionspinning} reads
\begin{equation}
    T_{MN} = -\frac{2}{\sqrt{-g}} \frac{\delta \mathcal{L}_\text{D6}}{\delta g^{MN}} = T_6\, e^{-\phi}\, \left(-g_{MN} \, \sqrt{1+ \frac{1}{2}g^{AB} g^{CD} \bar{f}_{AC} \bar{f}_{BD}} + \dfrac{ g^{AB}\,\bar{f}_{MA}\bar{f}_{NB}} {\sqrt{1+ \dfrac{1}{2}g^{AB} g^{CD} \bar{f}_{AC} \bar{f}_{BD}}}\right),
\end{equation}
where the indices $M,N,A,B,C,D$ run over the world volume coordinates of the D6-brane, and $g$ is the induced D6-brane metric. 
We are interested in the component
\begin{equation}
    T^{t\psi} =\dfrac{ T_6\,e^{-\phi}}{\sqrt{-g_{tt}g_{\psi\psi}g_{uu}}}\dfrac{\partial_u\bar{a}_t\partial_u\bar{a}_\psi}{\sqrt{-[g_{tt}(g_{\psi\psi}g_{uu}+(\partial_u\bar{a}_{\psi})^2)+g_{\psi\psi}(\partial_u\bar{a}_{t})^2]}}\,,
\end{equation}
from which we can compute the angular momentum. After some algebra, using the equation of motion \eqref{eq:defG}, we get
\begin{align}\label{eq:JD6}
    J = \dfrac{N}{2}\int du\,\partial_u\left[\left(a_\psi -\dfrac{\mathbf{k}_t}{3}\right)^2\right] 
    =\dfrac{N}{2}\int du\,\dfrac{\rho^2(u)}{f_T(u)}\partial_u\left[\left(a_t -\dfrac{\mathbf{k}_\psi}{3}\right)^2\right].
\end{align}

Notice that the results \eqref{eq:nbD6}, \eqref{qscharge}, and \eqref{eq:JD6} are general, and they hold for all kinds of D6-brane embeddings. We are going to apply them to the string loop embedding for which the $u$-integration runs from the horizon $u_T$ to the D8-brane tip $u_J$. We can look for a general relation between the baryon number \eqref{eq:nbD6} and the angular momentum \eqref{eq:JD6}. Both can be integrated to give
\begin{align}\label{eq:nbD6general}
    & n_B = a_\psi(u_T) - a_\psi(u_J)\,,\\ \label{eq:JD6general}
    & J = \dfrac{N}{2}\bigg[\left(a_\psi(u_J) -\dfrac{\mathbf{k}_t}{3}\right)^2-\left(a_\psi(u_T) -\dfrac{\mathbf{k}_t}{3}\right)^2\bigg]\,.
\end{align}
In the case that $a_\psi(u_T) = \mathbf{k}_t/3$, which is an allowed boundary condition, the following relation holds
\begin{equation}\label{eq:Jnbrelation}
    \boxed{J = \frac{N}{2}n_B^2}
\end{equation}
The quadratic relation of the charge with the angular momentum is the hallmark of anyonic systems, enforcing the result that these topological defects are excitations of a topological phase described by an $U(1)_N$ CS theory.

Notice that the boundary condition $a_\psi(u_T) = \mathbf{k}_t/3$ is in fact unique if (\ref{eq:Jnbrelation}) has to hold, in the sense that no other possibilities are compatible with the equations of motion for a (meta)stable charged string loop case. Indeed, by analyzing the equation of motion (\ref{eq:eqrho}), it turns out that the choice $a_\psi(u_J) = \mathbf{k}_t/3$ is in contradiction with the conditions $\rho'(u_J)=0$ and $\rho''(u_J)<0$, which give the desired profile.\footnote{Profiles with $\rho''(u_J)>0$ tend to describe particular DW configurations rather than vortons.}

Let us consider objects with a baryon number $n_B$ which satisfy the relation (\ref{eq:Jnbrelation}). 
The vorton, holographically described by a spinning D6-brane, \textit{i.e.}~a D6-brane  with $J \neq 0$, exists only in the state with $n_B$ and $J = + (N/2)n_B^2$. This is easily understandable from the FQHE's point of view. The presence of the CS term breaks explicitly the discrete symmetries $P$ (parity) and $T$ (time reversal), therefore given an object with total angular momentum $|J| = (N/2)n_B^2$, we can describe only the state with one particular projection of the angular momentum, in this case, $J=+(N/2)n_B^2$. In the literature, this phenomenon is well-established and understood in various contexts. See for example \cite{Floreanini:1987as,Jackiw:1990pr,Dunne:1990qe,Banerjee:1996np,Dunne:1998qy,Bolognesi:2007ez}. However, in general, we should be able to describe also the state with $J =-(N/2)n_B^2$. This can be done in terms of a $\overline{\text{D6}}$-brane since the opposite charge flips the sign of the CS term (equivalently the level of the CS theory), and as a result, only the state with a baryon number $n_B$ and angular momentum $J = -(N/2)n_B^2$ exists. In the latter case, taking now $a_\psi(u_T) = -\mathbf{k}_t/3$,\footnote{This relation is affected by the charge-flip and this is why it gets a minus sign w.r.t. the D6-brane case.} we have
\begin{equation}
n_B = a_\psi(u_J) + \dfrac{\mathbf{k}_t}{3}\,,\quad J = -\dfrac{N}{2}\left(a_\psi(u_J) + \dfrac{\mathbf{k}_t}{3}\right)^2\,.
\end{equation}

Importantly, altogether we can describe all the possible states with baryon number $n_B$ by means of D6 and $\overline{\text{D6}}$-branes.

Let us comment on the relation between $n_B$ and $q_s$. Going back to the variation of the CS term \eqref{eq:CSvariation}, we realize that we have to set $\delta a_\psi$ at $u_J$ to zero if we do not want to change the baryon number. Then, in order for the CS variation $\delta S_{CS}$ to vanish we require that $a_\psi(u_J) = 0$. 
This condition automatically leads to $\mathbf{k}_t/3 = n_B$ and hence to $q_s = Nn_B$ which is expected both from the gravity side and the dual quantum field theory interpretation. 

Let us outline that, in the charged domain wall case, where $u_E > u_T$, the position of the tip of the D6-branes, replaces $u_T$ in the equations \eqref{eq:nbD6general}, \eqref{eq:JD6general}, the relation $|J|=(N/2)n_B^2$ is automatically satisfied. In fact, from the equation of motion 
\eqref{eq:defktkpsi} and the regularity condition at the tip (where $\rho(u_E) = 0$) it follows that $a_{\psi}(u_E)=\mathbf{k}_t/3$. These solutions correspond to the axionic-baryons, ``a-baryons", introduced in \cite{Bigazzi:2022ylj}.
From the field theory point of view, the property that a baryon has spin $N/2$ comes from the fact that with a single flavor the spins of the quarks must be aligned.

These results have consequences on the $\lambda$-dependence of the fields in the vorton case. 
Since in this case the integration domain in (\ref{eq:nbD6}) is of order zero in $\lambda$, we need to have $a_{\psi}\sim n_B\lambda^0$. Moreover, in order for $J$ not to scale with $\lambda$, the fields $\rho$ and $a_t$ have to scale as
\begin{equation}\label{eq:scalingrhoat}
    \rho\sim\lambda^{-\gamma}\,,\,\,\,\,\,\,\,\,\,\,\,\,\,\,\,\,\,\,\,a_t\sim\lambda^\gamma\,,
\end{equation}
for some $\gamma \geq 0$. We rule out immediately the case $\gamma>1$ because it leads to an imaginary action.\footnote{Remember that we are working in the large-$\lambda$ limit.} Moreover, for $\gamma\in[0,1)$, in the large-$\lambda$ limit, the equations of motion do not admit solutions. 

Actually, from the equations of motion (\ref{eq:eqrho}), (\ref{eq:eqat}), (\ref{eq:eqapsi}), the condition $\rho'(u_J)=0$, the relations $\alpha'\sim\lambda^{-1}$ and $\mathbf{k}_t/3 = n_B$, and the above observation concerning the scaling of $a_{\psi}$, it follows that $a'_t(u_J)\sim \lambda$ and
\begin{equation}
l\equiv \rho(u_J)\sim \frac{n_B}{\lambda}\,.
\label{ellescaling}
\end{equation}
Thus, if a (meta)stable vorton solution exists, the vorton radius $l$ is expected to scale as above. 

It is interesting to notice that the above result also agrees with the following estimates. Let us approximate the vorton as a rigid non-relativistic rotor of radius $l$, with mass $M$, moment of inertia $I$, and angular momentum $J$ given by
\begin{equation}
    M = a\,\lambda^2\,N\,l,\hspace{1cm} I = b Ml^2,\hspace{1cm}J = \dfrac{N}{2}n_B^2,
\end{equation}
where the coefficients $a,b$ do not scale with $\lambda$ and the scaling of $M$ reproduces what we have found studying the energy of the D6-brane corresponding to the uncharged string loop. Then, the total energy of the rotor can be written as
\begin{equation}    
E_\text{TOT} = M + \,\dfrac{J^2}{2I} =a\,\lambda^2Nl + \dfrac{1}{8a b}\dfrac{Nn_B^4}{\lambda^2l^3}.
\end{equation}
The minimization over $l$ gives $l_\text{stable}\sim n_B\lambda^{-1}$ consistently with eq. (\ref{ellescaling}).

It is easy to realize that if  $n_B={\cal O}(\lambda^{0})$ the equations of motion for the embedding and the gauge field on the D6-brane, do not admit (meta)stable vorton solutions. In fact in this case we would need to have $\rho(u)\sim\lambda^{-1}$ for any $u$: in the large-$\lambda$ limit, this would lead to a solution to \eqref{eq:eqrho} with a constant radius that is inconsistent with the other equations of motion. 

Actually, if $n_B={\cal O}(\lambda^{0})$ it turns out that the gauge field is subleading w.r.t.~the profile $\rho(u)$, and as such it can be treated in a probe approximation. But such approximation does not change the embedding solution found in the previous section and thus, in particular, the charged string loop remains unstable: it is not possible to find a solution with $\rho'(u_J)=0$, which is the condition for stability at the tip of the D8-branes. We may still envisage a kind of first-order phase transition - happening when the D6-brane's radius $l=\rho(u_J)$ is of order zero in $\lambda$ - between this configuration and the charged DW one.  

The instability of the charged string loop solution with order-one baryon charge is easily seen also by looking at the free energy in the probe approximation.
The expansion up to quadratic order in the gauge field of the free energy reads
\begin{align}\nonumber
    &E_\text{D6} \approx N\lambda^2 \widetilde{C}\int dt\,du\,\rho(u)\,u\,D_0(u)+ \dfrac{N}{2}\int dt\,du\,\dfrac{\rho(u)\,u}{D_0(u)}\bigg[-(\partial_ua_t)^2+\dfrac{f_T(u)}{\rho(u)^2}(\partial_ua_\psi)^2\bigg]\\
    &\hspace{1cm}- \dfrac{N}{2}\int dt\,du\,(a_\psi\partial_u a_t -a_t\partial_u a_\psi)\,,
\end{align}
where
\begin{equation}
    \widetilde{C} \equiv \dfrac{1}{3\cdot 2^4\pi^2R^6M_{KK}^2}\,.
\end{equation}
From the above expression of the free energy, we can deduce that no stability radius $l$ can exist with scaling $\lambda^0$. The Maxwell-Chern-Simons free energy is of order $\lambda^0$ if $l\sim\lambda^0$, so it cannot counter-balance the energy due to the tension from the first term (which scales as $\lambda^2$), and no (meta)stable solution can be found. In order to be stable, the vorton radius should be suppressed with a power of $\lambda$, which is not possible as we have seen in the previous section (the minimal radius of the string loop was of order $\lambda^0$). 
Thus, if a string loop of radius of order $\lambda^0$ has an order-one baryon charge, it is still unstable.
It will evolve in a smaller configuration with the same charge by emitting axions, eventually decaying completely (for example in a charged DW) before its size can become parametrically small in $\lambda$.
Phenomenologically, this defect is expected to decay producing axions and gravitational waves (if one adds gravitational interactions to the story).

\subsubsection{Vortons with large baryon charge}
The conclusions about the instability of the vorton come from the requirement that such a solution should have  baryon number and angular momentum of order $\lambda^0$. As a consequence, the gauge field can be treated in the probe approximation and the vorton cannot be stabilized by the D6-brane spin or the Coulomb repulsion due to the electric field.

However, there is the case where the vorton solution can be at least metastable if we allow it to carry a large baryon charge of order $\lambda$. It can be shown, by a parametric analysis, that this is the only possible choice. Let us then consider 
\begin{equation}
    n_B = -\int^{u_J}_{u_T}du\,\partial_u a_\psi \sim \lambda\,,
\end{equation}
which, according to eq.~(\ref{ellescaling}), would correspond to a vorton radius $l$ of order $\lambda^{0}$. In this case $a_\psi\sim n_B\sim \lambda$. Then, by looking at the equation \eqref{eq:JD6}, if we want a vorton with $\rho\sim\lambda^0$ we must also have $a_t\sim\lambda$. These scalings for the fields completely factorize the $\lambda$ dependence out of the action. To simplify the latter, we rescale the coordinate $u$ and the fields as
\begin{equation}\label{resca}
	u\to u = u_J\,\hat{u}\,,\,\,\,\,\,\,\,\,\,\,\,\,\rho\to \rho=L\,\hat{\rho}\,,\,\,\,\,\,\,\,\,\,\,\,\,\bar{a}_t\to \bar{a}_t=u_J\,\hat{a}_t\,,\,\,\,\,\,\,\,\,\,\,\,\,\bar{a}_\psi\to \bar{a}_\psi=L\,u_J\,\hat{a}_\psi\,.
\end{equation}
The baryon number $n_B$ and the angular momentum $J$, using these rescaled fields, are written as
\begin{align}\label{eq:nblambda}
    &n_B = \dfrac{\lambda}{4\pi M_{KK}L}\,J_T^2(\tilde{b})\left(-\hat{a}_\psi(\hat{u}_J) + \dfrac{\hat{\mathbf{k}}_t}{3}\right),\\
    &J = \dfrac{N}{2}\,n_B^2 = \dfrac{\lambda^2N}{2^5\pi^2 M_{KK}^2L^2}\,J_T^4(\tilde{b})\left(-\hat{a}_\psi(\hat{u}_J) + \dfrac{\hat{\mathbf{k}}_t}{3}\right)^2\,,
\end{align}
where $\hat{\mathbf{k}}_t$ is rescaled as $a_t$ in (\ref{resca}). Dropping the hats the action then reads
\begin{equation}
 (CLu_J^2)^{-1}S_\text{D6} = -\int dt\,du\,\rho(u)\,u\,\hat{D}(u) \,+ \dfrac{3}{2}\int dt\,du\,(a_\psi\partial_u a_t -a_t\partial_u a_\psi)\,,
 \end{equation}
where
\begin{equation}
\hat{D}(u) = \sqrt{1+J_T(\tilde{b})^2f_T(u)u^3\rho'(u)^2 + \dfrac{f_T(u)}{\rho^2(u)}\,(\partial_ua_\psi)^2-(\partial_u a_t)^2}\,.
\end{equation}
The corresponding equations of motion read
    \begin{align}
  &J_T(\tilde{b})^2\partial_u\left(\dfrac{u^4\,\rho(u)}{\hat{D}(u)}f_T(u)\rho'(u)\right)- u\,\hat{D}(u) + \dfrac{u\,f_T(u)}{\hat{D}(u)}\dfrac{(\partial_ua_{\psi})^2}{\rho^2(u)}=0\,,\\
  & \dfrac{u\,\rho(u)}{\hat{D}(u)}\partial_ua_{t}+3a_\psi=\mathbf{k}_t\,,\\
  &\dfrac{u}{\hat{D}(u)}\dfrac{f_T(u)}{\rho(u)}\partial_ua_{\psi}+3a_t=\mathbf{k}_\psi\,.
    \end{align}

Solutions describing metastable vortons with a large baryonic charge $n_B\sim\lambda$ and spin $J\sim (N/2) \lambda^2$ have been found numerically shooting from the D8-brane tip and imposing the following boundary conditions for the \textit{hatted} fields: $\rho'(u_J) = 0$ to guarantee the local stability of the solution; $a_\psi(u_T) = \mathbf{k}_t/3$ to realize the relation \eqref{eq:Jnbrelation} between the baryon number and the angular momentum; $a_\psi(u_J) = -J_T({\tilde b})^{-2}+ \mathbf{k}_t /3$ to fix a particular $T$-independent value for the baryon number; orthogonality of the configuration at the horizon, again for local stability. 
The integration constants $\mathbf{k}_t$ and $\mathbf{k}_\psi$ are then fixed as follows: $\mathbf{k}_t/3 = J_T^{-2}(\tilde{b})$ in order to have the right number of fundamental strings corresponding to the baryon number $n_B$; $\mathbf{k}_\psi$ fixed so to enforce regularity of the gauge field component $a_t$ at the horizon. 

\begin{figure}
    \centering
    \begin{subfigure}{0.47\textwidth}
        \centering        \includegraphics[width=\textwidth]{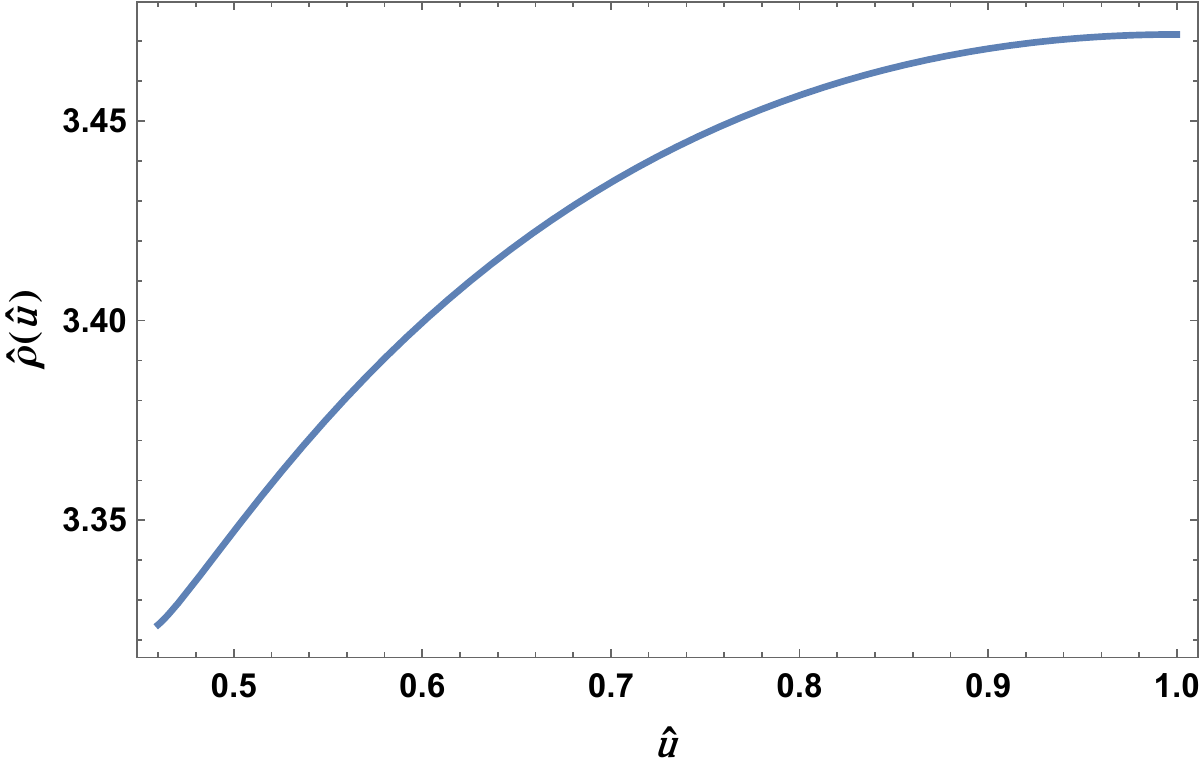}
        \caption{Plot of the solution $\hat{\rho}$ for $\tilde{b} = 0.45$.}
        \label{fig:largechargeprofile}
    \end{subfigure}
    \hfill
    \begin{subfigure}{0.47\textwidth}
        \centering
        \includegraphics[width=\textwidth]{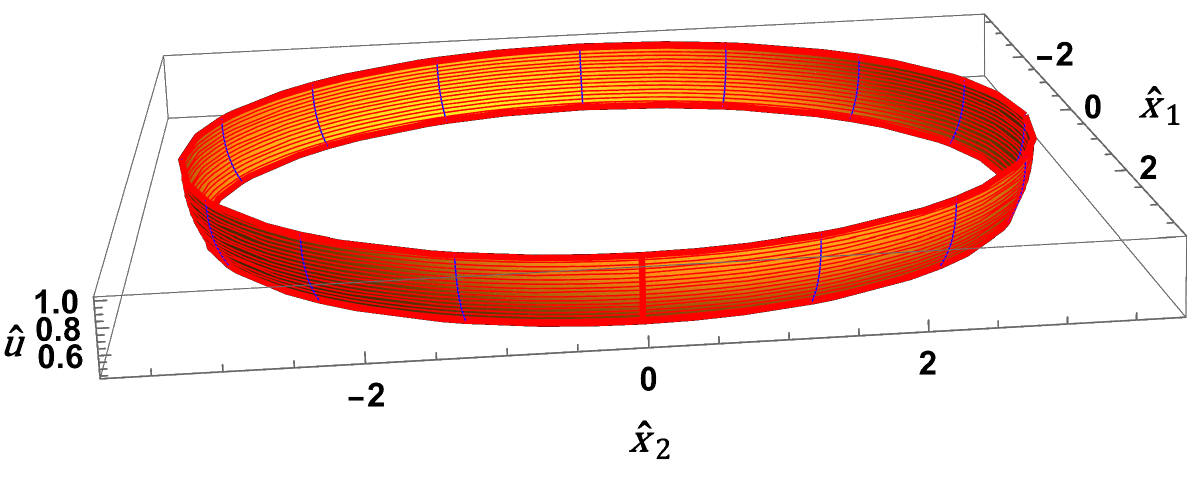}
        \caption{Revolution plot around the $\hat{u}$ axis of the embedding $\hat{\rho}$  for $\tilde{b} = 0.45$.}
        \label{fig:largechargerevolution}
    \end{subfigure}
    \caption{Numerical solution of $\hat{\rho}$ for $\tilde{b}=0.45$ as function of the holographic coordinate $\hat{u}$, describing the metastable vorton configuration with large charge. The profile derivative at $\hat{u} = 1$ is zero indicating the (meta)stability of this embedding. The $\hat{u}$ variable runs from the horizon $\tilde{b}=0.45$ to the D8-brane tip.}\label{fig:rholargecharge}
\end{figure}
\begin{figure}
    \centering
    \begin{subfigure}{0.47\textwidth}
        \centering        \includegraphics[width=\textwidth]{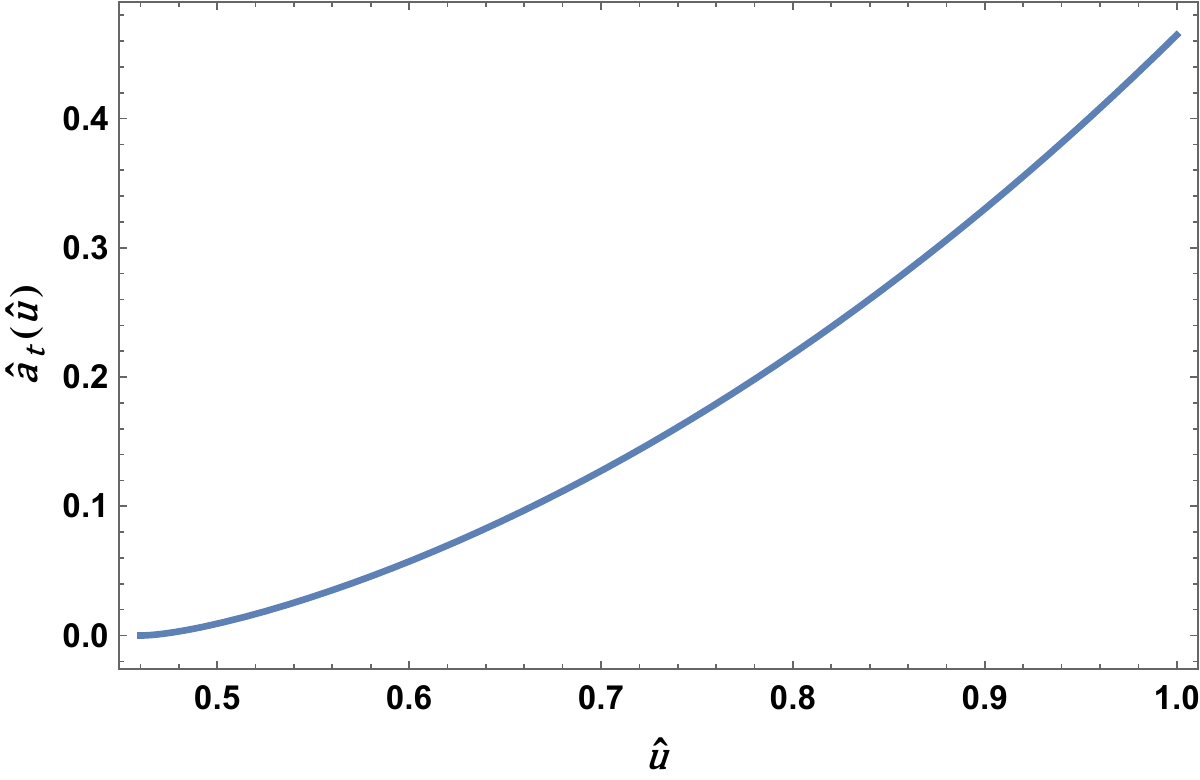}
        \caption{Numerical solution for $\hat{a}_t$ for $\tilde{b} = 0.45$. The gauge field component $\hat{a}_t$ goes to zero to enforce regularity at the horizon.}
        \label{fig:atlargecharge}
    \end{subfigure}
    \hfill
    \begin{subfigure}{0.47\textwidth}
        \centering
        \includegraphics[width=\textwidth]{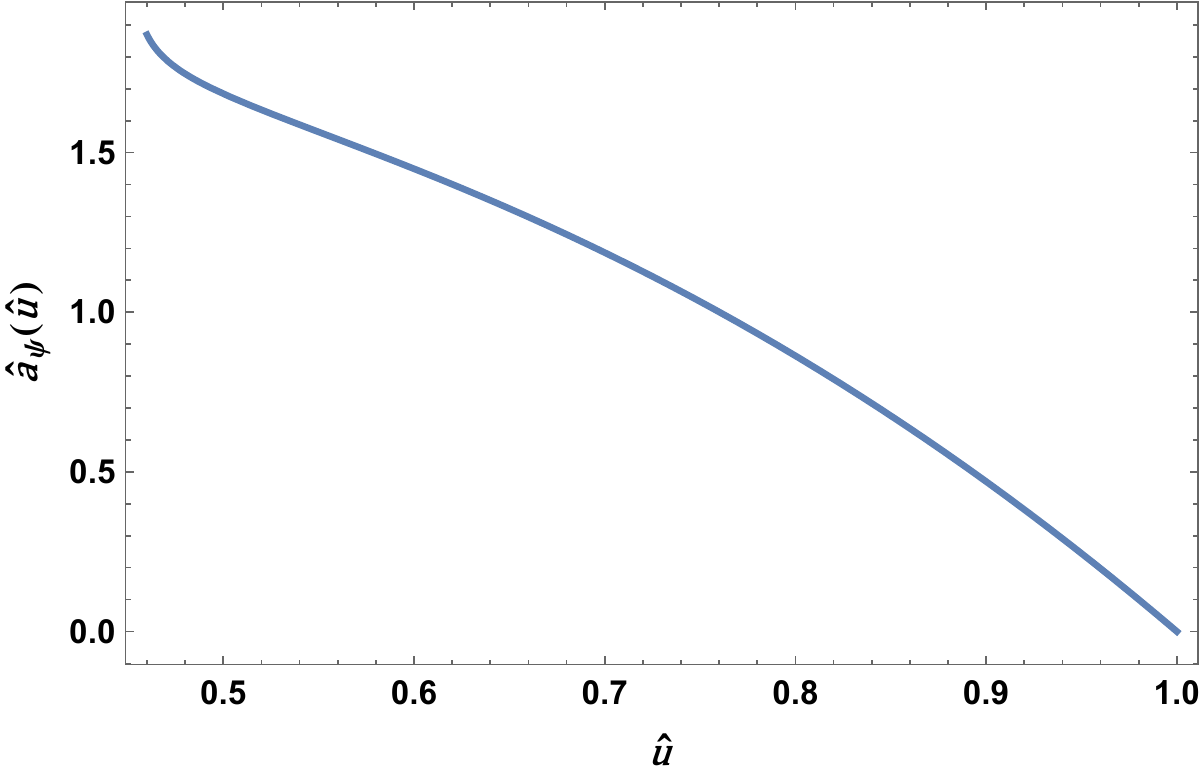}
        \caption{Numerical solution for $\hat{a}_\psi$ for $\tilde{b} = 0.45$. This solution has the right boundary conditions to give baryon number $\lambda$.}
        \label{fig:apsilargecharge}
    \end{subfigure}
    \caption{Numerical solutions for $\hat{a}_t$ and $\hat{a}_\psi$ for $\tilde{b}=0.45$ as function of the holographic coordinate $\hat{u}$.}\label{fig:atpsilargecharge}
\end{figure}

Examples of the related numerical solutions for the intermediate temperature corresponding to $\tilde{b} = 0.45$ are shown in figures \ref{fig:rholargecharge}, \ref{fig:atpsilargecharge}. Remember that these plots refer to the hatted fields, as indicated in the plots' labels. The plot of the profile $\rho(u)$ and its revolution plot around the $u$ axis of the profile are shown in figures \ref{fig:largechargeprofile} and \ref{fig:largechargerevolution} respectively. 
Figure \ref{fig:atlargecharge} shows the plot of $a_t(u)$, which goes to zero at the horizon consistently with regularity. The plot of the gauge field component $a_\psi(u)$ is shown in figure \ref{fig:apsilargecharge}. 

We find numerically that for very small values of $\tilde{b}$ such vorton solutions can no longer be found. This suggests the presence of a similar phase transition to the baryon phase as it happens in the uncharged case. We plan to clarify this phenomenon in a future work \cite{JL}.

In a sentence, these big vortons are stabilized by their huge charge, which balances their tension. 
Moreover, these configurations (representing a bound state of quarks) have smaller free energy than the corresponding-charge state of free quarks, i.e.~than a bunch of fundamental strings with string charge $q_s = N n_B$ extending from the D8-branes' tip to the horizon. The free energy of the latter can be written as
\begin{equation}
    E_\text{f-strings} = \dfrac{q_s}{2\pi\alpha'}\int du\sqrt{-g} = \dfrac{q_s}{2\pi\alpha'}(u_J-u_T)=\dfrac{\lambda^2N}{M_{KK}^2}\left(\dfrac{u_J}{R^3}\right)^{3/2}\dfrac{1-\tilde{b}}{2^4\pi^2J_T(\tilde{b})}\,.
\end{equation}
In the last equality, we have used $q_s = Nn_B$ with $n_B$ given by formula \eqref{eq:nblambda} together with the boundary condition $\hat{a}_\psi(u_J) -\mathbf{k}_t/3 = -J_T^{-2}(\tilde{b})$. The free energy associated with the large-charged vorton can be written, in terms of the hatted solutions discussed above, as
\begin{equation}
    E_\text{vorton} = \dfrac{\lambda^2N}{M_{KK}^2}\left(\dfrac{u_J}{R^3}\right)^{3/2}\dfrac{J_T(\tilde{b})}{3\cdot2^4\pi^2}\bigg[\int du\,\rho(u)\,u\,\hat{D}(u) \,- \dfrac{3}{2}\int du\,(a_\psi\partial_u a_t -a_t\partial_u a_\psi)\bigg].
\end{equation}
The energy comparison reduces to compare two functions of $\tilde{b}$. Numerically it can be shown that $E_\text{f-strings} >  E_\text{vorton}$ for every value of $\tilde{b}$ for which a vorton solution can be found. The relative difference $\text{rd} = |E_\text{vorton}-E_\text{f-string}|/E_\text{f-string}$ is a linear function of $\tilde{b}$. For $\tilde{b} = 0.45$, rd $\simeq 0.35$, and it decreases as $\tilde{b}$ grows.

Nevertheless, these highly-charged vortons could still be only metastable: there exist other configurations towards which the big vortons could decay, for example, a DW with the same baryonic charge or a bunch of small-charged DWs.
We are going to investigate this issue in a separate study \cite{JL}.

\section{D8-brane description of the defects}\label{sec:D8}

In this section, we study the charged string loop configurations from the point of view of the D8-brane world-volume theory.
The latter provides the description of the strings in terms of mesonic degrees of freedom, which are dominated by the (light) axion.

The analysis of the D8-D6 setup in the WSS model involves the action \eqref{sd8} and \eqref{cscompleto} for the U-shaped D8-branes and the action $S_{C_{7}}$ for the $C_7$ RR form in the presence of a localized D6-brane source. The latter reads
\begin{equation}
\label{c7inf}
S_{C_{7}} = -\dfrac{1}{4\pi}(2\pi l_{s})^{6}\int dC_{7}\wedge\star dC_{7} + \dfrac{1}{2\pi}\int C_{7} \wedge \frac{F}{\sqrt{2}}\wedge\omega_{1} + \int C_{7}\wedge\omega_{3}\,,
\end{equation}
where $F$ is the $U(1)$ gauge field on the D8-brane and $\omega_{1}$ and $\omega_3$ are, respectively, the differential forms representing the D8-brane and the D6-brane embeddings in the ten-dimensional spacetime.

As in the previous sections, we consider a D6-brane which wraps the internal background $S^4$ and is extended along the holographic coordinate $u$ and one Minkowski coordinate. 

The equation of motion for $C_7$ deduced from \eqref{c7inf} leads to a violation of the Bianchi identity for the D8-brane gauge field (see also \cite{Bigazzi:2022luo}). In particular, if, as in section \ref{subsec:straight}, the boundary of the D6-brane attached to the tip of the U-shaped D8-brane is a straight line along, say, the $x_1$ direction, then the non-trivial Bianchi identity reads
\begin{equation}
    dF = -2\pi\sqrt{2}\,\delta(x_2)\delta(x_3)\delta(z)dx_2\wedge dx_3\wedge dz\,,
\end{equation}
being $z$ the holographic coordinate as defined in \eqref{uzeta}. This setup has already been studied in \cite{Bigazzi:2022ylj}. 

If the boundary of the D6-brane is, as in section \ref{subsec:axldw}, a circular loop of radius $l$ in the Minkowski subspace $(x_1,x_2)$, then the Bianchi identity becomes
\begin{equation}
    dF = -2\pi\sqrt{2}\,\delta(\rho - l)\delta(x_3)\delta(z)d\rho\wedge dx_3\wedge dz\,,
\label{modb}
\end{equation}
where $\rho = \sqrt{x_1^2+x_2^2}$\,. We are going to focus on this setup in the present section.

In particular, we will study the solutions of the Maxwell-Bianchi problem for the D8-brane gauge field in various steps. First of all, we will look for a solution near the tip of the D8-brane, where the induced metric, in the subspace transverse to $S^4$, behaves effectively as a five-dimensional flat metric. This analysis aims to study the system with a non-zero baryon charge and to extract the stability radius $l_\text{stable}$ due to Coulomb repulsion mediated by the CS term. Then we will move to the asymptotic region where we can solve exactly the equations of motion if the electric potential is turned off; in the presence of the electric potential, we will find linearized solutions following the procedure outlined in \cite{Hashimoto:2008zw}.  

\subsection{A solution near the D8-brane tip }
Let us start by considering the D8-D6 setup in the region close to the tip of the U-shaped D8-brane. This is where the circular boundary of the D6-brane, acting as a monopole source for the magnetic field on the D8-brane according to \eqref{modb}, lies. The near-tip region is thus close to the source and it is expected to capture the main contribution to the full solution. As it will be clear in a moment, the DBI action for the D8-brane in this region effectively reduces to that of a D4-brane in a flat five-dimensional spacetime. In this effective description, the wrapped D6-brane source can be seen as a D2-brane with the same circular boundary. Hence the whole setup, in this limit, resembles that examined in \cite{Kruczenski:2002mn} in the context of the supersymmetric supertubes, with the crucial difference being the occurrence, in our case, of a CS term in the five-dimensional effective action.

The induced metric \eqref{ushapmet} on the U-shaped D8-brane near the tip $u=u_J$, hence using \eqref{uzeta} near $z=0$, reads 
\begin{align}\nonumber
ds^2 &= \left(\dfrac{u_J}{R}\right)^{3/2}\left(-\left(1-\tilde{b}^3\right)dt^2 + d\rho^2 + \rho^2d\psi^2 + dx_3^2 + \dfrac{4R^3}{3u_J}\dfrac{1}{(8-5\tilde{b}^3)}dz^2\right) + R^{3/2}u_J^{1/2}d\Omega_4^2=\\
    & = \left(\dfrac{u_J}{R}\right)^{3/2}\left(-\left(1-\Tilde{b}^3\right)dt^2 + d\rho^2 + \rho^2d\psi^2 + dr^2+ r^2 d\theta^2\right)+ R^{3/2}u_J^{1/2}d\Omega_4^2\,,
\end{align}
where we have introduced the polar coordinates $\rho,\psi$ in the $(x_1,x_2)$ plane where the circular source lies, and the polar coordinates $r,\theta$ in the orthogonal plane $(x_3,\omega)$ with 
\begin{equation}\label{rcoord}
\omega = \sqrt{\dfrac{4R^3}{3u_J}}\dfrac{z}{\sqrt{8-5\Tilde{b}^3}}\,,\qquad x_3= r\cos\theta\,,\qquad \omega = r\sin\theta\,.
\end{equation}

Let us now assume that the gauge field on the D8-brane only depends on the two radial directions $r$ and $\rho$. The non-zero field strength components can be chosen to be $F_{t\rho}$, $F_{tr}$, $F_{\rho\psi}$, $F_{\rho\theta}$, $F_{r\psi}$ and $F_{r\theta}$. Hence, the low energy D8-brane DBI action \eqref{sd8} in the near-tip region reduces to
\begin{align}\label{smaxr}
    S_\text{D8}^\text{DBI} = -\dfrac{\lambda N}{M_{KK}}\dfrac{u_J\sqrt{1-\Tilde{b}^3}}{3\cdot 2^6\pi^3R^3}&\int dt\, \rho \,d\rho \,d\psi\, r \,dr\, d\theta\left[-\dfrac{\left(F_{t\rho}^2+F_{tr}^2\right)}{1-\Tilde{b}^3}+\dfrac{F_{\rho\psi}^2}{\rho^2} +\dfrac{F_{\psi r}^2}{\rho^2} +\dfrac{F_{\rho\theta}^2}{r^2} +\dfrac{F_{r\theta}^2}{r^2}\right].
\end{align}
The CS part of the D8-brane action reads as in \eqref{cscompleto} 
\begin{align}\label{eq:CS4}
    S_{CS} &= \dfrac{N}{24\pi^2\sqrt{2}}\epsilon^{tMNPQ}\int d^5x\left(\dfrac{1}{4}A_tF_{MN}F_{PQ} - A_M F_{tN}F_{PQ}\right),
\end{align}
with $M,N=t,\rho,\psi,r,\theta$. If the gauge field is not time-dependent as in our ansatz, then we can integrate by parts the second term in the above integral to get
\begin{equation}
    S_{CS} = \dfrac{N}{32\pi^2\sqrt{2}}\epsilon^{tMNPQ}\int d^5x\,A_tF_{MN}F_{PQ}\,.
\end{equation}

The equations of motion for the gauge field $A$ resulting from the sum of the above DBI and CS action terms read
\begin{align}
    &\rho\partial_r\left(r F_{rt}\right)+r\partial_\rho\left(\rho F_{\rho t}\right) -\tau\left(F_{\rho\psi}F_{r\theta}-F_{r\psi}F_{\rho\theta}\right) = 0\,, \label{equat}\\
    &   r\partial_\rho\left(\dfrac{1}{\rho}F_{\rho\psi}\right)+\dfrac{1}{\rho}\partial_r\left(r F_{r\psi}\right) + \dfrac{\tau}{1-\Tilde{b}^3}\left(F_{t\rho}F_{r\theta}-F_{tr}F_{\rho\theta}\right) =0\,,\label{apsieqori}\\
    & \dfrac{1}{r}\partial_\rho\left(\rho F_{\rho\theta}\right)+\rho\partial_r\left(\dfrac{1}{r}F_{r\theta}\right) + \dfrac{\tau}{1-\Tilde{b}^3}\left(F_{tr}F_{\rho\psi}-F_{t\rho}F_{r\psi}\right) = 0\,,
\end{align}
where
\begin{equation}
\tau\equiv\dfrac{24\pi M_{KK}\,\sqrt{1-\Tilde{b}^3}}{\lambda\,\sqrt{2}}\dfrac{R^3}{u_J}\,.
\label{deftau}
\end{equation}
These equations have to be supplemented by the modified Bianchi identity \eqref{modb}, with the magnetically charged loop of radius $l$ acting as a source. In polar coordinates it reads
\begin{equation}\label{eq:bianchi}
dF= - 2\pi\sqrt{2}\,\delta(\rho-l)\dfrac{\delta(r)}{2\pi r}d\rho\wedge dr\wedge rd\theta\,.
\end{equation}
A similar system of equations dealing with five-dimensional CS terms has been studied in the literature, see for example \cite{Freed:1998tg,Becker:1999kh,Boyarsky:2002ck,Harvey:2005it}.

We have not found an analytic solution to the above set of differential equations.\footnote{Unfortunately this prevents us from studying the large charge configurations analytically.} To simplify the problem we consider the parametric dependence on $\lambda$ of the gauge field. As for the standard Hata-Sakai-Sugimoto-Yamato baryon \cite{Hata:2007mb}, we want to get an action for a particle with baryonic charge on the D8-branes. This corresponds to the charge of a D4-brane wrapped on $S^4$ which as we know can be carried by the D6-brane, as discussed in section \ref{subsec:spinning}. Now, defining $a_t(t) = A_t/\sqrt{2}$ and treating it as a time-dependent perturbation, the CS action \eqref{eq:CS4} reduces to
\begin{equation}
    \dfrac{N}{12\pi^2}\int dt\,a_t(t)\int d^4x\left(F_{\rho\psi}F_{r\theta} - F_{r\psi}F_{\rho\theta}\right)\,,
\end{equation}
from which we realize that a baryonic solution has to be some kind of ``Abelian instanton" with baryon number $n_B$ equal to the instanton number
\begin{align}\label{eq:nbD8}
    &n_B = \dfrac{1}{12\pi^2}\int d^4x\left(F_{\rho\psi}F_{r\theta} - F_{r\psi}F_{\rho\theta}\right) =\dfrac{1}{3}\int dr\,d\rho\,I(r,\rho;l)\,,
\end{align}
where 
\begin{align}\label{eq:instanton}
    I(r,\rho;l) = F_{\rho\psi}F_{r\theta} - F_{r\psi}F_{\rho\theta}\,.
\end{align}
We are looking for a solution with a baryon number of order $\lambda^0$ supported by the Abelian instanton density $I(r,\rho;l)$, so the magnetic components of the gauge field do not have to scale with $\lambda$ at leading order:
\begin{equation}\label{eq:scalingfields}
F_{r\theta},\,\,\,F_{\rho\theta},\,\,\,F_{r\psi},\,\,\,F_{\rho\psi}\,\,\sim\,\,\mathcal{O}\left(\lambda^0\right).
\end{equation}
With this observation in mind, let us look at the equation of motion \eqref{equat} for the electric potential $A_t$ taking into account that $\tau$ as defined in \eqref{deftau} scales like $\lambda^{-1}$.
This equation tells us that the electric potential should scale as $\lambda^{-1}$ at leading order.\footnote{As we will see, also the zero mode will scale as $\lambda^{-1}$, and this observation is consistent with a solution with angular momentum of order $\lambda^0$. This request agrees with the relation between baryon number and angular momentum found for a charged D6-brane in section \ref{subsec:spinning}. In particular, we will look for a solution with baryon number $n_B=1$ and angular momentum $J=N/2$.} As a consequence, the CS term in the equations of motion for the magnetic components of the field strength is negligible due to the $1/\lambda$ suppression. Therefore, the equations of motion for the gauge field $A$ in the flat limit can be approximated as
\begin{align}\label{mono}
   &\dfrac{1}{r}\,\partial_\rho\left(\rho\, F_{\rho\theta}\right) + \rho\,\partial_r\left(\dfrac{F_{r\theta}}{r}\right) +\mathcal{O}\left(\lambda^{-2}\right)= 0\,,\\  \label{apsi}
    &\dfrac{1}{\rho}\,\partial_r\left(r\, F_{r\psi}\right) + r\,\partial_\rho\left(\dfrac{F_{\rho\psi}}{\rho}\right) +\mathcal{O}\left(\lambda^{-2}\right)= 0\,,\\ \label{eomat}
      &\rho\,\partial_r\left(r\,F_{tr}\right) + r\,\partial_\rho\left(\rho \,F_{t\rho}\right) + \tau(F_{\rho\psi}F_{r\theta} - F_{r\psi}F_{\rho\theta}) +\mathcal{O}\left(\lambda^{-3}\right)= 0\,,
\end{align}
ignoring $\mathcal{O}\left(\lambda^{-2}\right)$ and further subleading terms. It can be checked that a correction to the solution from these terms gives a correction of the same order to the action. As a matter of fact, at leading order, the CS term enters only in the equation of motion for $A_t$. The Bianchi identity \eqref{eq:bianchi} is topological and it contains only the magnetic components of the field strength so there is no $1/\lambda$ expansion to perform. 

The equation \eqref{mono} is automatically solved by the ansatz
\begin{equation}
    F_{r\theta} = \dfrac{r}{\rho}\partial_\rho H\,,\,\,\,\,\,\,\,\,\,\,\,\,\, F_{\rho\theta} = -\dfrac{r}{\rho}\partial_r H\,,
\label{ansaH}
\end{equation}
where $H = H(r,\rho;l)$ is a function that has to be determined by solving the Bianchi identity \eqref{eq:bianchi}, which becomes
\begin{equation}\label{flatvortonmono}
    \dfrac{1}{\rho}\,\dfrac{1}{r}\,\partial_r(r\,\partial_r H) + \partial_\rho \left(\dfrac{1}{\rho}\,\partial_\rho H\right) = -2\pi\sqrt{2}\,\delta(\rho - l)\dfrac{\delta(r)}{2\pi r}\,.
\end{equation}
To solve the differential equation for $H$ we use the machinery of two-dimensional Fourier transform in polar coordinates. Let us define the Fourier transform of $H$, namely $\widetilde{H}$, as
\begin{align}
    H(r,\rho;l) &= \dfrac{1}{(2\pi)^2}\int^\infty_0 k\,dk\,\widetilde{H}(k,\rho;l)\int^{2\pi}_0 d\phi\,e^{ikr\cos\phi}= \dfrac{1}{2\pi}\int^\infty_0k\,dk\,\widetilde{H}(k,\rho;l)J_0(kr),\\
    \widetilde{H}(k,\rho;l) &= \int^\infty_0 r\,dr\,H(r,\rho;l)\int^{2\pi}_0 d\phi\,e^{-ikr\cos\phi} = 2\pi\int^\infty_0 r\,dr\,H(r,\rho;l)J_0(kr)\,,
\end{align}
where we have used the radial symmetry of the problem. $J_0$ is the $0^{th}$ order Bessel function. The equation \eqref{flatvortonmono} can be Fourier transformed to get
\begin{equation}
      \partial_\rho \left(\dfrac{1}{\rho}\,\partial_\rho \widetilde{H}\right) -k^2\,\dfrac{\widetilde{H}}{\rho} = -2\pi\sqrt{2}\,\delta(\rho - l)\,.
\end{equation}
This differential equation can be solved with the Green function method and the solution reads\footnote{We have to impose the continuity of $\widetilde{H}$ at $\rho = l$ and the discontinuity, coming from the $\delta$ source, of the $\rho$-derivative $\partial_\rho\widetilde{H}$ at $\rho = l$.}
\begin{equation}
    \widetilde{H}(k,\rho;l)=\begin{cases}
&2\pi\sqrt{2}\,l\rho\,I_1\left(lk\right) K_1\left(\rho k\right)\,\,\,\,\,\,\,\, \rho>l\,,\\
&2\pi\sqrt{2}\,l\,\rho\,K_1\left(lk\right) I_1\left(\rho k\right)\,\,\,\,\,\,\, \rho<l\,.
\end{cases}
\end{equation}
Using the results from \cite{GradshteynRyzhik2000} we can perform the Hankel transform to get
\begin{equation}\label{eq:H}
    H(r,\rho;l) = \pi\sqrt{2}\left[\dfrac{\rho^2 + l^2 + r^2}{\sqrt{\left(l^2 + \rho ^2+r^2\right)^2 - 4\rho^2 l^2}} - 1\right].
\end{equation}
The corresponding field strength components, from \eqref{ansaH}, are thus
\begin{align}
    &F_{r\theta} = \dfrac{4\pi \sqrt{2}\, l^2 r \left(l^2-\rho ^2+r^2\right)}{\left(\left(l^2 + \rho ^2+r^2\right)^2 - 4\rho^2 l^2\right)^{3/2}}\,,\\
    &F_{\rho\theta} = \dfrac{8\pi \sqrt{2}\, l^2 r^2\rho}{\left(\left(l^2 + \rho ^2+r^2\right)^2 - 4\rho^2 l^2\right)^{3/2}}\,.
\end{align}

Now we have to solve the equation of motion \eqref{apsi} for $A_\psi$. This equation is homogeneous in $A_\psi$ and this field does not enter the Bianchi identity, therefore the solution will be defined up to a constant pre-factor. Remember that the full equation of motion for $A_\psi$ \eqref{apsieqori} has a source term due to the CS contribution, but this is suppressed in the large $\lambda$ expansion we are considering. To leading order, one can, however, argue that $A_{\psi}$ is sourced by a current $j_{\psi}\sim\delta(\rho-l)\delta(r)$, flowing along the circular loop, which is (proportional to) the spatial four-dimensional Hodge dual of the three-form magnetic current entering the modified Bianchi identity \eqref{eq:bianchi}. In the supersymmetric supertubes context \cite{Kruczenski:2002mn} the proportionality constant is fixed since the magnetic gauge field is taken to be self-dual w.r.t.~the four-dimensional Hodge transform. In our non-supersymmetric setup with CS terms, we will stay agnostic and leave the proportionality constant undetermined. The equation of motion with such a source term looks like \eqref{flatvortonmono}, the differential equation for $H$, hence the solution reads
\begin{equation}
    A_\psi = \mathcal{N}\left(\dfrac{\rho^2 + l^2 + r^2}{\sqrt{\left(l^2 + \rho ^2+r^2\right)^2 - 4\rho^2 l^2}}-1\right),
\end{equation}
where $\mathcal{N}$ is a constant. The corresponding field strength components are
\begin{align}
    &F_{\rho\psi} = 4\mathcal{N}\,\dfrac{l^2 \rho \left(l^2-\rho ^2+r^2\right)}{\left(\left(l^2 + \rho ^2+r^2\right)^2 - 4\rho^2 l^2\right)^{3/2}}\,,\\
    &F_{r\psi} = -8\mathcal{N}\,\dfrac{l^2 r\rho^2}{\left(\left(l^2 + \rho ^2+r^2\right)^2 - 4\rho^2 l^2\right)^{3/2}}\,.
\end{align}
The overall factor $\mathcal{N}$ can be fixed by requiring that the above configuration supports baryon number $n_B = 1$. However, it is easy to realize, using the above solutions, that the integral of $I(r,\rho;l)$ in \eqref{eq:nbD8}, giving $n_B$, diverges close to the circular source, and hence it has to be regularized. We can parametrize the divergence by introducing a small-distance cutoff $\epsilon$ along the $r$ integration.

Before going on, let us point out that up to this moment, we have neglected another crucial ingredient that is necessary for the D8-brane description to fully capture the attached D6-brane setup. It is well-known that magnetic charges on a D(p+2)-brane correspond to the endpoints of a Dp-brane \cite{Callan:1997kz,Thorlacius:1997zd,Gibbons:1997xz,Lee:1997xh,Hashimoto:1997px}, and the D8-D6 system studied in this work is just a particular example within this class. Given this correspondence, we should be able to read off the D6-brane energy from the energy of the monopole configuration on the D8-branes by turning on a scalar field fluctuation which pulls the D8-branes towards the horizon. Let us thus consider a scalar field $X = \pi\alpha'\,\Phi$, representing the fluctuation along the $u$ direction transverse to the brane at $z\sim 0$. The factor $\pi\alpha'$ is the required normalization for the scalar field to be of the same order as the gauge field $A$. Within the quadratic low energy approximation of the DBI action we are considering in this section,\footnote{As stated in \cite{Callan:1997kz} in the non-BPS case, the quadratic limit of the DBI action necessarily leads to quantitatively incorrect results and we should make use of the full non-linear action. However, we will use it to study qualitatively the D8-brane embedding around the tip.} the action for the scalar field near the D8-brane tip reads
\begin{equation}
    S_X = -\dfrac{\lambda N}{M_{KK}}\dfrac{u_J}{R^3}\dfrac{1}{3\cdot 2^7\pi^3\sqrt{1-\tilde{b}^3}}\int dt\, \rho \,d\rho \,d\psi\, r \,dr\, d\theta \bigg[(\partial_r\Phi)^2+(\partial_\rho\Phi)^2\bigg].
\end{equation}
The equation of motion for $\Phi$, sourced by the charge distribution corresponding to the local density of strings ending on the circle of radius $l$ \cite{Kruczenski:2002mn} at the tip of the D8-branes  is
\begin{equation}
    \dfrac{1}{r}\,\partial_r\left(r\,\partial_r \Phi\right) + \dfrac{1}{\rho}\,\partial_\rho\left(\rho \,\partial_\rho \Phi\right) \sim \delta(\rho-l)\dfrac{\delta(r)}{2\pi r}\,.
\end{equation}
The solution reads
\begin{equation}
    \Phi=\dfrac{l\,c}{\sqrt{\left(l^2 + \rho ^2+r^2\right)^2 - 4\rho^2 l^2}}\,,
\label{solphi}
\end{equation}
where we leave the constant $c$ unfixed. In the supersymmetric case of the supertube, the proportionality constant is given in terms of the total charge of the fundamental strings ending on the loop, which in our case would be $N$. As it will be clear in the following, the scalar field given in \eqref{solphi} is proportional to the zero mode of the electric potential $A_t$ (\textit{i.e.} to the solution of the related equation of motion obtained neglecting the CS term contribution), as pointed out in \cite{Kruczenski:2002mn}. We will return to this point in a moment.

Let us observe that if we go close to the source, $\rho = l$ and $r = \epsilon$ with $\epsilon \to 0$, the scalar field behaves as the usual transverse field representing 
a magnetic source on a D(p+2)-brane \cite{Callan:1997kz} where now $\epsilon$ is the distance from the source
\begin{equation}
    X = \pi\alpha'\Phi = \dfrac{\pi\alpha'c}{2}\dfrac{1}{\epsilon} + \mathcal{O}(\epsilon).
\end{equation}

In the quadratic approximation of the DBI action, the action for the scalar field sums to the Maxwell action for the monopole gauge field $A$.
Therefore, the total energy of the D8-D6 branes system from the D8-brane perspective is
\begin{align}\nonumber
    E_\text{D8}=E&_{\text{monopole}} + E_\text{scalar}=\dfrac{\lambda N}{M_{KK}}\dfrac{u_J}{R^3}\dfrac{\sqrt{1-b^3}}{3\cdot 2^4 \pi}\int dr\,d\rho\,\bigg\{\dfrac{r}{\rho}\left[\left(\partial_{\rho}H\right)^2+\left(\partial_{r}H\right)^2\right]+\\ \nonumber
& \hspace{8cm} + \dfrac{r\,\rho}{2(1-\tilde{b}^3)}\left[\left(\partial_{\rho}\Phi\right)^2+\left(\partial_{r}\Phi\right)^2\right]\bigg\}= \\ \nonumber
&= \dfrac{\lambda\,N}{M_{KK}}\dfrac{u_J}{R^3}\dfrac{\pi  \sqrt{1-b^3}}{96\left(\chi+\chi^3\right)}\bigg\{\left(1+\dfrac{c^2}{16\pi^2(1-\tilde{b}^3)}\right)\left(\pi  \chi^2+2 \left(\chi^2+1\right) \arccot\left(\frac{2 \chi}{1-\chi^2}\right)+\pi \right)+\\ \nonumber
    &\hspace{8cm}-4\chi\left(1-\dfrac{c^2}{16\pi^2(1-\tilde{b}^3)}\right)\bigg\} =\\
    &= \dfrac{\lambda\,N}{M_{KK}}\dfrac{u_J}{R^3}\dfrac{g(c,\tilde{b},\chi)}{3\cdot 2^4}\,,
    \label{emono}
\end{align}
where we have introduced the dimensionless quantity $\chi = \epsilon/l$ and we have defined the function  
\begin{align}\nonumber
    g(c,\tilde{b},\chi) = &\dfrac{\pi  \sqrt{1-b^3}}{2\left(\chi+\chi^3\right)}\bigg\{\left(1+\dfrac{c^2}{16\pi^2(1-\tilde{b}^3)}\right)\left(\pi  \chi^2+2 \left(\chi^2+1\right) \arccot\left(\frac{2 \chi}{1-\chi^2}\right)+\pi \right)+\\
    &-4\chi\left(1-\dfrac{c^2}{16\pi^2(1-\tilde{b}^3)}\right)\bigg\}.
\end{align}

In the limit $\chi = \epsilon/l\to 0$ with $\epsilon\to 0$, we can expand the energy to get
\begin{equation}
    E_\text{D8} = \dfrac{\lambda N}{M_{KK}}\dfrac{u_J}{R^3}\dfrac{\pi^2\sqrt{1-\tilde{b}^3}}{48}\left(1+\dfrac{c^2}{16\pi^2\left(1-\tilde{b}^3\right)}\right)\dfrac{l}{\epsilon} + ...
\end{equation}
From this expansion we can retrieve the usual formula for the energy for the D(p+2)-Dp system of branes \cite{Callan:1997kz}, using the expansion of the transverse scalar close to the source
\begin{align}
&E_\text{D8} \sim \dfrac{\lambda^2 N}{M_{KK}^2}\dfrac{u_J}{R^6}\dfrac{\pi\sqrt{1-\tilde{b}^3}}{48c}\left(1+\dfrac{c^2}{16\pi^2\left(1-\tilde{b}^3\right)}\right)l\,X(\epsilon)\big|_\text{div.},\\
&\text{with}\,\,\,\,\,X(\epsilon)\big|_\text{div.}=\dfrac{\pi\alpha'c}{2}\dfrac{1}{\epsilon}\,,
\end{align}
where the prefactor reproduces the tension, in a flat spacetime, of a D6-brane with radius $l$ extending from the D8-brane's tip along the $u$ direction. We stress that this result becomes increasingly reliable as the D6-brane extends further along the $u$ direction. In other words, it is the limiting scenario of a BIon in flat space which is infinitely extended along the transverse direction to the D(p+2)-brane. We also remember that when $\epsilon\to 0$, higher order terms in the full DBI Lagrangian that we dropped become important, and one has to replace the Maxwell action with the Born-Infeld action. We will not investigate these corrections in this work.

Now we are going to compute the baryon number, the angular momentum, and the energy of our system giving the results as functions of the parameter $\chi$ and $\tilde{b}$. Then we will analyze them carefully from various perspectives to extract the parametric dependence of the stability radius of the string loop.

We start by performing the integration of the Abelian instanton density to get the baryon charge \eqref{eq:nbD8}. The result reads
\begin{align*}
 n_B=\dfrac{1}{3}\int dr\,d\rho\,I(r,\rho;l) =  \dfrac{\mathcal{N}\,\pi\left(\pi\chi^2+2 \left(\chi^2+1\right) \arccot\left(\frac{2 \chi}{1-\chi^2}\right)-4 \chi+\pi \right)}{6 \sqrt{2} \left(\chi^3+\chi\right)}\,.
\end{align*}
Hence, if we want to focus on solutions with baryon number $n_B=1$, the normalization of the gauge field component $A_\psi$ has to be fixed as
\begin{equation}\label{enne}
    \mathcal{N} =\dfrac{6 \sqrt{2} \left(\chi^3+\chi\right)}{\pi  \left(\pi  \chi^2+2 \left(\chi^2+1\right)\arccot\left(\frac{2 \chi}{1-\chi^2}\right)-4 \chi+\pi \right)}\,.
\end{equation}

The equation of motion \eqref{eomat} for the electric potential $A_t$ can be solved since we know explicitly the source term (see appendix \ref{app:green}), and the solution is
\begin{equation}
    A_t = \dfrac{d}{\sqrt{\left(l^2 + \rho^2+r^2\right)^2 - 4\rho^2 l^2}}+\dfrac{2\sqrt{2}\,\pi\mathcal{N}\,\tau\left(\rho^2+r^2\right)}{\left(l^2 + \rho^2+r^2\right)^2 - 4\rho^2 l^2}\,,
\end{equation}
where the first term is the zero mode of the differential operator applied to $A_t$ and,\footnote{As shown in appendix \ref{app:green}, the Laplacian operator in duo-polar coordinates has two zero modes. We set one of them to zero because it leads to unphysical solutions.} as anticipated, it is proportional to the scalar field $\Phi$ as given in \eqref{solphi}. This relation is to be expected from the supersymmetric configuration first studied in \cite{Callan:1997kz} and then in \cite{Kruczenski:2002mn} in the context of supertubes. In this picture, the presence of $A_t$ accounts for the fundamental strings as constituents of the D6-brane which source the D8-brane gauge field. This interpretation is clear also from the D6-brane perspective from which we can read the fundamental string charge $q_s$ from the D6-brane gauge theory.\footnote{The fundamental string charge can be matched with the canonical conjugate momentum to the electric potential $a_t$ on the D6-brane.} More precisely, the charge density sourcing the scalar field equation of motion which has support on the loop corresponds to the local density of strings ending on the loop. Note that, in our system, the presence of the CS term prevents a direct identification of the scalar field with $A_t$. In the supersymmetric case, the constant $d$ would be fixed in terms of $c$.

The coefficient $d$ in our case can be fixed by requiring that our solution with baryon number $n_B=1$ has angular momentum
\begin{equation}
    J= \dfrac{\lambda N}{M_{KK}}\dfrac{u_J}{R^3}\dfrac{1}{3\cdot 2^5\pi^3\sqrt{1-\Tilde{b}^3}}\int\rho\,d\rho\,d\psi\,r\,dr\,d\theta\left[F_{t\rho}F_{\rho\psi}+F_{tr}F_{r\psi}\right]\,,
\end{equation}
equal to $N/2$. 
The above request fixes the coefficient $d$ to be
\begin{align}\nonumber
    d =& -\dfrac{8\, \sqrt{2}\,R^3M_{KK} \sqrt{1- \tilde{b}^3}  }{\lambda\,u_J}\bigg[-8 \pi ^2 \left(\chi^3+\chi\right)+16 \pi  \chi^2+\pi ^3 \left(\chi^2+1\right)^2+96 \left(\chi^2+1\right)^2+\\ \nonumber
    &+4 \pi  \left(\chi^2+1\right) \arccot\left(\frac{2 \chi}{1-\chi^2}\right) \left(\pi  \chi^2+\left(\chi^2+1\right) \arccot\left(\frac{2 \chi}{1-\chi^2}\right)-4 \chi+\pi \right)\bigg]\cdot\\
    & \cdot\bigg[\left(\chi^2+1\right) \left(\pi  \chi^2+2 \left(\chi^2+1\right) \arccot\left(\frac{2 \chi}{1-\chi^2}\right)-4 \chi+\pi \right)^2\bigg]^{-1}\,.
\end{align}
Using these results we can verify that the solution we have found has the desired scaling with $\lambda$ that we assumed at the beginning of this section.

Now that we have found the solution that solves the leading-order equations of motion for the gauge field, we can compute the total free energy of the configuration, defined as $S_\text{TOT} = -\int dt\, E_\text{TOT}$. It can be written as
\begin{align*}
E_\text{TOT} = E_{\text{magn.}} +E_\text{scalar}+E_{\text{el.}}\,,
\end{align*}
where $E_{\text{mag.}}$ and $E_\text{scalar}$ are of order $\lambda$ while $E_{\text{el.}}$ scales as $\lambda^{-1}$ and they are associated respectively to the magnetic, scalar, and electric energy of the configuration. They read
\begin{align} 
    &E_{\text{magn.}} = \dfrac{\lambda N}{M_{KK}}\dfrac{u_J\sqrt{1-\Tilde{b}^3}}{3\cdot 2^6\pi^3R^3}\int \rho \,d\rho \,d\psi\, r \,dr\, d\theta\left[\dfrac{F_{\rho\psi}^2}{\rho^2} +\dfrac{F_{\psi r}^2}{\rho^2} +\dfrac{F_{\rho\theta}^2}{r^2} +\dfrac{F_{r\theta}^2}{r^2}\right],\\ &E_\text{scalar} = \dfrac{\lambda N}{M_{KK}}\dfrac{u_J}{3\cdot 2^7\pi^3R^3\sqrt{1-\Tilde{b}^3}}\int \rho \,d\rho \,d\psi\, r \,dr\, d\theta\left[\left(\partial_{\rho}\Phi\right)^2+\left(\partial_{r}\Phi\right)^2\right],\\\nonumber
    &E_{\text{el.}} = -\dfrac{\lambda N}{M_{KK}}\dfrac{u_J}{3\cdot 2^6\pi^3R^3\sqrt{1-\Tilde{b}^3}}\int \rho \,d\rho \,d\psi\, r \,dr\, d\theta\,\left[F_{t\rho}^2+F_{tr}^2\right]\\ 
    &\,\,\,\,\,\,\,\,\,\,\,\,\,\, - \dfrac{N}{4\pi^2\sqrt{2}}\int d^5x\,A_t\left[F_{\rho\psi}F_{r\theta} - F_{\rho\theta}F_{r\psi}\right].
\end{align}
 
\begin{figure}[h!]
    \centering
    \begin{subfigure}{0.4\textwidth}
        \centering
        \includegraphics[width=\textwidth]{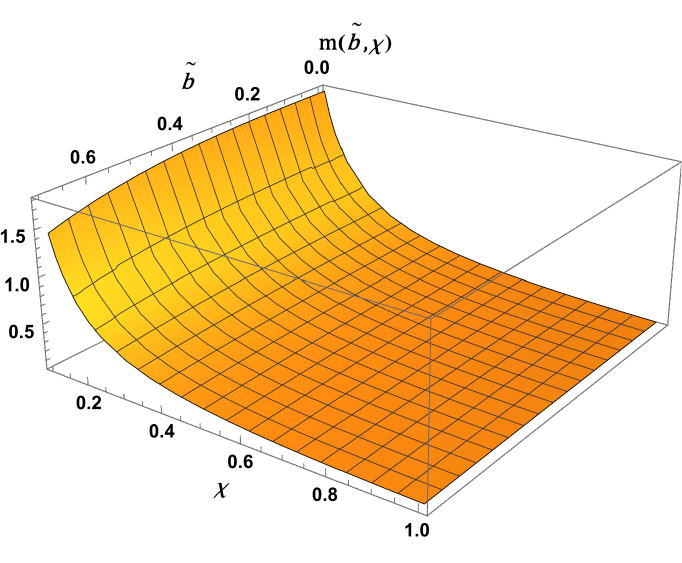}
        \caption{Plot of $m(\tilde{b},\chi)$.} \label{fig:mfunction}
    \end{subfigure}
    \hfill
    \begin{subfigure}{0.5\textwidth}
        \centering
        \includegraphics[width=\textwidth]{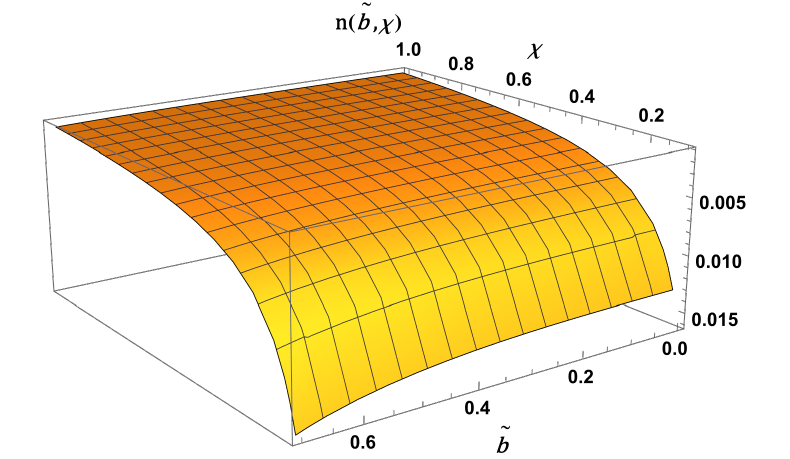}
        \caption{Plot of $n(\tilde{b},\chi)$.} \label{fig:nfunction}
    \end{subfigure}

    \begin{subfigure}{0.5\textwidth}
        \centering
        \includegraphics[width=\textwidth]{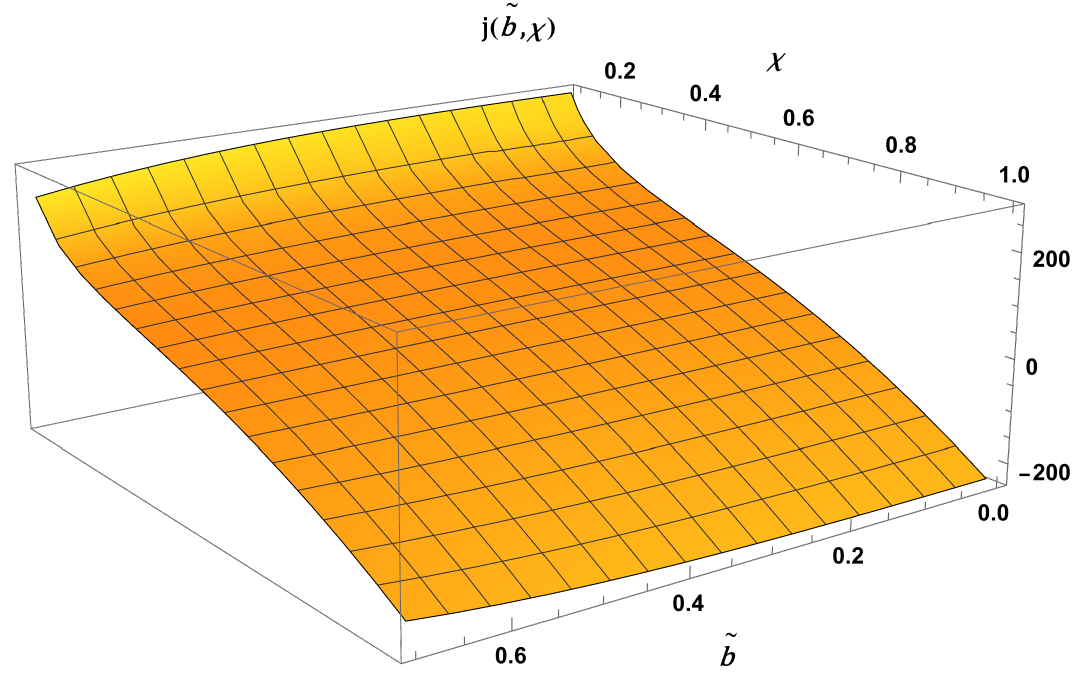}
        \caption{Plot of $j(\tilde{b},\chi)$.} \label{fig:jfunction}
    \end{subfigure}

    \caption{Plots of the energy contributions as functions of $\tilde{b}$ and $\chi$. $\tilde{b}$ spans the whole range of values for which chiral symmetry is broken, while the variable $\chi$ spans the range from 0.1 to 1.} \label{fig:mnjfunction}
\end{figure}

Using the solutions to the equations of motion, we find
\begin{align*} 
    E_{\text{magn.}}=\dfrac{\lambda N}{M_{KK}}\dfrac{u_J}{R^3}\,m(\tilde{b},\chi)\,,
\end{align*}
where
\begin{align}\nonumber
  m(\tilde{b},\chi) = &\dfrac{\sqrt{1-\tilde{b}^3}}{48\pi}\bigg[\dfrac{\pi ^3}{2 \chi}-\dfrac{2 \pi ^2}{\chi^2+1}+\dfrac{\pi ^2 \arccot\left(\frac{2 \chi}{1-\chi^2}\right)}{\chi}+\\
  &+\dfrac{18 \left(\chi^3+\chi\right)}{\pi ^2 \left(\pi  \chi^2+2 \left(\chi^2+1\right) \arccot\left(\frac{2 \chi}{1-\chi^2}\right)-4 \chi+\pi \right)}\bigg]\,,
\end{align}
is a function of $\tilde{b}$ and $\chi$.
The plot of the function $m(\tilde{b},\chi)$ is shown in figure \ref{fig:mfunction}. It encodes the dependence of the magnetic energy on $\tilde{b}$ and $\chi$.

Then, the contribution coming from the transverse scalar field is 
\begin{equation*}
    E_\text{scalar} = \dfrac{\lambda N}{M_{KK}}\dfrac{u_J}{R^3}\, c^2\,n(\tilde{b},\chi)\,,
\end{equation*}
where 
\begin{equation}
    n(\tilde{b},\chi) = \dfrac{1}{1536 \pi  \left(\chi+\chi^3\right)\sqrt{1-\tilde{b}^3}}\,\left(\pi  \chi^2-2 \left(\chi^2+1\right) \arccot\left(\dfrac{2 \chi}{\chi^2-1}\right)+4 \chi+\pi \right)\,,
\end{equation}
is a function of $\tilde{b}$ and $\chi$ whose plot is shown in figure \ref{fig:nfunction}.

Finally, the electric contribution to the energy $E_{\text{el.}}$ reads
\begin{align*}
    E_{\text{el.}} =\dfrac{N\,M_{KK} R^3}{\lambda\,u_J}\dfrac{j(\tilde{b},\chi)}{l^2}\,,
\end{align*}
with 
\begin{align}
  &j(\tilde{b},\chi) = \bigg\{-2 \left(\chi^2+1\right)^3 \arccot\left(\dfrac{2 \chi}{1-\chi^2}\right) \left(\hat{d}\,^2 \chi^2+1152 \pi ^4 \left(1-\tilde{b}^3\right) \mathcal{N}\,^2 \left(\chi^2+1\right)\right)\\ \nonumber
  &-256 \pi ^2\hat{d}\,\mathcal{N}\,\chi \sqrt{1-\tilde{b}^3}\, \left(3 \chi^6+6 \chi^4+5 \chi^2+2\right)-384 \pi ^4\mathcal{N}\,^2\left(1-\tilde{b}^3\right) \Bigl(3 \pi  \left(\chi^2+1\right)^4 +\\ 
  &+ 4 \chi \left(9 x^6+9 \chi^4+11 \chi^2+3\right)\Bigr) 
  -\hat{d}\,^2 \chi^2 \left(\pi \chi^2+4 \chi+\pi \right) \left(\chi^2+1\right)^2\bigg\}\left[768\,\pi\sqrt{1-\tilde{b}^3}\left(\chi^3+\chi\right)^3\right]^{-1}\,,\nonumber
\end{align}
a function of $\tilde{b}$ and $\chi$. We remember that $\mathcal{N}$ is also a function of $\chi$ and here we have defined $\hat{d} = d\,R^3 M_{KK}/u_J$ which is again a function of $\tilde{b}$ and $\chi$. See figure \ref{fig:jfunction} for a plot of $j(\tilde{b},\chi)$.

At this point, we can analyze the total energy in the limit of a very long spike which is $\chi = \epsilon/l \to 0$ with $\epsilon\to 0$. The total energy expanded at leading order reads
\begin{align}\nonumber
    E_\text{TOT}^\text{on-shell} = &\dfrac{\lambda\, N}{M_{KK}R^3}\dfrac{\sqrt{1-\tilde{b}^3}}{\pi c}\,X(\epsilon)\big|_\text{div.}\bigg\{\dfrac{\lambda}{M_{KK}}\dfrac{u_J}{R^3}\dfrac{\pi}{48}\left(1+\dfrac{c^2}{16\pi^2\left(1-\tilde{b}^3\right)}\right)l + \\
    &+ \dfrac{M_{KK}}{\lambda}\dfrac{R^3}{u_J}\dfrac{1728-\pi^2(162-48\pi+\pi^4)}{3\pi^2}\dfrac{1}{l}\bigg\} + ...
\end{align}
The dependence on the transverse scalar field factorizes and hence we can minimize the action over the free parameter $l$, representing the loop size. The minimization gives
\begin{eqnarray}
    l_\text{stable} = \dfrac{M_{KK}}{\lambda}\dfrac{u_J}{R^3}\dfrac{16}{\pi}\sqrt{\dfrac{\left(1-\tilde{b}^3\right)(1728-\pi^2(162-48\pi+\pi^4))}{c^2+16\pi^2\left(1-\tilde{b}^3\right)}}\,\,.
\label{elleD8}   
\end{eqnarray}
From the above result, we can deduce that the stability radius scales like $\lambda^{-1}$. Hence the string loop radius in the flat region is suppressed similarly to the instanton size of the standard baryon in the WSS model but with a different power of $\lambda$. 

It is useful to give the result for the stability radius also in terms of the chiral symmetry breaking critical temperature $T_a$
\begin{equation}
    l_{\text{stable}} =  \dfrac{M_{KK}}{\lambda\,T_a^2}\dfrac{16(0.154)^2}{\pi \,J_T^2(\tilde{b})}\sqrt{\dfrac{\left(1-\tilde{b}^3\right)(1728-\pi^2(162-48\pi+\pi^4))}{c^2+16\pi^2\left(1-\tilde{b}^3\right)}}.
\end{equation}

It is also interesting to write the stability radius in terms of the more common phenomenological quantity $f_a$, the axion decay constant. Remembering the relation between $f_a$ and $T_a$ (see e.g.~\cite{Bigazzi:2019eks})
\begin{equation}
    f_a^2 = \dfrac{N\lambda}{16\pi^3}\dfrac{J_T^3(\Tilde{b})}{(0.154)^3\,I_T(\Tilde{b})}\dfrac{T_a^3}{M_{KK}}\,,
\end{equation}
where
\begin{equation}
I_T(\Tilde{b})	 = \int^1_0dy\dfrac{\sqrt{1-\tilde{b}^3y}}{\sqrt{y\left(1-\tilde{b}^3y-\left(1-\tilde{b}^3\right)y^{8/3}\right)}} \,,
\end{equation}
we get
\begin{equation}\label{eq:lstable}
    l_{\text{stable}} = \dfrac{N\,T_a}{f_a^2}\dfrac{J_T(\tilde{b})}{0.154\,\pi^4I_T(\tilde{b})}\sqrt{\dfrac{\left(1-\tilde{b}^3\right)(1728-\pi^2(162-48\pi+\pi^4))}{c^2+16\pi^2\left(1-\tilde{b}^3\right)}}\,.
\end{equation} 

The analysis can be repeated by considering an object with baryon number $n_B$ (of order $\lambda^0$). This leads to the following redefinition of the integration constants $\mathcal{N}$ and $d$ as
\begin{equation}
\mathcal{N}\to n_B\,\mathcal{N}\,,\quad d\to n_B\,d\,.
\end{equation}
This propagates to the energy, leading to the following result for its minimization with respect to $l$:
\begin{equation}
    l_\text{stable} = n_B\,l_{\text{stable}}^{n_B=1}\,,
\end{equation}
where $l_{\text{stable}}^{n_B=1}$ is the result written in equation \eqref{elleD8} or equivalently  in \eqref{eq:lstable}. Therefore, from the D8-brane analysis in flat space, we recover the D6-brane result \eqref{ellescaling}
\begin{equation}
    l_\text{stable} \sim \dfrac{n_B}{\lambda}\,.
\end{equation}
This confirms that the flat limit gives the correct result for the stability radius scaling, even if it is not able to capture if the solution exists in the full spacetime.

Let us make some observations about the cutoff $\epsilon$. The result $l_\text{stable}\sim\lambda^{-1}$ leads to observe that in order to have $n_B \sim\lambda^0$ and the scaling \eqref{eq:scalingfields} for the magnetic fields, it follows that $\epsilon\sim \lambda^{-1}$. Therefore, the ratio $\chi = \epsilon/l$ does not scale with $\lambda$. This statement is expected to be true also from other perspectives. Going back to the transverse scalar, this represents the extension of the D6-brane along $u$ and we know from the D6-brane perspective that this length is of order $(\alpha')^0$ in the vorton case. Since $X = \pi\alpha'\Phi \sim \pi\alpha'c/(2\epsilon)$ has to go like $(\alpha')^0$, we get $\epsilon\sim\alpha'\sim\lambda^{-1}$. This identification can be written qualitatively as
\begin{equation}\label{eq:eps1}
    \epsilon = y(\tilde{b},c)\,\dfrac{R^3}{u_J}\dfrac{M_{KK}}{\lambda},
\end{equation}
where $y(\tilde{b},c)$ is some function of $\tilde{b}$ and $c$ which cannot be fixed at this level.

\begin{figure}
\center
\includegraphics[height = 5 cm]{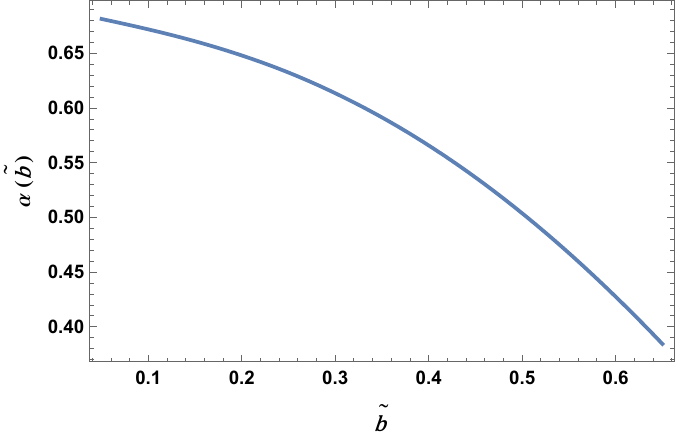}\caption{Plot of the slope $\alpha(\tilde{b})$ of the D6-brane energy for the string loop configuration as a function of $l$.}
\label{fig:alphaslope}
\end{figure}

Another argument that leads to the same qualitative result comes from equating the energy of the transverse scalar plus the energy of the monopole with the energy of the D6-brane. According to our results in section \ref{subsubphase}, the energy of the D6-brane that describes the uncharged string loop can be written as
\begin{align}
    E_\text{D6} = \dfrac{\lambda^2 N}{3\cdot 2^4\pi^2}\dfrac{J_T^5(\tilde{b})T_a^4}{(0.15)^4M_{KK}^2}\alpha(\tilde{b})l\,,
\label{ed6l}
\end{align}
where $\alpha(\tilde{b})$ is the slope of the linear dependence of the energy with $l$, which can be found numerically as a function of $\tilde{b}$ and it is shown in figure \ref{fig:alphaslope}.

Then, an estimate for the cutoff is obtained by equating the two energies \eqref{ed6l} and \eqref{emono}:
\begin{equation}
    \epsilon = y(c,\tilde{b},\chi)\dfrac{R^3}{u_J}\dfrac{M_{KK}}{\lambda}\,,
\label{eqepsilon}
\end{equation}
where
\begin{equation}
    y(c,\tilde{b},\chi) \sim \dfrac{\pi^2}{J_T(\tilde{b})\alpha(\tilde{b})}\,\chi \,g(c,\tilde{b},\chi)\,.
\end{equation}
Once again assuming that the quantity $\chi$ does not scale with $\lambda$, we get $\epsilon\sim\lambda^{-1}$. The coefficient $y(c,\tilde{b},\chi)$ is a function of $\tilde{b}$ and $\chi$, and its exact value cannot be fixed within our approximations since we cannot compare the flat D8-brane result with the D6-brane result valid in the whole spacetime. This function is more involved than the one defined in \eqref{eq:eps1} but since we are not able to fix either of them we will keep generically the cutoff $\epsilon$ depending on a function $y$. The important result relies on the parametric dependence of $\epsilon$ with $\lambda$.

We are going to show that also in this case we will obtain the same qualitative result for the stability radius. We can express $l$ in terms of $\chi$, using the expression \eqref{eqepsilon} for $\epsilon$, 
\begin{equation}\label{eq:lversuschi}
    l = \dfrac{M_{KK}}{\lambda}\dfrac{R^3}{u_J}\dfrac{y(c,\tilde{b},\chi)}{\chi}\,.
\end{equation}
We can substitute the above relation in the electric contribution to the energy $E_{\text{el.}}$ to give
\begin{equation}
    E_{\text{el.}} = \dfrac{\lambda N}{M_{KK}}\dfrac{u_J}{R^3}\,\dfrac{\chi^2}{y^2(c,\tilde{b},\chi)}\,j(\tilde{b},\chi).
\end{equation}

Remarkably, this term becomes of the same order in the parameters as $E_\text{magn.}$ and $E_\text{scalar}$. Eventually, the total energy depends only on $\chi$, which has not yet been fixed, and all the dimensionful parameters, together with $\lambda$ and $N$, factorize
\begin{equation}
    E^{\text{on-shell}}_{\text{TOT}} = \dfrac{\lambda N}{M_{KK}}\dfrac{u_J}{R^3}\left(m(\tilde{b},\chi)+ c^2\,n(\tilde{b},\chi)+\dfrac{\chi^2}{y^2(c,\tilde{b},\chi)}\,j(\tilde{b},\chi)\right).
\end{equation}
We have to minimize the above action with respect to $\chi$. This procedure will return a function for $\chi_{\text{stable}}(\tilde{b},c)$ that does not scale with $\lambda$. The result of the minimization
\begin{equation}
    \partial_\chi m(\tilde{b},\chi)+c^2\,\partial_\chi n(\tilde{b},\chi)+ \partial_\chi\left(\dfrac{\chi^2\,j(\tilde{b},\chi)}{y^2(c,\tilde{b},\chi)}\right) = 0\,,
\end{equation}
depends on $c$ and on the explicit expression for $y(c,\tilde{b},\chi)$ and it is in general a function of the temperature $\tilde{b}$.

We can conclude, from the relation \eqref{eq:lversuschi}, that $l_\text{stable}$ scales like $\lambda^{-1}$. More precisely 
\begin{equation}
    l_{\text{stable}} =  \dfrac{M_{KK}}{\lambda\,T_a^2}\dfrac{(0.154)^2e(c,\tilde{b},\chi_\text{stable})}{J_T^2(\tilde{b})\chi_\text{stable}(\tilde{b},c)} = p(\tilde{b},c,\chi_\text{stable})\,\dfrac{M_{KK}}{\lambda\,T_a^2} = \dfrac{p(\tilde{b},c,\chi_\text{stable})\,J_T^3(\tilde{b})}{16\pi^3(0.15)^3I_T(\tilde{b})}\dfrac{N\,T_a}{f_a^2}\,,
\end{equation}
where
\begin{equation}
    p(\tilde{b},c,\chi_\text{stable}) = \dfrac{(0.154)^2y(c,\tilde{b},\chi_\text{stable})}{J_T^2(\tilde{b})\chi_{\text{stable}}(\tilde{b},c)}\,.
\end{equation}

All in all we can conclude that the local D8-brane analysis around the tip at $u=u_J$, provides a result on the stability radius which is perfectly consistent with what we have found from the D6-brane perspective. In that case, in fact, we have outlined how any (meta)-stable vorton solution would have a radius $l$ scaling as $n_B\lambda^{-1}$, see eq. (\ref{ellescaling}) and related comments. As we have seen, the actual existence of such a solution can be eventually confirmed only by a global analysis of the full equations of motion. In particular, we have found that if $n_B\sim\lambda^{0}$, as in the case considered in the present section, no (meta)stable vorton solution actually exists. In order for the same conclusion to be drawn from the D8-brane perspective we should go far beyond the near tip limit and consider the full equations of motion of the D8 on the whole curved background, something which for the moment we are not able to do. In other words, the flat limit in the D8-brane perspective captures only the local result near $u=u_J$ of the complementary D6-brane description.

\subsection{Analysis in the asymptotic region}
In this section, we analyze the equations of motion for the D8-brane gauge field on the curved background \eqref{eq:bkgmetric}.\footnote{In the deconfined phase, the contribution from the on-shell action of the $C_7$ RR form is zero. See \cite{Bergman:2006xn} for details. In the confined phase this term is not zero and the authors considered its contribution in \cite{Bigazzi:2022luo} to study a similar D6-D8-brane system.} As in the previous subsection, the D6-brane plays the role of a magnetic source for the flavor gauge field. First, we switch off the electric potential and we solve exactly the Maxwell-Bianchi system in the curved background. Then, we solve the linearized equations of motion with non-zero $A_t$ and $A_\psi$ following \cite{Hashimoto:2008zw}. Such solutions are written in terms of the curved metric Green functions that can be smoothly connected with the flat space solutions. Finally, we compute the one-point function of flavor currents in the deconfined phase of the WSS and we use the result to compute the baryon number of the solution to the linearized equations of motion.

The D6-brane-monopole source for the D8-brane gauge field leads to the modified Bianchi identity \eqref{modb}, 
which complements the equations of motion for the D8-branes gauge field. Turning off for a moment the potentials $A_t$ and $A_{\psi}$, the only field strength components that we need to consider are $F_{\rho x_3}$, $F_{\rho z}$ and $F_{x_3 z}$. With this gauge choice, there is no contribution from the five-dimensional CS term of flavor brane. 

The action \eqref{eq:d8action} in the present case reduces to
\begin{align*}
    S^\text{DBI}_\text{D8} = &-\dfrac{3T_8V_4R^{3/2}u_J^{3/2}(2\pi\alpha')^2}{8g_s}\int d^4x\,dz \,\bigg\{\dfrac{k^{3/2}(z)}{|z|}\sqrt{\gamma_T(z)f_T(z)}\left[F_{\rho z}^2 + F_{x_3 z}^2\right]\\
    & + \dfrac{4R^3}{9u_J}\dfrac{|z|}{k^{5/6}(z)}\sqrt{\dfrac{f_T(z)}{\gamma_T(z)}}\,F_{\rho x_3}^2\bigg\}\,,
\end{align*}
where, again, $\rho = \sqrt{x_1^2 + x_2^2}$ and we have used the equation \eqref{uzeta}. Moreover, $d^4x = dt\rho d\rho d\psi dx_3$. 
The resulting equations of motion
\begin{align}
    &\partial_\rho\left(\rho F_{\rho z}\right) + \rho\,\partial_{x_3}F_{x_3 z} = 0\,,\\
    & \dfrac{4R^3}{9u_J}\dfrac{|z|}{k(z)^{5/6}}\sqrt{\dfrac{f_T(z)}{\gamma_T(z)}}\,\partial_{x_3} F_{\rho x_3} + \partial_z\left(\dfrac{k^{3/2}(z)}{|z|}\sqrt{\gamma_T(z)f_T(z)}\,F_{\rho z}\right) = 0\,,\\
    & \dfrac{4R^3}{9u_J}\dfrac{|z|}{k^{5/6}(z)}\sqrt{\dfrac{f_T(z)}{\gamma_T(z)}}\,\partial_\rho\left(\rho F_{\rho x_3}\right) -\rho\,\partial_z\left(\dfrac{k^{3/2}(z)}{|z|}\sqrt{\gamma_T(z)f_T(z)}\,F_{x_3 z}\right) = 0\,,
\end{align}
can be solved by the following ansatz
\begin{align}\label{v1}
&F_{\rho x_3} = \dfrac{9u_J}{4R^3}\dfrac{k^{5/6}(z)}{|z|}\sqrt{\dfrac{\gamma_T(z)}{f_T(z)}}\dfrac{\partial_z H^{(C)}}{\rho}\,,\\ \label{v2}
&F_{x_3z} = \dfrac{|z|}{k^{3/2}(z)}\dfrac{\partial_\rho H^{(C)}}{\rho\sqrt{f_T(z)\gamma_T(z)}}\,,\\ \label{v3}
&F_{\rho z} = -\dfrac{|z|}{k^{3/2}(z)}\dfrac{\partial_{x_3} H^{(C)}}{\rho\sqrt{f_T(z)\gamma_T(z)}}\,,
\end{align}
where $H^{(C)}$ is an unknown function that has to be determined by requiring that the above components of the field strength satisfy the Bianchi identity \eqref{modb},
\begin{align*}
&- 2\pi\sqrt{2}\,\delta(\rho-l)\delta(x_3)\delta(z) = \partial_\rho F_{x_3z} + \partial_zF_{\rho x_3} - \partial_{x_3}F_{\rho z} =\\
& = \dfrac{|z|}{k^{3/2}(z)\sqrt{f_T(z)\gamma_T(z)}}\left[\partial_\rho\left(\dfrac{\partial_\rho H^{(C)}}{\rho}\right)+ \dfrac{\partial_{x_3}^2H^{(C)}}{\rho}\right]+ \dfrac{9u_J}{4R^3}\dfrac{1}{\rho}\partial_z\left(\dfrac{k^{5/6}(z)}{|z|}\sqrt{\dfrac{\gamma_T(z)}{f_T(z)}}\partial_z H^{(C)}\right).
\end{align*}
The function $H^{(C)}$ is the curved space generalization of the function $H$ introduced in the formula \eqref{eq:H}. We look for a solution by series such as
\begin{equation}
H^{(C)}(\rho,x_3,z;l) = \sum\limits_{n=0}^\infty \zeta_{n}(0)\zeta_{n}(z)J_n\left(\rho,x_3;l\right),
\end{equation}
where the center of mass moduli $(X_3,Z)$ are taken to be zero for simplicity. The Bianchi identity becomes
\begin{align*}
&\sum\limits_{n=0}^\infty\zeta_{n}(0)\bigg[\dfrac{9u_J}{4R^3}\dfrac{J_n}{\rho}\partial_z\left(\dfrac{k^{5/6}(z)}{|z|}\sqrt{\dfrac{\gamma_T(z)}{f_T(z)}}\partial_z \zeta_{n}(z)\right) + \dfrac{|z|\zeta_{n}(z)}{k^{3/2}(z)\sqrt{f_T(z)\gamma_T(z)}}\left(\partial_\rho\left(\dfrac{\partial_\rho J_n}{r}\right)+ \dfrac{\partial_{x_3}^2J_n}{\rho}\right)\bigg] \\
&= - 2\pi\sqrt{2}\,\delta(\rho-l)\delta(x_3)\delta(z)\,.
\end{align*}
This is solved by the following infinite set of equations
\begin{align}\label{eigenv}
    &\partial_z\left(\dfrac{k^{5/6}(z)}{|z|}\sqrt{\dfrac{\gamma_T(z)}{f_T(z)}}\partial_z \zeta_{n}(z)\right) + \dfrac{|z|}{k^{3/2}(z)\sqrt{f_T(z)\gamma_T(z)}}\,\lambda_{n}\,\zeta_{n}(z)=0\,,\\ \label{green}
    &\partial_\rho\left(\dfrac{\partial_\rho J_n}{\rho}\right)+ \dfrac{\partial_{x_3}^2J_n}{\rho}- \dfrac{9u_J}{4R^3}\dfrac{J_n}{\rho} = -2\pi\sqrt{2}\,\delta(\rho - l)\delta(x_3)\,,
\end{align}
supported by the completeness condition
\begin{equation}\label{eq:zetacomplete}
    \sum\limits_{n=0}^\infty\dfrac{|z|}{k^{3/2}(z)\sqrt{f_T(z)\gamma_T(z)}}\zeta_{n}(0)\zeta_{n}(z) = \delta(z)\,,
\end{equation}
satisfied by the functions $\zeta_n$ with orthonormality condition given by
\begin{equation}
    \int^{+\infty}_{-\infty}dz\dfrac{|z|}{k(z)^{3/2}}\dfrac{\zeta_n(z)\zeta_m(z)}{\sqrt{f_T(z)\gamma_T(z)}} = \delta_{nm}\,.
\end{equation}
We have numerically solved the first set of equations \eqref{eigenv}, determining the eigenvalues $\lambda_{n}$, by requiring that the derivatives of the functions $\zeta_{n}$ vanish at $z\to\pm\infty$ for $n>0$,
so that they are orthogonal to the even part of the zero-mode $\zeta_{0}$, which is just a constant given by
\begin{equation}
    \zeta_{0} = \left(\int^{+\infty}_{-\infty}dz\dfrac{|z|}{k(z)^{3/2}}\dfrac{1}{\sqrt{f_T(z)\gamma_T(z)}}\right)^{-1/2}.
\end{equation}
The eigenvalues of the first even modes computed at $\tilde{b} = 0.1$ from $\eqref{eigenv}$ are 
\begin{equation}
    \lambda_{n} = 0,\,2.11,\,6.20,\,12.31,\,20.43,\,30.59,\,...
\end{equation}
In general, the eigenvalues $\lambda_n$ and the solutions $\zeta_n$ depend on the temperature which is encoded in $\tilde{b}$. However, this dependence is very weak and hence the $\lambda_n$'s and the $\zeta_n$'s profile are approximately constant with $\tilde{b}$.
\\We have checked that the numerical solutions form a complete set according to \eqref{eq:zetacomplete}. Notice that since the functions $\zeta_n$ have definite parity, so that $\zeta_n(Z = 0) = 0$ for odd parity, the function $H^{(C)}$ is even in $z$. 

\begin{figure}
\center
\includegraphics[height = 6 cm]{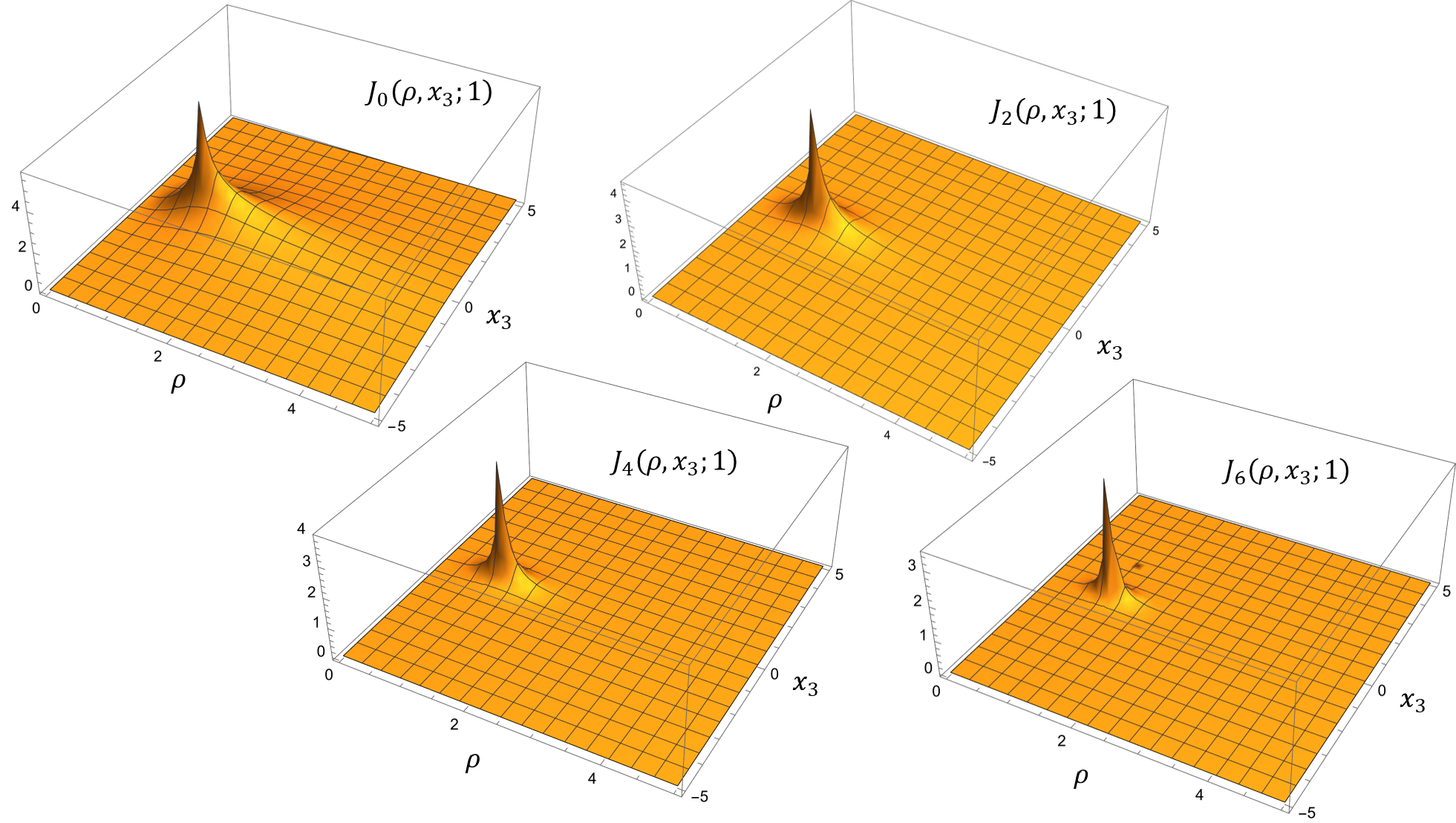}\caption{The first modes of $J_n(\rho,x_3;l)$ for $l=1$, $9u_J /(4R^3) = 1$ and $\tilde{b} = 0.1$. 
}
\label{Jn}
\end{figure}

The solution to \eqref{green} can be found by performing the Fourier transform along the coordinate $x_3$
\begin{equation}\label{eq:Jnmono}
J_n(\rho,x_3;l) = \int \dfrac{dk}{2\pi}e^{ikx_3}\hat{J}_n(\rho,k;l)\,,
\end{equation}
and then by solving the following equation for $\hat{J}_n$ through the Green function method
\begin{equation}
\partial_\rho\left(\dfrac{\partial_\rho \hat{J}_n(\rho,k;l)}{\rho}\right) - \left(k^2 + \dfrac{9u_J}{4R^3}\lambda_{n}\right)\dfrac{\hat{J}_n(\rho,k;l)}{\rho} = -2\pi\sqrt{2}\,\delta(\rho - l)\,.
\end{equation}
Using the Green theorem we can write the solutions as
\begin{equation}
\hat{J}_n(\rho,k;l)=\begin{cases}
&2\pi\sqrt{2}\,lI_1\left(l\sqrt{k^2 + \dfrac{9u_J}{4R^3}\lambda_{n}}\right)\rho K_1\left(\rho\sqrt{k^2 + \dfrac{9u_J}{4R^3}\lambda_{n}}\right)\,\,\,\,\,\,\,\,\,\, \rho>l\,,\\
& 2\pi\sqrt{2}\,lK_1\left(l\sqrt{k^2 + \dfrac{9u_J}{4R^3}\lambda_{n}}\right)\rho I_1\left(\rho\sqrt{k^2 + \dfrac{9u_J}{4R^3}\lambda_{n}}\right)\,\,\,\,\,\,\,\,\,\, \rho<l\,.
\end{cases}
\end{equation}
In order to obtain the solutions $J_n(\rho,x_3;l)$ we have to perform the Fourier transform \eqref{eq:Jnmono} numerically. The first modes $J_0$, $J_2$, $J_4$ and $J_6$ are shown in figure \ref{Jn}, for $l=1$, $9u_J/(4R^3) = 1$ and $\tilde{b} = 0.1$.

We can now evaluate the on-shell D8-branes action,
\begin{align*}
S^\text{DBI}_\text{D8} &= -\dfrac{\kappa_J}{2}\int d^4x\,dz \,\bigg\{\dfrac{k^{3/2}(z)}{|z|}\sqrt{\gamma_T(z)f_T(z)}\left[F_{\rho z}^2 + F_{x_3 z}^2\right]+ \dfrac{4R^3}{9u_J}\dfrac{|z|}{k^{5/6}(z)}\sqrt{\dfrac{f_T(z)}{\gamma_T(z)}}\,F_{\rho x_3}^2\bigg\}=\\
& =-\dfrac{\kappa_J}{2}\sum\limits_{n=0}^\infty\sum\limits_{m=0}^\infty\zeta_{n}(0)\zeta_{m}(0)\int \dfrac{d^4x}{\rho^2}\,dz \,\bigg\{\dfrac{|z|}{k^{3/2}(z)}\dfrac{\zeta_{n}(z)\zeta_{m}(z)}{\sqrt{\gamma_T(z)f_T(z)}}\left[\left(\partial_\rho J_n\right)^2 + \left(\partial_{x_3} J_n\right)^2\right] +\\
& \,\,\,\,\,\,+ \dfrac{9u_J}{4R^3}\sqrt{\dfrac{\gamma_T(z)}{f_T(z)}}\dfrac{k^{5/6}(z)}{|z|}\partial_z\zeta_{n}(z)\partial_z\zeta_{m}(z)J_n J_m\bigg\} =\\
&= -\dfrac{\kappa_J}{2}\sum\limits_{n=0}^\infty\zeta_{n}^2(0)\int \dfrac{d^4x}{\rho^2}\,\bigg\{\left(\partial_\rho J_n\right)^2 + \left(\partial_{x_3} J_n\right)^2 + \dfrac{9u_J}{4R^3}\lambda_{n} J_n^2\bigg\}\,,
\end{align*}
where we have introduced the constant
\begin{equation}
    \kappa_J = \dfrac{\lambda N}{2^6\pi^3 M_{KK}}\left(\dfrac{u_J}{R^3}\right)^{3/2}.
\label{cappaj}
\end{equation}
Defining 
\begin{equation}
\phi_n(\rho,x_3;l) = \sqrt{\kappa_J}\,\zeta_{n}(0)\dfrac{J_n(\rho,x_3;l)}{\rho}\,,
\end{equation}
we get the canonically normalized action
\begin{equation*}
S^\text{DBI}_\text{D8} = -\dfrac{1}{2}\sum\limits_{n=0}^{\infty}\int d^4x\,\bigg\{(\partial_\rho \phi_n)^2 + (\partial_{x_3} \phi_n)^2 + \dfrac{\phi_n^2}{\rho^2} + 2\dfrac{\phi_n}{\rho}\partial_\rho\phi_n + \dfrac{9u_J}{4R^3}\lambda_{n}\phi_n^2\bigg\}\,.
\end{equation*}
The interaction term $\phi\partial\phi$ can be eliminated by integrating by parts so that finally we get
\begin{equation}
S^\text{DBI}_\text{D8} = -\dfrac{1}{2}\sum\limits_{n=0}^{\infty}\int d^4x\,\bigg\{(\partial_\rho \phi_n)^2 + (\partial_{x_3} \phi_n)^2 + \left(\dfrac{1}{\rho^2} +\dfrac{9u_J}{4R^3}\lambda_{n}\right)\phi_n^2\bigg\}\,.
\end{equation}
Upon extending the derivatives to the four coordinates and adding the source term
\begin{equation}
	S_\text{source} = \int d^4x\, 2\pi\sqrt{2}\,\delta(\rho - l)\delta(x_3) \frac{\rho}{\sqrt{\kappa_J}\,\zeta_{n}(0)}\phi_n\,,
\end{equation}
this can be read as the four-dimensional effective action for the uncharged string loop, with the eigenvalues $\lambda_n$ being related to mass parameters. 
The fields in this action come from the mesonic modes at finite temperature, the lightest one being the axion (the zero-mode with $\lambda_0=0$). 
More precisely, the fields $\phi_n$ come from the D8 gauge field modes ``transverse'' to the circular source, i.e.~$A_z$ (including the axion) and $A_{\rho}, A_{x_3}$ (two components of the (tower of) vector mesons), with a specific identification, coming from the five-dimensional equations of motion, reducing to the single tower $\phi_n$ the number of independent fields. 

The free energy of the string loop, defined as $S^\text{DBI}_\text{D8} = -\int dt\, E_{\text{loop}}$, scales linearly with $N$. 
The energy is divergent at the location $x_3 = 0$, $\rho = l$ of the string while it does not diverge at large distances, \textit{i.e.}~for $x_3\to\pm\infty$ and $\rho\to\infty$, as expected from the literature \cite{Kibble:1976sj,Vilenkin:1984ib}. One must remember that the solution is derived under the hypothesis that the field strength is small (compared to $l_s^{-1}$) so to consider the quadratic expansion of the DBI action. Thus, the solution we found cannot be trusted close to the source, where the divergence is expected to be resolved in the full DBI analysis.

\subsubsection{The charged mode}
If we turn on the gauge field components $A_t$ and $A_\psi$ we also have contributions from the CS term in the action. The equations of motion, in this case, are\footnote{Under the reasonable assumption that the gauge field still depends only on $(\rho,x_3,z)$ and the parameter $l$.}
\begin{align}\nonumber
    &\bullet\,\, \kappa_J\rho\,\bigg\{\dfrac{4R^3}{9u_J}\dfrac{|z|}{k^{5/6}(z)}\sqrt{\dfrac{f_T(z)}{\gamma_T(z)}}\,\partial_{x_3} F_{\rho x_3} + \partial_z\left(\dfrac{k^{3/2}(z)}{|z|}\sqrt{\gamma_T(z)f_T(z)}\,F_{\rho z}\right)\bigg\} +\\
    &\,\,\,\,\,\,\,\,\,\,+\dfrac{N\sqrt{2}}{68\pi^2}\epsilon^{\rho MNPQ}F_{MN}F_{PQ} = 0\,.
\end{align}

\begin{align}\nonumber
    &\bullet\,\, -\kappa_J\,\bigg\{\dfrac{4R^3}{9u_J}\dfrac{|z|}{k^{5/6}(z)}\sqrt{\dfrac{f_T(z)}{\gamma_T(z)}}\,\partial_\rho\left(\rho F_{\rho x_3}\right) -\rho\,\partial_z\left(\dfrac{k^{3/2}(z)}{|z|}\sqrt{\gamma_T(z)f_T(z)}\,F_{x_3 z}\right)\bigg\} +\\
    &\,\,\,\,\,\,\,\,\,\,+\dfrac{N\sqrt{2}}{68\pi^2}\epsilon^{3MNPQ}F_{MN}F_{PQ} = 0\,.
\end{align}

\begin{align}
    &\bullet\,\, -\kappa_J\dfrac{k^{3/2}(z)}{|z|}\sqrt{\gamma_T(z)f_T(z)}\,\bigg\{\partial_\rho\left(\rho F_{\rho z}\right) + \rho\,\partial_{x_3}F_{x_3 z}\bigg\}+\dfrac{N\sqrt{2}}{68\pi^2}\epsilon^{zMNPQ}F_{MN}F_{PQ} = 0\,.
\end{align}

\begin{align}\nonumber
    &\bullet\,\, -\kappa_J\,\bigg\{\dfrac{4R^3}{9u_J}\dfrac{|z|}{k^{5/6}(z)}\sqrt{\dfrac{1}{f_T(z)\gamma_T(z)}}\left[\partial_\rho\left(\rho F_{t\rho}\right) + \rho\,\partial_{x_3}F_{t x_3}\right] + \rho\,\partial_z\left(\dfrac{k^{3/2}(z)}{|z|}\sqrt{\dfrac{\gamma_T(z)}{f_T(z)}}\,F_{tz}\right)\bigg\} +\\ \label{atcurvo}
    &\,\,\,\,\,\,\,\,\,\,+\dfrac{N\sqrt{2}}{68\pi^2}\epsilon^{tMNPQ}F_{MN}F_{PQ} = 0\,. 
\end{align}

\begin{align}\nonumber
    &\bullet\,\, -\kappa_J\,\bigg\{\dfrac{4R^3}{9u_J}\dfrac{|z|}{k^{5/6}(z)}\sqrt{\dfrac{f_T(z)}{\gamma_T(z)}}\left[\partial_\rho\left(\dfrac{1}{\rho} F_{\rho\psi}\right) - \dfrac{1}{\rho}\partial_{x_3}F_{\psi x_3}\right] - \dfrac{1}{\rho}\,\partial_z\left(\dfrac{k^{3/2}(z)}{|z|}\sqrt{\gamma_T(z)f_T(z)}\,F_{\psi z}\right)\bigg\} +\\
    &\,\,\,\,\,\,\,\,\,\,+\dfrac{N\sqrt{2}}{68\pi^2}\epsilon^{\psi MNPQ}F_{MN}F_{PQ} = 0\,. \label{apsicurvo}
\end{align}

This system of equations is very hard to solve in general. Following \cite{Hashimoto:2008zw}, we can linearize the equations of motion, neglecting the non-linear terms since they are small for large $\lambda$ and because we look for solutions in the large-$z$ region. In this regime, the monopole solution found in the previous section is not spoiled. We can thus study the linearized equations of motion for $A_t$ and $A_\psi$ \eqref{atcurvo}, \eqref{apsicurvo} searching for curved space Green functions sourced by a circular loop distribution
\begin{align}\nonumber
    \dfrac{4R^3}{9u_J}\dfrac{|z|}{k^{5/6}(z)}&\sqrt{\dfrac{1}{f_T(z)\gamma_T(z)}}
    \left[\partial_\rho\left(\rho \,\partial_\rho A_t\right) + \rho\,\partial_{x_3}^2A_t\right] +\\
    &\rho\,\partial_z\left(\dfrac{k^{3/2}(z)}{|z|}\sqrt{\dfrac{\gamma_T(z)}{f_T(z)}}\,\partial_zA_t\right) \sim \delta(x_3)\delta(\rho-l)\delta(z)\,,\\ \nonumber
    \dfrac{4R^3}{9u_J}\dfrac{|z|}{k^{5/6}(z)}&\sqrt{\dfrac{f_T(z)}{\gamma_T(z)}}\left[\partial_\rho\left(\dfrac{1}{\rho}\partial_\rho A_\psi\right) + \dfrac{1}{\rho}\partial_{x_3}^2 A_\psi\right] +\\
    &\dfrac{1}{\rho}\,\partial_z\left(\dfrac{k^{3/2}(z)}{|z|}\sqrt{\gamma_T(z)f_T(z)}\,\partial_z A_\psi\right) \sim\delta(x_3)\delta(\rho-l)\delta(z)\,.
\end{align}
To solve these equations we define $A_t$ and $A_\psi$, similarly to $H^{(C)}$, as Green's functions in curved space,\footnote{As for $H^{(C)}$, we set to zero the center of mass moduli $(X_3,Z)$.}
\begin{align}\label{eq:A0inf}
    &A_t(\rho,x_3,z;l) \sim \sum\limits_{n=1}^\infty\alpha_n(0)\alpha_n(z)E_n(\rho,x_3;l)\,,\\
    &A_\psi(\rho,x_3,z;l)\sim  \sum\limits_{n=1}^\infty\beta_n(0)\beta_n(z)F_n(\rho,x_3;l)\,.
\end{align}
The functions $\alpha_n$ and $\beta_n$ are solutions of the eigenvalue equations
\begin{align}\label{alphaeq}
    &\partial_z\left(\dfrac{k^{3/2}(z)}{|z|}\sqrt{\dfrac{\gamma_T(z)}{f_T(z)}}\,\partial_z \alpha_n(z)\right) + \dfrac{|z|}{k^{5/6}(z)}\dfrac{1}{\sqrt{f_T(z)\gamma_T(z)}}a_n\alpha_n(z) = 0\,,\\ \label{betaeq}
    &\partial_z\left(\dfrac{k^{3/2}(z)}{|z|}\sqrt{f_T(z)\gamma_T(z)}\,\partial_z \beta_n(z)\right) + \dfrac{|z|}{k^{5/6}(z)}\sqrt{\dfrac{f_T(z)}{\gamma_T(z)}}b_n\beta_n(z) = 0\,,
\end{align}
supported by the completeness conditions
\begin{align}\label{alphacompl}
    &\sum\limits_{n=1}\dfrac{|z|}{k^{5/6}(z)}\dfrac{\alpha_n(0)\alpha_n(z)}{\sqrt{f_T(z)\gamma_T(z)}} = \delta(z)\,,\\ \label{betacompl}
    &\sum\limits_{n=1}\dfrac{|z|}{k^{5/6}(z)}\sqrt{\dfrac{f_T(z)}{\gamma_T(z)}}\beta_n(0)\beta_n(z) = \delta(z)\,,
\end{align}
and the orthonormality conditions given by
\begin{align}
    &\int^{+\infty}_{-\infty}dz\dfrac{|z|}{k^{5/6}(z)}\dfrac{\alpha_n(z)\alpha_m(z)}{\sqrt{f_T(z)\gamma_T(z)}} = \delta_{nm}\,,\\
    &\int^{+\infty}_{-\infty}dz\dfrac{|z|}{k^{5/6}(z)}\sqrt{\dfrac{f_T(z)}{\gamma_T(z)}}\beta_n(z)\beta_m(z) = \delta_{nm}\,.
\end{align}
We have numerically solved the equations \eqref{alphaeq} and \eqref{betaeq}, determining the eigenvalues $a_{n}$ and $b_n$, by requiring that the functions $\alpha_{n}$ and $\beta_n$ vanish at $z\to\pm\infty$. The first even eigenvalues $a_n$ and $b_n$ are:
\begin{align}
    &a_n=0.92,\,1.96,\,3.74,\,5.84,\,8.59,\,11.79,\,15.58,\,...\\
    &b_n = 0.92,\,1.96,\,3.74,\,5.84,\,8.59,\,11.79,\,15.59,\,...
\end{align}
The two sets of eigenvalues appear equal only because we truncated their values to two significant digits. In reality, they differ, albeit slightly. Notice that the eigenvalue equations \eqref{alphaeq} and \eqref{betaeq} do not admit a normalizable zero mode.

The modes $E_n$ and $F_n$ are the solutions of the infinite set of equations
\begin{align}\label{eqE}
    &\dfrac{1}{\rho}\partial_\rho\left(\rho\,\partial_\rho E_n(\rho,x_3;l)\right) + \partial_{x_3}^2E_n(\rho,x_3;l) - \dfrac{9u_J}{4R^3}a_nE_n(\rho,x_3;l) = \delta(\rho-l)\delta(x_3)\,,\\
    &\rho\,\partial_\rho\left(\dfrac{1}{\rho}\partial_\rho F_n(\rho,x_3;l)\right) + \partial_{x_3}^2F_n(\rho,x_3;l) - \dfrac{9u_J}{4R^3}b_nF_n(\rho,x_3;l) = \delta(\rho-l)\delta(x_3)\,.
\end{align}
In order to solve these equations we have to perform the Fourier transform and use the Green theorem
\begin{align}\label{eq:E}
    &E_n(\rho,x_3;l)=\int^{+\infty}_{-\infty}\dfrac{dk}{2\pi}e^{ikx_3}\hat{E}_n(\rho,k;l)\,,\\ \label{eq:F}
    &F_n(\rho,x_3;l)=\int^{+\infty}_{-\infty}\dfrac{dk}{2\pi}e^{ikx_3}\hat{F}_n(\rho,k;l)\,.
\end{align}
Following the same computations we have done for the monopole solution we get
\begin{align}
    &\dfrac{1}{\rho}\partial_\rho\left(\rho\,\partial_\rho\hat{E}_n\right) -\left(k^2 + \dfrac{9u_J}{4R^3}a_n\right)\hat{E}_n = \delta(\rho-l)\,,\\
    &\rho\,\partial_\rho\left(\dfrac{1}{\rho}\partial_\rho\hat{F}_n\right) - \left(k^2 + \dfrac{9u_J}{4R^3}b_n\right)\hat{F}_n = \delta(\rho-l)\,,
\end{align}
whose solutions are
\begin{align}
    &\hat{E}_n(\rho,k,l)=\begin{cases}
&-I_0\left(l\sqrt{k^2 + \dfrac{9u_J}{4R^3}a_n}\right) K_0\left(\rho\sqrt{k^2 + \dfrac{9u_J}{4R^3}a_n}\right)\,\,\,\,\,\,\,\,\,\, \rho>l\,,\\
&-K_0\left(l\sqrt{k^2 + \dfrac{9u_J}{4R^3}a_n}\right) I_0\left(\rho\sqrt{k^2 + \dfrac{9u_J}{4R^3}a_n}\right)\,\,\,\,\,\,\, \rho<l\,.
\end{cases}\\
&\hat{F}_n(\rho,k,l)=\begin{cases}
&-l\rho I_1\left(l\sqrt{k^2 + \dfrac{9u_J}{4R^3}b_n}\right) K_1\left(\rho\sqrt{k^2 + \dfrac{9u_J}{4R^3}b_n}\right)\,\,\,\,\,\,\,\,\,\, \rho>l\,,\\
&-l\rho K_1\left(l\sqrt{k^2 + \dfrac{9u_J}{4R^3}b_n}\right) I_1\left(\rho\sqrt{k^2 + \dfrac{9u_J}{4R^3}b_n}\right)\,\,\,\,\,\,\, \rho<l\,.
\end{cases}
\end{align}

\begin{figure}
\center
\includegraphics[height = 6 cm]{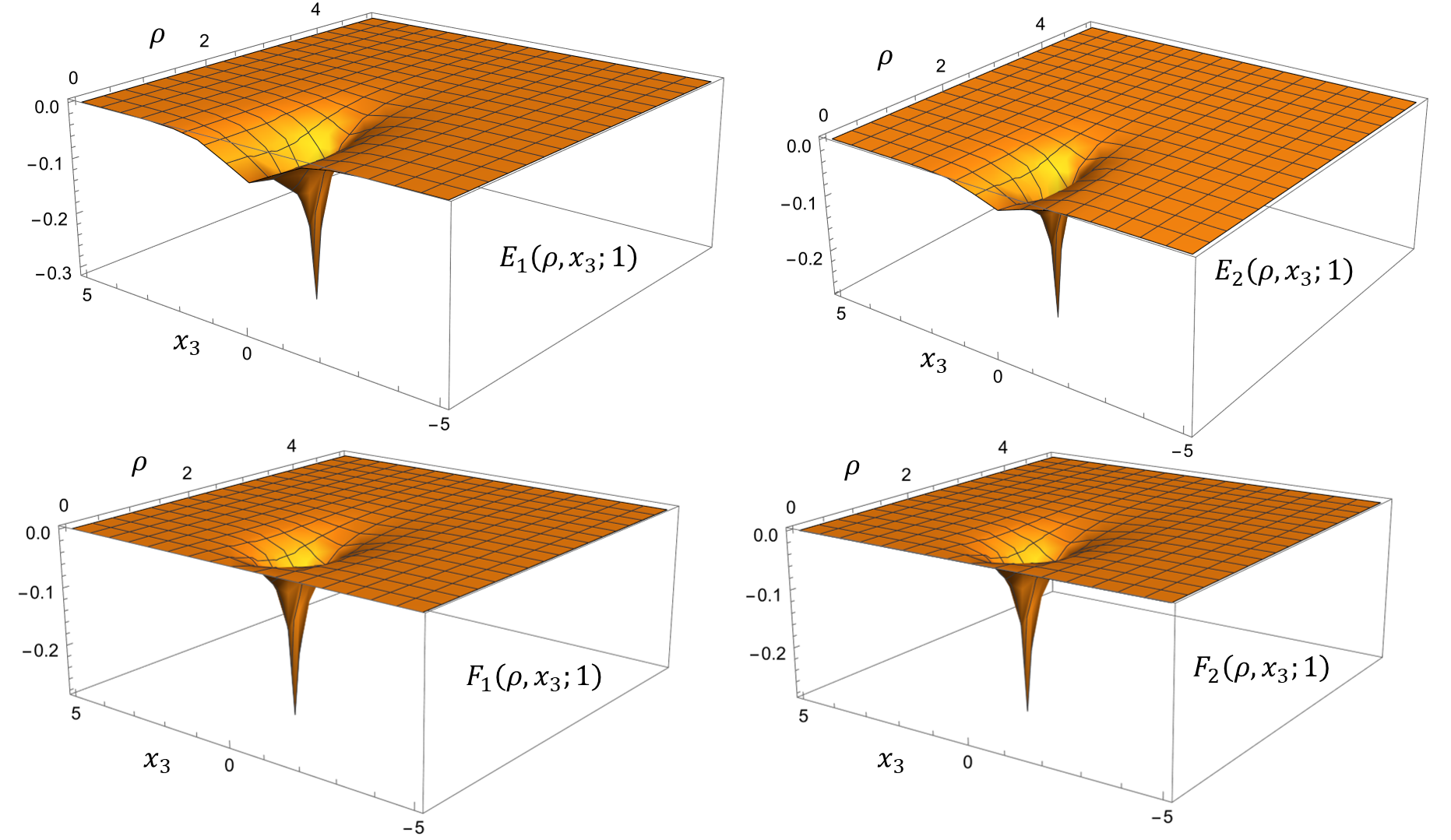}\caption{The first modes of $E_n(\rho,x_3;l)$ and $F_n(\rho,x_3;l)$ for $l=1$, $9u_J /(4R^3) = 1$ and $\tilde{b} = 0.1$. 
}
\label{EFn}
\end{figure}

Figure \ref{EFn} shows the plots of the first modes $E_n$ and $F_n$ for $l=1$, $9u_J /(4R^3) = 1$ and $\tilde{b} = 0.1$. They are obtained by performing numerically the Fourier transforms \eqref{eq:E} and \eqref{eq:F}.
Physically they are the excitation modes of the $t$ and $\psi$ components of the field theory vector mesons, whose non-trivial profiles encompass the charge of the string loop.

We conclude this section by writing the effective action arising from the on-shell Maxwellian DBI action. For $A_t$ we have
\begin{align}\nonumber
    S^\text{DBI}_{A_t} &\sim \dfrac{\lambda N}{9\cdot 2^5\pi^3 M_{KK}}\sqrt{\dfrac{u_J}{R^3}}\int d^4x\,dz\bigg\{\dfrac{|z|}{k^{5/6}(z)\sqrt{\gamma_T(z)f_T(z)}}\big[F_{tx_3}^2+F_{t\rho}^2\big]+\dfrac{9u_J}{4R^3}\dfrac{k^{3/2}(z)}{|z|}\sqrt{
    \dfrac{\gamma_T(z)}{f_T(z)}}\,F_{tz}^2\bigg\} =\\ \nonumber
    &= \dfrac{\lambda N}{9\cdot 2^5\pi^3 M_{KK}}\sqrt{\dfrac{u_J}{R^3}}\sum\limits_{n=1}^{\infty}\sum\limits_{m=1}^{\infty}\alpha_n(0)\alpha_m(0)\int d^4x\,dz\bigg\{\dfrac{|z|\,\alpha_n(z)\alpha_m(z)}{k^{5/6}(z)\sqrt{\gamma_T(z)f_T(z)}}\big[(\partial_{x_3} E_n)^2+(\partial_{\rho}E_n)^2\big]+\\ \nonumber
    &\,\,\,\,\,\,\,\,+\dfrac{9u_J}{4R^3}\dfrac{k^{3/2}(z)}{|z|}\sqrt{
    \dfrac{\gamma_T(z)}{f_T(z)}}\,\partial{_z\alpha_n(z)\partial_z\alpha_m(z)}E_nE_m\bigg\} =\\
    &=\dfrac{\lambda N}{9\cdot 2^5\pi^3 M_{KK}}\sqrt{\dfrac{u_J}{R^3}}\sum\limits_{n=1}^{\infty}\alpha_n^2(0)\int d^4x\bigg\{(\partial_{x_3} E_n)^2+(\partial_{\rho}E_n)^2+\dfrac{9u_J}{4R^3}a_nE_n^2\bigg\}.
\end{align}
We can define 
\begin{equation}
    \varepsilon_n(\rho,x_3;l) = \sqrt{\dfrac{\lambda N}{9\cdot 2^4\pi^3 M_{KK}}\sqrt{\dfrac{u_J}{R^3}}}\,\alpha_n(0)E_n(\rho,x_3;l),
\end{equation}
to get
\begin{align}\label{acat}
    S_{A_t}^\text{DBI} \sim \dfrac{1}{2}\sum\limits_{n=1}^{\infty}\int d^4x\bigg\{(\partial_{x_3} \varepsilon_n)^2+(\partial_{\rho}\varepsilon_n)^2+\dfrac{9u_J}{4R^3}a_n\varepsilon_n^2\bigg\}.
\end{align}
We can do the same thing for $A_\psi$:
\begin{align} \nonumber
    S^\text{DBI}_{A_\psi} &\sim -\dfrac{\lambda N}{9\cdot 2^5\pi^3 M_{KK}}\sqrt{\dfrac{u_J}{R^3}}\int \dfrac{d^4x}{\rho^2}\,dz\bigg\{\dfrac{|z|}{k^{5/6}(z)}\sqrt{
    \dfrac{f_T(z)}{\gamma_T(z)}}\big[F_{\psi x_3}^2+F_{\psi\rho}^2\big]+\dfrac{9u_J}{4R^3}\dfrac{k^{3/2}(z)}{|z|}\sqrt{\gamma_T(z)f_T(z)}\,F_{\psi z}^2\bigg\} =\\ \nonumber
    &= -\dfrac{\lambda N}{9\cdot 2^5\pi^3 M_{KK}}\sqrt{\dfrac{u_J}{R^3}}\sum\limits_{n=1}^{\infty}\sum\limits_{m=1}^{\infty}\alpha_n(0)\alpha_m(0)\int \dfrac{d^4x}{\rho^2}\,dz\bigg\{\dfrac{|z|}{k^{5/6}(z)}\sqrt{
    \dfrac{f_T(z)}{\gamma_T(z)}}\,\beta_n(z)\beta_m(z)\big[(\partial_{x_3} F_n)^2+(\partial_{\rho}F_n)^2\big]+\\ \nonumber
    &\,\,\,\,\,\,\,\,+\dfrac{9u_J}{4R^3}\dfrac{k^{3/2}(z)}{|z|}\sqrt{\gamma_T(z)f_T(z)}\,\partial{_z\beta_n(z)\partial_z\beta_m(z)}F_nF_m\bigg\} =\\
    &=-\dfrac{\lambda N}{9\cdot 2^5\pi^3 M_{KK}}\sqrt{\dfrac{u_J}{R^3}}\sum\limits_{n=1}^{\infty}\beta_n^2(0)\int \dfrac{d^4x}{\rho^2}\bigg\{(\partial_{x_3} F_n)^2+(\partial_{\rho}F_n)^2+\dfrac{9u_J}{4R^3}b_nF_n^2\bigg\},
\end{align}
and define
\begin{equation}
    \varsigma_n(\rho,x_3;l) = \sqrt{\dfrac{\lambda N}{9\cdot 2^4\pi^3 M_{KK}}\sqrt{\dfrac{u_J}{R^3}}}\,\beta_n(0)\dfrac{F_n(\rho,x_3;l)}{\rho},
\end{equation}
so that we can write
\begin{align}\label{acapsi}
    S_{A_\psi}^\text{DBI} \sim -\dfrac{1}{2}\sum\limits_{n=1}^{\infty}\int d^4x\bigg\{(\partial_{x_3} \varsigma_n)^2+(\partial_{\rho}\varsigma_n)^2+\left(\dfrac{1}{\rho^2}+\dfrac{9u_J}{4R^3}b_n\right)\varsigma_n^2\bigg\}.
\end{align}

Promoting the derivatives to the four coordinates, the actions (\ref{acat}), (\ref{acapsi}) represent the four-dimensional effective actions for the charged modes of the vortons.
They are nothing else that the reduced actions, in cylindrical coordinates, for the D8 gauge field components ``longitudinal'' to the circular source, i.e.~$A_t$ and $A_{\psi}$, corresponding to two components of the (tower of) vector mesons.\footnote{The masses appearing in the four-dimensional effective actions are different for the various components due to the breaking of Lorentz invariance by the temperature and the choice of cylindrical coordinates.}

\subsubsection{Currents in the deconfined phase of the WSS model}\label{currents}
In this section, we compute the currents associated with the WSS flavor symmetry in the deconfined phase.
The procedure is analogous to the computation of the chiral currents in the confined phase of the WSS model \cite{Hashimoto:2008zw}. In order to study the flavor currents, it is useful to introduce external gauge fields $\mathcal{A}_{L\,\mu}$, $\mathcal{A}_{R\,\mu}$, associated to the chiral symmetry $U(N_f)_L\times U(N_f)_R$ in Minkowski space. The chiral currents $\mathcal{J}_L^\mu$, $\mathcal{J}_R^\mu$ can be read off from the action term that is linear in the external gauge fields:
\begin{equation}\label{eq:prescription}
    S_{\text{eff}}\big[\mathcal{A}_L,\mathcal{A}_R\big] = -2\int d^4x\,\text{tr}\left(\mathcal{A}_{L\,\mu}\mathcal{J}_L^\mu + \mathcal{A}_{R\,\mu}\mathcal{J}_R^\mu\right).
\end{equation}
We can trade the left and right components for the vectorial and axial components of the current
\begin{equation}
     S_\text{eff}\big[\mathcal{A}_L,\mathcal{A}_R\big] = -2\int d^4x\,\text{tr}\left(\mathcal{V}_{\mu}^{(+)}\mathcal{J}_V^\mu + \mathcal{V}_{\mu}^{(-)}\mathcal{J}_A^\mu\right),
\end{equation}
where
\begin{equation}
    \mathcal{J}_V^\mu =\mathcal{J}_L^\mu  +\mathcal{J}_R^\mu\,,\,\,\,\,\,\,\,\,\,\,\,\,\,   \mathcal{J}_A^\mu =\mathcal{J}_L^\mu - \mathcal{J}_R^\mu\,,\,\,\,\,\,\,\,\,\,\,\,\,\, \mathcal{V}_{\mu}^{(\pm)} = \dfrac{\mathcal{A}_{\mu,L} \pm\mathcal{A}_{\mu,R}}{2}\,.
\end{equation}
In the WSS model the chiral symmetry is identified with the constant gauge transformation at $z=\pm\infty$, and the external gauge fields $\mathcal{A}_{L\,\mu}$, $\mathcal{A}_{R\,\mu}$ can be introduced by the following boundary conditions for the five-dimensional fields
\begin{equation}
    \mathcal{A}_{\mu}(x^\mu,z\to +\infty) =\mathcal{A}_{\mu,L}(x^\mu)\,, \,\,\,\,\,\,\,\,\,\,\,\,\,\,\,\,\mathcal{A}_{\mu}(x^\mu,z\to -\infty) =\mathcal{A}_{\mu,R}(x^\mu)\,.
\end{equation}
In order to write down an expression for the current we need to compute the first variation of the action. Let us consider the connection
\begin{equation}
    \mathcal{A}_M(x^\mu,z) =\mathcal{A}_M^\text{cl}(x^\mu,z) + \delta \mathcal{A}_M(x^\mu,z)\,,
\end{equation}
such that $\mathcal{A}_M^\text{cl}(x^\mu,z)$ is a classical solution to the equations of motion with boundary conditions $\mathcal{A}_M^\text{cl}(x^\mu,z\to\pm\infty)=0$  while $\delta \mathcal{A}_M(x^\mu,z)$ is a small deviation from the classical solution such that
\begin{equation}\label{bc}
     \delta\mathcal{A}_{\mu}(x^\mu,z\to +\infty) =\mathcal{A}_{L\,\mu}(x^\mu), \,\,\,\,\,\,\,\,\,\,\,\,\,\,\,\,\delta\mathcal{A}_{\mu}(x^\mu,z\to -\infty) =\mathcal{A}_{R\,\mu}(x^\mu)\,.
\end{equation}
Remember that for the one-flavored case $\mathcal{A} = A/\sqrt{2}$, with $A$ the $U(1)$ gauge field. 

The background metric in the deconfined phase breaks the covariance in the Minkowski space but it is straightforward to find the currents as in the confined background. The DBI action for the D8-brane expanded at quadratic order is
\begin{align*}
    S^\text{DBI}_\text{D8} = &-\dfrac{T_8V_4R^{3/2}(2\pi\alpha')^2}{4g_s}\int d^4x\,dz \,u^{5/2}\sqrt{\dfrac{f_T(u)}{\gamma_T(u)}}\bigg|\dfrac{\partial u}{\partial z}\bigg|\bigg[\dfrac{1}{2}\left(\dfrac{R}{u}\right)^{3}F_{ij}^2 - \left(\dfrac{R}{u}\right)^{3}\dfrac{F_{ti}^2}{f_T(u)} +\\
    &  \bigg|\dfrac{\partial u}{\partial z}\bigg|^{-2}\dfrac{\gamma_T(u)}{f_T(u)}F_{tz}^2 + \bigg|\dfrac{\partial u}{\partial z}\bigg|^{-2}\gamma_T(u)F_{iz}^2\bigg].
\end{align*}
The CS term leads to a vanishing boundary term (at $z\to\pm\infty$) and hence it does not contribute to the current computation. This is why we will not consider the CS term in this section even if it does contribute to the bulk action and hence to the equations of motion for the gauge field.

We consider a small deviation $\delta A_M$ from the classical solution to the equation of motion obeying the boundary conditions \eqref{bc}. The field strength components can be written as
\begin{align*}
    &F_{ij}^2 = \left(F_{ij}^\text{cl}\right)^2 + 2\eta^{ik}\eta^{jl}F_{kl}^\text{cl}\left(\partial_i\delta A_j - \partial_j \delta A_i\right),\\
    &F_{ti}^2 = \left(F_{ti}^\text{cl}\right)^2 + 2F_{ti}^\text{cl}\left(\partial_t\delta A_i - \partial_i \delta A_t\right),\\
    &F_{tz}^2 = \left(F_{tz}^\text{cl}\right)^2 + 2F_{tz}^\text{cl}\left(\partial_t\delta A_z - \partial_z \delta A_t\right),\\
    &F_{iz}^2 = \left(F_{iz}^\text{cl}\right)^2 + 2F_{iz}^\text{cl}\left(\partial_i\delta A_z - \partial_z \delta A_i\right),
\end{align*}
where the coefficients $\eta_{MN}$ ($M,N = t, x_1, x_2,x_3,z$) refer to the five-dimensional flat metric.

In order to read off the flavor currents we look at the terms with derivatives w.r.t.~the holographic direction $z$ since they give the boundary terms of the action:
\begin{align*}
    S_\text{bdry} = &\dfrac{T_8V_4R^{3/2}(2\pi\alpha')^2}{2g_s}\int d^4x\,dz \,\partial_z\left[u^{5/2}\bigg|\dfrac{\partial u}{\partial z}\bigg|^{-1}\left(\sqrt{\dfrac{\gamma_T(u)}{f_T(u)}}\,\eta^{tt}F_{tz}^{cl}\delta A_t + \sqrt{\gamma_T(u)f_T(u)}\,
    \eta^{ij}F_{iz}^{cl}\delta A_{j}\right)\right]\\
    & \hspace{-0.4cm}= \dfrac{T_8V_4R^{3/2}(2\pi\alpha')^2}{2g_s}\int d^4x\left[u^{5/2}\bigg|\dfrac{\partial u}{\partial z}\bigg|^{-1}\left(\sqrt{\dfrac{\gamma_T(u)}{f_T(u)}}\,\eta^{tt}F_{tz}^{cl}\delta A_t + \sqrt{\gamma_T(u)f_T(u)}\,
    \eta^{ij}F_{iz}^{cl}\delta A_{j}\right)\right]\Bigg|^{z=+\infty}_{z=-\infty}\,\,\,.
\end{align*}
Following the prescription \eqref{eq:prescription}, the chiral currents can be written as
\begin{align}
    &J^t_{L} = 
    -\kappa_J\bigg[\dfrac{k(z)^{3/2}}{|z|}\sqrt{\dfrac{\gamma_T(z)}{f_T(z)}}\,\eta^{tt}F_{tz}^\text{cl}\bigg]\bigg|_{z=+\infty}\,\,\,,\\
    &J^t_{R} = 
    \kappa_J\bigg[\dfrac{k(z)^{3/2}}{|z|}\sqrt{\dfrac{\gamma_T(z)}{f_T(z)}}\,\eta^{tt}F_{tz}^\text{cl}\bigg]\bigg|_{z=-\infty}\,\,\,,\\
    &J^i_{L} = 
    -\kappa_J\bigg[\dfrac{k(z)^{3/2}}{|z|}\sqrt{\gamma_T(z)f_T(z)}\,\eta^{ij}F_{jz}^\text{cl}\bigg]\bigg|_{z=+\infty}\,\,\,,\\
    &J^i_{R} = 
    \kappa_J\bigg[\dfrac{k(z)^{3/2}}{|z|}\sqrt{\gamma_T(z)f_T(z)}\,\eta^{ij}F_{jz}^\text{cl}\bigg]\bigg|_{z=-\infty}\,\,\,,
\end{align}
where $\kappa_J$ has been defined in \eqref{cappaj}. 

From these expressions, we can extract the vectorial and axial components of the currents:
\begin{align}
    &J^t_{V} = -\kappa_J\bigg[\dfrac{k(z)^{3/2}}{|z|}\sqrt{\dfrac{\gamma_T(z)}{f_T(z)}}\,\eta^{tt}F_{tz}^\text{cl}\bigg]\bigg|_{z=-\infty}^{z=+\infty}\,\,\,,\\
    &J^t_{A} = -\kappa_J\bigg[\dfrac{\psi_0(z)k(z)^{3/2}}{|z|}\sqrt{\dfrac{\gamma_T(z)}{f_T(z)}}\,\eta^{tt}F_{tz}^\text{cl}\bigg]\bigg|_{z=-\infty}^{z=+\infty}\,\,\,,\\
    &J^i_{V} = -\kappa_J\bigg[\dfrac{k(z)^{3/2}}{|z|}\sqrt{\gamma_T(z)f_T(z)}\,\eta^{ij}F_{jz}^\text{cl}\bigg]\bigg|_{z=-\infty}^{z=+\infty}\,\,\,,\\
    &J^i_{A} = -\kappa_J\bigg[\dfrac{\psi_0(z)k(z)^{3/2}}{|z|}\sqrt{\gamma_T(z)f_T(z)}\,\eta^{ij}F_{jz}^\text{cl}\bigg]\bigg|_{z=-\infty}^{z=+\infty}\,\,\,,
\end{align}
where $\psi_0(z) = \frac{2}{\pi}\arctan(z)$. We will drop the apex ``cl.'' from now on.

We can use these results to compute the baryon number current (and in general all the other components of the currents $J_R$ and $J_L$) associated with the electric potential $A_t$. This fixes the proportionality constant in front of the solution \eqref{eq:A0inf} for $A_t$ to $K = -(3^2 2^3\pi^3R^{3/2})/(u_J^{1/2}M_{KK}\lambda)$:
\begin{align}\nonumber
    J^t_B &= \dfrac{2}{N}J^t_V = -\dfrac{2}{N}\kappa_J\left[\dfrac{k^{3/2}(z)}{|z|}\sqrt{\dfrac{\gamma_T(z)}{f_T(z)}}\eta^{tt}F_{tz}\right]\bigg|^{z=+\infty}_{z=-\infty} \\ \nonumber
    &=K\kappa_J\sum\limits_{n=1}^\infty\alpha_n(0)E_n(\rho,x_3;l)\left[\dfrac{k^{3/2}(z)}{|z|}\sqrt{\dfrac{\gamma_T(z)}{f_T(z)}}\partial_z\alpha_n(z)\right]\bigg|^{z=+\infty}_{z=-\infty} =\\
    &= K\kappa_J\sum\limits_{n=1}^\infty g_{V^n}^\alpha\alpha_{2n-1}(0)E_{2n-1}(\rho,x_3;l)\,,
\end{align}
where we have defined the decay constants associated with the eigenfunctions $\alpha_n(z)$ as
\begin{equation*}
    g_{V^n}^\alpha = a_{2n-1}\int dz\,\dfrac{|z|}{k^{5/6}(z)}\dfrac{\alpha_{2n-1}(z)}{\sqrt{\gamma_T(z)f_T(z)}}\,,\,\,\,\,\,\,\, \quad  g_{A^n}^\alpha = a_{2n}\int dz\,\dfrac{|z|}{k^{5/6}(z)}\dfrac{\alpha_{2n}(z)\psi_0(z)}{\sqrt{\gamma_T(z)f_T(z)}}\,.
\end{equation*}

A direct application of this formula is to compute the baryon number charge 
\begin{align}\nonumber
    n_B &= \int^{+\infty}_{-\infty}dx_3\int^\infty_0 \rho\,d\psi\,d\rho\,\,\langle J^t_B\rangle = 2\pi\int^{+\infty}_{-\infty}dx_3\int^\infty_0 \rho\,d\rho\,\,\langle J^t_B\rangle \\\nonumber
    &=2\pi\dfrac{9u_J}{4R^3}\sum\limits_{n=1}^\infty g_{V^n}^\alpha\langle\alpha_{2n-1}(0)\rangle\int^{+\infty}_{-\infty}dx_3\int^\infty_0 \rho\,d\rho\,E_{2n-1}(\rho,x_3;l)=\\
    &=\sum\limits_{n=1}^\infty \dfrac{g_{V^n}^\alpha}{a_{2n-1}}\langle\alpha_{2n-1}(0)\rangle = 1\,,
\end{align}
where we used the properties of the Fourier transform and the addition sum rule of Bessel functions. Note that this definition of baryon number is equivalent to the one given in terms of the CS term in equation \eqref{eq:nbD8} (plus total derivatives) by using the equations of motion.

\section{Conclusions and outlook}\label{sec:conclusions}

In this work, we provided a detailed description of cosmic topological defects within the holographic dual of an $SU(N)$ gauge theory coupled to one fundamental flavor at large $N$ and finite temperature. At low energies, the effective theory is governed by an axion, which emerges as a pseudo-Goldstone boson associated with breaking the global $U(1)_A$ flavor symmetry. The cosmological evolution of this system can lead to the spontaneous formation of topological defects. These defects arise as non-trivial configurations of the axion field and other mesons, and their properties depend on the specific symmetry-breaking pattern and the thermal history of the Universe.

We examined various topological defects, including straight axionic strings, axionic string loops, axionic vortons, and axionic DWs, through (wrapped) D6-branes in the WSS model. A notable feature of this construction is the emergence of a Chern-Simons theory on the (reduced) (2+1)-dimensional world volume of the D6-branes.  The presence of a CS theory on $SU(N)$ DWs has been studied extensively in \cite{Gaiotto:2017tne}, and the interplay between topological field theories like the CS one and axionic defects has been recently explored in \cite{Brennan:2023kpw}. Hereby, this work positions itself at the intersection of several recent studies that explore the topological aspects of theories involving an $SU(N)$ gauge group coupled with fundamental flavors. 

Our analysis first focused on the D6-brane description of the defects, followed by a description in terms of the flavor degrees of freedom living on the D8-brane. Initially, we reviewed the description of straight axionic strings and computed their tension. Subsequently, we analyzed the D6-brane embedding for string loops and DWs. The embedding of the D6-brane with a circular boundary on the D8-brane revealed the existence of a first-order phase transition\footnote{As discussed earlier, this transition is not the typical first-order phase transition between two equilibrium phases.} between the string loop embedding and the DW embedding. We computed the critical radius (of the D6-brane boundary on the D8-brane) at which this transition occurs, finding that it is a non-trivial function of the background temperature. The WSS model is well known for its rich phase structure, with recent studies addressing phase transitions and bubble nucleation in this model (see, for example, \cite{Bigazzi:2020avc,Bigazzi:2020phm,Bigazzi:2021ucw}). However, this phase transition has not been considered previously in the literature, as DWs are absent in the deconfined phase of conventional $SU(N)$ gauge theory. The reason can be found in the behavior of the WSS at high temperatures. In this regime, the deconfined phase of the WSS model starts to exhibit properties reflecting the nature of the underlying string theory setup diverging from standard $SU(N)$ gauge theories.\footnote{It has been argued in \cite{Mandal:2011ws} that the deconfined phase of WSS is not in the same universality class as that of $SU(N)$ Yang-Mills since some discrete symmetries do not match.} This behavior has consequences for axion physics, for example, the axion mass does not go to zero for large temperatures but grows instead \cite{Bigazzi:2019eks}. Since in standard phenomenological axion models the appearance of axionic DWs is tied to the emergence of an effective axion mass, one can argue for the presence of axionic DWs in the deconfined phase of WSS to be related to these high-temperature features.

Recent works have also highlighted the relevance of phase transitions seeded by cosmic topological defects \cite{Blasi:2022woz,Blasi:2023rqi,Blasi:2024mtc}. 
It would be interesting to investigate if and how the D6-brane defects influence the first-order phase transition associated with chiral symmetry breaking in the WSS model.

We also studied vortons, which are described as charged (wrapped) D6-branes. As we have seen in section \ref{subsec:spinning}, the D6-brane supports both a fundamental string charge and a D4-brane charge. The first is associated with the number of constituent quarks while the latter is associated with a baryon vertex in the WSS model. Moreover, the D6-brane possesses non-zero angular momentum induced by the non-zero gauge field living on its world volume. Since our solutions are time-independent, this angular momentum is actually interpreted as a spin in the dual QFT. We have shown that, provided certain boundary conditions are satisfied, the spin $J$ of vortons satisfies the relation $|J| = n_B^2\,N/2$, where $n_B$ is the baryon charge (associated with the D4-brane charge), with the (2+1)-dimensional $U(1)_N$ CS theory playing a crucial role in the derivation. This quadratic relation between charge and angular momentum is a hallmark of anyonic systems, reinforcing the interpretation of these cosmic topological defects as excitations of a topological phase governed by the $U(1)_N$ CS term.

The stability of charged string loops has been the focus of extensive literature, with renewed interest in recent years (see e.g.~\cite{Carter:1993wu,Brandenberger:1996zp,Martins:1998gb,Martins:1998th,Carter:1999an,Agrawal:2020euj,Ibe:2021ctf,Abe:2022rrh}). These studies argue that charged fermions present on the string can exert pressure to counteract shrinking, potentially leading to stable configurations. Typically, vortons are considered charged under electromagnetic symmetry, and their decay into Standard Model particles often destabilizes them. Our analysis indicates that vortons with baryon charges of order $\lambda^0$ are unstable, but increasing the charge can render them (meta)stable.

We further explored uncharged string loops and vortons through the D8-brane gauge field, \textit{i.e.}, mesonic modes. The D6-brane boundary on the D8-brane acts as a source for the D8-brane gauge field, leading to a non-trivial profile for low-energy mesonic modes. 
In particular, we derived the mesonic profile and the four-dimensional effective action for the string loop. Then, we turned on the electric potential on the D8-branes, associated with the baryon charge, which gives rise to a Coulomb repulsion mediated by the CS term that can stabilize the loop. 
A non-trivial electric potential corresponds to non-trivial profiles for the vector mesons.
Analyzing the system locally around the D8-brane tip at $u=u_J$, i.e.~in a flat limit of the WSS background, we found that the stability radius of string loops with baryon charge $n_B = 1$ and spin $J = N/2$, should be parametrically small in $\lambda$, in perfect agreement with the D6-brane analysis. However, the existence of such meta(stable) solutions cannot be verified in the flat space limit. Actually, the D6-brane embedding study suggests that vortons with the above quantum numbers cannot be stable. Unfortunately, we are not able to attach the problem from the D8-brane perspective in curved spacetime.

Finally, we emphasized the need to study charged DWs and their stability in both the confined and deconfined phases of the WSS model \cite{JL}. As discussed in section \ref{sec:D6}, charged DWs, described as spinning D6-branes, represent baryonic excitations of the dual theory. In the deconfined phase, since these D6-branes do not terminate at the horizon, they avoid direct interaction with the $SU(N)$ plasma, resulting in a decay rate that is parametrically suppressed in $\lambda$.  
In a cosmological context, these objects may serve as novel dark matter candidates, as recently proposed in \cite{Bigazzi:2022ylj} under the name ``\textit{axionic baryons}''. These charged DWs could persist in the present Universe, potentially contributing to the dark matter content. Moreover, they might leave observable imprints, such as through interactions with the cosmic microwave background or gravitational waves. Investigating these imprints could open new avenues for dark matter detection and shed light on dark QCD phases.

Future work should also aim to provide a coherent and detailed description of charged domain walls in terms of mesonic modes, specifically the excitations of the D8-brane gauge fields. This approach could illuminate the underlying mechanisms that stabilize these charged DWs, as well as their potential interactions with other mesonic excitations. Such a description would not only enrich our understanding of baryons as topologically protected excitations but also extend the holographic picture of these objects to scenarios reminiscent of Hall droplets in Chern-Simons theories. This analogy draws on the idea that baryons, viewed as solitonic objects, could emerge as collective excitations within a topologically non-trivial phase of the gauge theory. This would complete the holographic picture of baryons as \textit{quantum Hall droplets} initiated in \cite{Bigazzi:2022luo} and inspired by the seminal ideas in \cite{Komargodski:2018odf}.
In this context, extending the model to include multiple quark flavors is crucial for making it relevant to real-world hadronic interactions. Instantons initially
used to describe baryons in holographic models as solitons of four-dimensional QCD, can also be connected to the dynamics of (2+1)-dimensional solitons in CS theories.

\section*{Acknowledgments}
We thank Sergio Cacciatori, Alessio Caddeo, Fabrizio Canfora, Andrea Cappelli, Carlos Hoyos and Jean Loup Raymond for suggestions, comments, and very helpful discussions. This project has been partially supported by the grant PRIN 20227S3M3B,  ``Bubble Dynamics in Cosmological Phase Transitions".

\appendix

\section{Green function in duo-polar coordinates}\label{app:green}
In this appendix, we analyze a more general version of the equation of motion for $A_t$ in flat space which is
\begin{equation}
    \dfrac{1}{r}\,\partial_r\left(r\,\partial_r A_t\right) + \dfrac{1}{\rho}\,\partial_\rho\left(\rho \,\partial_\rho A_t\right) +\tau \,I(r,\rho)= 0\,,
\end{equation}
where $I(r,\rho)$ is a generic instanton density 
and $\tau$ is a constant. We use the Green function method which 
consists in finding the Green function associated with the operator that acts on $A_t$ and then calculating the convolution with $I(r,\rho)$. In other words, the solution can be written as
\begin{equation}\label{conv}
    A_t = \tau\int d^4x'\, G(r,\rho,r',\rho')\,I(r,\rho)\,,
\end{equation}
where $G$ is the Green function satisfying the equation
\begin{equation}\label{Geq}
    \dfrac{1}{r}\,\partial_r\left(r\,\partial_r G\right) + \dfrac{1}{\rho}\,\partial_\rho\left(\rho \,\partial_\rho G\right) = \dfrac{\delta(\rho-\rho')}{2\pi\rho'}\dfrac{\delta(r-r')}{2\pi r'}\,.
\end{equation}
One way to solve the above equation is to interpret $G$ as the potential generated by a charge distribution located at $\rho = \rho', r= r'$:
\begin{align*}
    G &= \int^{2\pi}_0\int^{2\pi}_0\dfrac{d\psi\, d\theta}{\rho^2 + \rho'^2 + r^2 + r'^2 - 2\rho\rho'\cos(\psi) - 2rr'\cos(\theta)} = \\
    &= 2\pi\int^{2\pi}_0\dfrac{d\theta}{\sqrt{(\rho^2 + \rho'^2 + r^2 + r'^2 - 2rr'\cos(\theta))^2 - 4\rho^2\rho'^2}}\,.
\end{align*}
We use the following trigonometric substitution in the integral:
\begin{equation*}
    t = \tan\left(\dfrac{\theta}{2}\right)\,,\,\,\,\, d\theta = \dfrac{2}{1+t^2}dt\,,\,\,\,\,\,\, \cos(\theta) = \dfrac{1-t^2}{1+t^2}\,,
\end{equation*}
that is defined for $\theta\in (-\pi,\pi)$. Therefore, we have to shift the angular variable, and then use the trigonometric substitution. After some calculations, we end up with
\begin{equation}
     G = \dfrac{4\pi}{\sqrt{mp}}\int^{+\infty}_{-\infty}\dfrac{dt}{\sqrt{\left(t^2+\dfrac{n}{m}\right)\left(t^2+\dfrac{q}{p}\right)}}\,, 
\end{equation}
where
\begin{align*}
    &m=(\rho-\rho')^2+(r-r')^2,\,\,\,\,\,\,\,\,\,\,\,\,\,\,\,\,\,\,\,\,n=(\rho-\rho')^2+(r+r')^2\,,\\
    &p=(\rho+\rho')^2+(r-r')^2,\,\,\,\,\,\,\,\,\,\,\,\,\,\,\,\,\,\,\,\,q= (\rho+\rho')^2+(r+r')^2\,.
\end{align*}
We can go to the complex plane and solve the integral using the Jordan Lemma taking care of the four simple poles and the two branch cuts. The integral admits an analytic solution in terms of the Elliptic integral of the first kind, which we will indicate as $\mathbf{K}(x)$:
\begin{align}
    G&= \dfrac{4\pi}{\sqrt{mp}}\int^{+\infty}_{-\infty}\dfrac{dt}{\sqrt{\left(t^2+\dfrac{n}{m}\right)\left(t^2+\dfrac{q}{p}\right)}} =\dfrac{4\pi}{\sqrt{mp}}\dfrac{2}{\sqrt{q/p}}\mathbf{K}\left(1-\dfrac{n}{m}\dfrac{p}{q}\right) =\nonumber \\
    &= \dfrac{8\pi}{\sqrt{((\rho-\rho')^2+(r-r')^2)((\rho+\rho')^2+(r+r')^2)}} \cdot \nonumber\\
    &\hspace{5cm}\cdot\mathbf{K}\left(1-\dfrac{(\rho-\rho')^2+(r+r')^2}{(\rho-\rho')^2+(r-r')^2}\cdot\dfrac{(\rho+\rho')^2+(r-r')^2}{(\rho+\rho')^2+(r+r')^2}\right)\,.
\end{align}
One can check that this is indeed a solution of \eqref{Geq}. To get the solution for $A_t$ we have to perform the convolution \eqref{conv} with $I(r,\rho;l)$. In general, the convolution will not have an analytic result and the integration has to be performed only numerically. In this work, we have found an analytic solution for the convolution with the instanton density
\begin{equation}
    I(r,\rho) =\dfrac{16\pi\sqrt{2}\,\mathcal{N}\,r\rho \,l^4}{\left[(l^2+\rho^2+r^2)^2-4\rho^2l^2\right]^2}\,,
\end{equation}
coming from the definition \eqref{eq:instanton}. The result of the convolution reads
\begin{equation}
    A_t = \dfrac{2\pi\sqrt{2}\,\mathcal{N}\,\tau(\rho^2+r^2)}{(l^2+\rho^2+r^2)^2-4\rho^2l^2} + A_t^{(0)}\,,
\end{equation}
where we have the freedom to add the zero modes of the Laplacian operator in duo-polar coordinates:
\begin{equation}
    A_t^{(0)} = \dfrac{c_0}{\sqrt{(l^2+\rho^2+r^2)^2-4\rho^2l^2}} + \dfrac{c_1\,l^2(\rho^2-r^2-l^2)}{\left[(l^2+\rho^2+r^2)^2-4\rho^2l^2\right]^{3/2}}\,.
\end{equation}

\bibliographystyle{JHEP}
\bibliography{Ref.bib}

\end{document}